\documentclass[journal]{new-aiaa}
\usepackage[utf8]{inputenc}
\usepackage{textcomp}

\usepackage{graphicx}
\usepackage{amsmath}
\usepackage[version=4]{mhchem}
\usepackage{siunitx}
\usepackage{longtable,tabularx}
\usepackage{multirow}
\usepackage{booktabs}
\usepackage[labelformat=simple]{caption,subcaption}

\usepackage{appendix}
\usepackage{bbding}
\usepackage{appendix}
\usepackage{tikz}
\usepackage[backend=biber, citestyle=numeric-comp, bibstyle=ieee, maxnames=10]{biblatex}
\usetikzlibrary{patterns,decorations.pathmorphing,positioning, arrows.meta,intersections,through,calc,backgrounds, shadows, mindmap, trees}

\title{Non-Deterministic Extension of Plasma Wind Tunnel Data Calibrated Model Predictions to Flight Conditions}

\author{Przemyslaw Rostkowski\footnote[2]{PhD, Department of Aerospace Engineering, University of Illinois at Urbana-Champaign, 302E Talbot Lab, 104 S. Wright St., Urbana, 61801 IL. E-mail: rstkwsk2@illinois.edu}}
\affil{University of Illinois at Urbana-Champaign, Urbana, IL, USA}

\author{Jeremie B.E. Meurisse\footnote[3]{Research Scientist, Thermal Protection Materials Branch, Analytical Mechanics Associates Inc. at NASA Ames Research Center. E-mail: jeremie.b.meurisse@nasa.gov}}
\affil{Analytical Mechanics Associates Inc. at NASA Ames Research Center, Moffett Field, CA, USA}

\author{Marco Panesi\footnote[4]{Professor, Corresponding Author, Department of Aerospace Engineering, University of Illinois at Urbana-Champaign, 314 Talbot Lab, 104 S. Wright St., Urbana, 61801 IL. E-mail: mpanesi@illinois.edu}\footnote[1]{Corresponding Author}}
\affil{University of Illinois at Urbana-Champaign, Urbana, IL, USA}

\begin{document}

\maketitle

\begin{abstract}

This work proposes a novel approach for non-deterministic extension of experimental data that considers structural model inadequacy for conditions other than the calibration scenario while simultaneously resolving any significant prior-data discrepancy with information extracted from flight measurements. This functionality is achieved through methodical utilization of model error emulators and Bayesian model averaging studies with available response data. The outlined approach does not require prior flight data availability and introduces straightforward mechanisms for their assimilation in future predictions. Application of the methodology is demonstrated herein by extending material performance data captured at the HyMETS facility to the MSL scenario, where the described process yields results that exhibit significantly improved capacity for predictive uncertainty quantification studies. This work also investigates limitations associated with straightforward uncertainty propagation procedures onto calibrated model predictions for the flight scenario and manages computational requirements with sensitivity analysis and surrogate modeling techniques.

\end{abstract}

\section{Introduction}
\label{sec:introduction}

\lettrine{T}hermal Protection System (TPS) architecture is a critical factor of spacecraft design in mission scenarios involving hypersonic atmospheric entries characterized by high-heating conditions \cite{anderson:2006}. When re-usability is not one of the defined mission requirements, materials like the Phenolic Impregnated Carbon Ablator (PICA) that exhibit internal decomposition and surface ablation are used to construct large-scale vehicle heat shields that attenuate conveyed thermal energy \cite{amar:2006,tran:1997,weng:phdt}. Anticipated surface heating rates spanning from upper to lower atmospheric layers that dictate the manufactured thickness of these materials are estimated through combined trajectory and aerothermal environment investigations based on Computational Fluid Dynamics (CFD) studies \cite{borner:2018,wright:2009,lebeau:2001}. The development of non-equilibrium chemistry models for use in CFD frameworks consisting of both deterministic and probabilistic approaches leveraging machine learning and classical techniques is, in particular, an active area of research in the hypersonics field \cite{mikietal:2012,panesi:2012,venturietal:2020_a,venturietal:2020_b,sharmaetal:2019,sharmaetal:2020}. Derived solutions for the flow environment surrounding the entry vehicle are then used to define necessary boundary conditions for use with material response solvers in TPS sizing studies \cite{wrightetal:2014,meurisse:2018,robinetal:2014}. 

Ground facility testing strategies offer a practical alternative to otherwise prohibitively expensive high-speed flight test studies concerning candidate TPS material performance based on CFD-predicted environments. Investigations concerning gas-surface interaction phenomena and TPS material performance are delegated to high-enthalpy plasma wind tunnels like the Hypersonic Materials Environmental Test System (HyMETS) that can sustain necessary heating conditions for long test periods \cite{diazetal:2021,splinteretal:2011, chazotetal:2015,barbanteetal:2006, barbante:2009, valetal:2021, valetal:2022}. Conducted studies primarily focus on utilizing boundary layer replication approaches like the Local Heat Transfer Simulation (LHTS) methodology, where key boundary layer parameters corresponding to the stagnation point are roughly reproduced to achieve target heat flux values on the test sample surface \cite{barbanteetal:2006,barbante:2009,kolesnikov:1993,turchi:2021, viladegutetal:2018}. Still, these facilities often cannot replicate all high-speed flow conditions simultaneously and are largely decoupled from the effects of radiation phenomena while limited to testing of small test articles immersed in static environments and flow conditions instead of the time-dependent conditions encountered in non-synthetic scenarios over the length of a trajectory.  Numerical tools are subsequently calibrated with obtained data at these facilities, and numerical output is applied beyond their respective testing envelopes to perform TPS sizing investigations and performance predictions, as well as predictive uncertainty quantification analysis \cite{robinetal:2014}. 

This extension of quantified uncertainty with ground facility data to flight scenarios is investigated in this effort in the Bayesian framework concerning the material temperature response of charring ablators. The utilized Bayesian inference approach allows for quantification of uncertainty stemming from parametric, modeling, and experimental sources that can subsequently be propagated onto the response in the predictive inference process where non-deterministic predictions for the quantity of interest are obtained \cite{tarantola:2005,siviaetal:2006,oliveretal:2015, rostkowski:2019}. A large body of existing literature, in contrast, focuses on sampling-based propagation of a priori parametric uncertainty in combination with ad-hoc approaches for estimation of effects of model inadequacy \cite{mahzarietal:2015,wrightetal:2014,wright:2007,mahzarietal:2011,rivier:2019,copelandetal:2012,turchietal:2017}.  The possible extent of current predictive uncertainty quantification methodologies, excluding any ad-hoc approaches, is compared through a series of predictive inference exercises based on applications of Bayesian inference with thermocouple data captured in the HyMETS facility and by those installed in the heatshield of the MSL entry vehicle. The quantified degree of uncertainty for material temperature response predictions that leverages MSL data is used as the reference solution in this work that is attempted to be reconstructed based solely on statistical exercises utilizing ground facility data and the prior state of knowledge. 

The observed deviations of HyMETS-based predictions for the total uncertainty from the reference solution are subsequently alleviated through a coupled procedure that addresses the extension of uncertainty stemming from material response modeling errors and the degree of applicability of calibration scenario data given limitations of plasma wind tunnel complexes. A methodology is herein proposed for non-deterministic prediction of material performance at flight conditions with a novel approach that combines generic models for structural model inadequacy and model averaging methodologies. The former utilizes statistical representation for the predicted residual that is grounded on an augmentation of the Kennedy and O'Hagan \cite{kennedyetal:2001} model that is outlined by Xu and Ballocchi \cite{xuetal:2015, xuetal:2017} in the study of groundwater models that considers all relevant input quantities, including, but not limited to, its inputs and any information that the model can provide. A similar approach was employed by Neufcourt et al. \cite{neufcourtetal:2018} in the examination of two-neutron separation energies, where a model for the residual was constructed using Gaussian Process and Bayesian Neural Network approaches with a substantial training set of known masses as a function of particle numbers and additional methodology inputs. The extension allows for quantified degree of uncertainty due to modeling inadequacy to be applied for scenarios no longer limited to the environments from which calibration data were extracted.

Furthermore, the model averaging aspect of the approach addresses situations where calibration with data from the prediction scenario and those obtained experimentally can often lead to inconsistent conclusions concerning optimal-fit parameter values. This is particularly relevant in this effort, where calibration with data obtained in plasma wind tunnel facilities yields information at odds with that obtained with MSL data. Prior-data conflicts like these have recently been addressed in the pediatrics field and clinical drug trials, where the mixture prior concept in conjunction with Bayesian inference is employed \cite{schmidletal:2014,wadswortgetal:2016,siebers:2017}. Information extracted from adult drug trial data in corresponding references was applied to the child population through dilution of adult population-based findings with information extracted from a vague (flat) distribution. The mixture prior approach in the present context is explored as a limiting case of a much broader Bayesian model averaging process \cite{hoetingetal:1999}. These capabilities are demonstrated in alleviating disparate information obtained with ground facility and flight data concerning material and inadequacy model inputs based on the applicability of the calibration scenario environment to the one encountered in flight. As is demonstrated herein through an abstract problem description, it can also be used to apply calibration data that in initial assessments may not in a straightforward application be utilized in the calibration of model input quantities. The referenced approach is here further modified to dilute information obtained through Bayesian updates based on experimental data with prior, unmodified information instead of a vague distribution that is believed to precede the availability of experimental observations. This paradigm leads to the utilization of prior information to a greater degree when experimental data are increasingly irrelevant for target conditions. As noted in the discussion, the approach does not call for the availability of flight data and allows for their implementation for future predictions through subsequent applications of Bayesian inference.

The remainder of this work adheres to the following structure. Details concerning the non-deterministic framework are introduced in Section \ref{sec:nondeterministic_framework}. The ground facility and flight scenarios utilized here are described in Section \ref{sec:scenarios}. Preliminary sensitivity analysis and surrogate modeling studies are conducted in Section \ref{sec:sa_sm}, and a discussion of obtained results with the extension framework described herein is given in Section \ref{sec:discussion}. Conclusions and a summary of efforts undertaken as part of this work are given in Section \ref{sec:conclusions}. Applications of statistical methodologies in this work are performed with the in-house developed Stochastic Modeling and Uncertainty Quantification (SMUQ) toolbox \cite{rostkowskietal:2022, rostkowski:2017,rostkowski:2019,venturi:2016}.

\section{Non-Deterministic Extension Framework}
\label{sec:nondeterministic_framework}

This section gives a general description of the novel extension approach for an abstract problem description. Details of the process, when applied to plasma wind tunnel data, are given following the introduction of the methodology. 

It is essential to recognize as part of this work that the commonly encountered ``extrapolation'' term in charring ablator literature and prediction of flight performance based on experimentally obtained data is in contrast with the definition typically attributed in works rooted in probability theory. The traditional usage of the word in charring ablator literature is found in applying extracted information from experimentally obtained measurements and models calibrated with said data to the flight scenario of interest \cite{boseetal:2009}.  However, works focusing on probability theory and applications thereof consider ``extrapolation'' in the prediction of system response outside the envelope established by previously observed conditions \cite{smith:2014}. The latter definition is here considered due to this work's focus on statistical analysis procedures. Application of calibrated models with ground facility data thus constitutes either extension of these data to flight conditions or an extrapolation of the conceptual model for scenarios characterized by physical phenomena not observed in environments generated in plasma wind tunnel facilities.

\subsection{Abstract Model Generalization}

The discussion here draws from the work performed by Oliver et al. \cite{oliveretal:2015} in the context of validating predictions of unobservable quantities. Quality predictions are made using reliable physical theory whose efficacy is not in question and is believed to generically not impose significant deficiencies in predicted quantities. These physical concepts can be written as per referenced literature as function operators yielding the following formulae
\begin{subequations}
\label{e:abstract_model}
\begin{align}
  & \mathcal{R} \left( \boldsymbol{u}, \tau; \boldsymbol{r} \right) = 0 \label{e:abstract_model_a} \\
  & \boldsymbol{y} = \mathcal{Y} \left( \boldsymbol{u}, \tau; \boldsymbol{r} \right) \label{e:abstract_model_b}  
\end{align}
\end{subequations}
where $\mathcal{R}$ is the operator expressing the physical theory like the partial differential operator for conservation of mass, momentum, and energy for viscous fluid motion. Accordingly, $\boldsymbol{u} \in \mathbb{R}^N$ is the state variable vector solution, and $\boldsymbol{r}\in \mathbb{R}^N_r$ is a set of scenario parameters. The quantity $\tau$ closes the system and allows for the solution of Equation \ref{e:abstract_model_a} and application of the operator $\mathcal{Y}$ given its inputs to determine response values and prediction for observed response $\boldsymbol{y} \in \mathbb{R}^\nu$ considering negligible observational error. However, in most scenarios, the exact relationship of $\tau$ with $\boldsymbol{u}$ and $\boldsymbol{r}$ is often unknown and is instead approximated as follows
\begin{equation}
  \tau \approx \tau_{m} \left( \boldsymbol{u}; \boldsymbol{\theta}, \boldsymbol{s} \right)  
\end{equation}
where $\tau_{m}$ is an embedded model with corresponding input parameters $\boldsymbol{\theta} \in \mathbb{R}^p$ and scenario parameters $\boldsymbol{s} \in \mathbb{R}^{N_s}$ that need not necessarily correspond to those of the core physical theory \cite{oliveretal:2015}.

An embedded model $\tau_{m}$ is typically an approximation of physical phenomena that is less reliable than the physical theory utilized in the definition of the $\mathcal{R}$ and $\mathcal{Y}$ operators. The introduced structural errors in computed quantities are also likely to be compounded by uncertainty in values of $\boldsymbol{\theta}$ parameters. In their seminal work, Kennedy and O'Hagan \cite{kennedyetal:2001} outlined a statistical model addressing parametric and modeling sources of uncertainty concerning observable quantities. The discussion reformulates the system in Equation \ref{e:abstract_model} concerning real response values as
\begin{subequations}
  \label{e:abstract_ko}
  \begin{align}
    & \mathcal{R} \left( \boldsymbol{u}, \tau_m; \boldsymbol{r} \right) = 0 \label{e:abstract_ko_a} \\
    & \boldsymbol{y} = \mathcal{Y} \left( \boldsymbol{u}, \tau_m; \boldsymbol{r} \right) + \boldsymbol{\delta}(\boldsymbol{r},\boldsymbol{\alpha}) \label{e:abstract_ko_b}  
  \end{align}
  \end{subequations}
where nomenclature set forth by Oliver et al. \cite{oliveretal:2015} is utilized to express the developed concept with the $\boldsymbol{\delta}(\boldsymbol{r},\boldsymbol{\alpha}) \in \mathbb{R}^\nu$ term capturing contribution to prediction error stemming from model discrepancy. A similar expression was derived based on these works in Venturi \cite{venturi:2016} in the context of non-equilibrium dominated hypersonic flows. Any experimental uncertainty when addressing measured response values can also be captured through an additional term $\boldsymbol{\epsilon}  \in \mathbb{R}^{\nu}$, often estimated a priori, while the structural model error term in Equation \ref{e:abstract_ko_b} introduces an additional set of hyperparameters $\boldsymbol{\alpha} \in \mathbb{R}^{p_h}$ to be determined together with inputs $\boldsymbol{\theta}$ of the embedded model. 

The preceding formulation can be extended similarly to Oliver et al. \cite{oliveretal:2015} for standard general prediction and validation scenarios. Foremost, approximation of additional quantities $\tau_1, \tau_2, \ldots$ is typically required for simulation of multi-physics phenomena for the prediction scenario. These terms are, in turn, approximated with embedded model approximations $\tau_{m,1}, \tau_{m,2}, \ldots$, where each embedded  $\tau_{m,j}$ component introduces its own set of parameters $\boldsymbol{\theta}$ to be determined given an $\boldsymbol{s}$ set of scenario parameters as well as its own degree of structural inadequacy to the general composite approach. It is also possible for an experimental scenario $i$ to diverge from the predictive one and warrant the use of alternative reliable $\mathcal{R}$ and $\mathcal{Y}$ physical theories for the $\boldsymbol{y}$ observable quantity of interest. These operators, which are more appropriate for the experiment scenario, are characterized by their state variables $\boldsymbol{u}$ and scenario parameters $\boldsymbol{r}$ in addition to a set of modeled quantities that necessarily include at least one of the $\tau_j$ terms in the utilized subset $\{\tau\}^i$  of prediction environment quantities $\tau_{1}, \tau_{2}, \ldots$ to be relevant in the study. Still, it is likely that the alternative theory also necessitates the inclusion of experiment-specific embedded models $\sigma_{m}$ for corresponding quantities that are required for closure of the system. These models further introduce an additional set of model parameters to be determined and consider an additional set of scenario parameter inputs \cite{oliveretal:2015}. In turn, scenario-specific model inadequacy term must also be considered following the defined statistical model, each of which introduces hyperparameters to be determined. These generalizations force the system in Equation \ref{e:abstract_ko} to be rewritten as
\begin{subequations}
  \label{e:generalized_ko}
  \begin{align}
    & \mathcal{R}^i \left( \boldsymbol{u}^i, \{\tau_m\}^i, \{\sigma_m\}^i; \boldsymbol{r}^i \right) = 0 \\
    & \boldsymbol{y}^i = \mathcal{Y}^i \left( \boldsymbol{u}, \{\tau_m\}^i, \{\sigma_m\}^i; \boldsymbol{r}^i \right) + \boldsymbol{\delta}_{\boldsymbol{y}^i}(\boldsymbol{r}^i,\boldsymbol{\alpha}^i)
  \end{align}
\end{subequations}
where impacts of varied experimental scenarios, relevant theory operators, and multiple component systems are acknowledged. The general specification of the state variable quantity solution of the system $\boldsymbol{u}$ may require definition of additional operators, in general, to translate state vector solutions to appropriate input quantities of embedded models \cite{oliveretal:2015}.

It is important to note that the above formulation does not prohibit the use of embedded uncertainty models $\epsilon_{m}$ and $\delta_{m}$ for embedded models $\tau_{m}$ and $\sigma_m$ detailed in the discussion by Oliver et al. \cite{oliveretal:2015} for use with unobservable quantities with observable response data. Indeed, if embedded uncertainty models are sufficient to capture modeling inadequacy, then the importance of the discrete model uncertainty term $\boldsymbol{\delta}$ should diminish, and eventually, its contribution deemed negligible. As specified by the authors, the formulation of these error models is highly model-dependent but can take the form of intrusive additive models or black-box approaches, e.g., uncertainty in input parameter distributions for embedded model $\tau_{m}$ is exaggerated to overcompensate for model inadequacy. Interested readers are directed to the referenced work for further discussion on validating predictions for observable and unobservable quantities.

\subsection{Model Discrepancy Emulator}

Proper quantification of uncertainty in numerically obtained predictions requires suitable treatment of model inadequacy in addition to parametric uncertainty and variability. In their seminal work, Kennedy and O'Hagan \cite{kennedyetal:2001} approached the calibration and prediction uncertainty analysis problem through a Bayesian calibration framework that allowed for quantification of uncertainty due to modeling inadequacy. The generalized situation was formulated in the system expressed by Equation \ref{e:generalized_ko}. However, the corresponding definition of the statistical model stipulates that the model inadequacy term is specific to the scenario and only applied to a particular system defined by physical theory and embedded models.

The preceding formulation for the model residual was extended by Xu and Valocchi \cite{xuetal:2015, xuetal:2017} to include additional inputs. The relationship between model output and observables in the corresponding references was rewritten, following here the notation set forth by Oliver et al. \cite{oliveretal:2015}, as
\begin{equation}
  \label{e:ko_extension}
  \boldsymbol{y}^i = \mathcal{Y}^i \left( \boldsymbol{u}, \{\tau_m\}^i, \{\sigma_m\}^i; \boldsymbol{r}^i \right) + \boldsymbol{\delta}\left( \boldsymbol{\beta}, \boldsymbol{\phi} \right)
\end{equation}
where restrictions on the inputs of the model inadequacy term are relaxed, and dependence on hyperparameters $\boldsymbol{\phi} \in \mathbb{R}^{p_{h'}}$ and general inputs $\boldsymbol{\beta}$ is established. The inputs of the model inadequacy term can include the physical model output of the operator $\mathcal{Y}$ and other relevant knowledge in contrast to the Kennedy and O'Hagan \cite{kennedyetal:2001} formulation. Given the scope of possible information to be passed under this paradigm to the model inadequacy term, it is then possible to write a general model emulator term $\boldsymbol{\delta}\left( \boldsymbol{\beta}, \boldsymbol{\phi} \right) \in \mathbb{R}^\nu$ that considers the physics models being utilized, embedded models, the response quantity, and the scenario parameters \cite{xuetal:2015, xuetal:2017}. While the challenge of specifying such a model is not to be understated, the difficulty is significantly diminished once the task is simplified to consideration of identical operators $\mathcal{R}$ and $\mathcal{Y}$ between scenarios being analyzed where only scenario parameters are possibly allowed to vary.

The formulation of the general inadequacy model based on the preceding arguments can consist of physical and statistical formulations, where inclusion of physical theory in the inadequacy term can yield better, more consistent extrapolative performance. Alternatively, the inadequacy model can be dominated by statistical reasoning and can take on the form of Gaussian processes (GP), among others. This approach was taken by Nefcourt et al. \cite{neufcourtetal:2018}, where both GP and Bayesian Neural Network (BNN) approaches were utilized in the extrapolation of model predictions beyond various training sets in the context of two-neutron separation energies. A statistical model is utilized similarly in this work to model the discrepancy term and generalize its use among various available material response frameworks. However, caution must be exercised when utilizing these statistical formulations in predictive exercises, especially with low-quantity data sets. The performance of statistical emulators like Gaussian processes is well known to deteriorate rapidly at significant departures from the data set and should ideally be reserved for interpolative predictions. Thus, when using these approaches, it should ideally be shown that predictions are largely interpolative. Obtained flight data can be used to further extend the envelope of the training set, even if somewhat sparse in between.

\subsection{Calibration of Model Quantities}

Embedded prediction models and the model discrepancy term introduce parameters $\boldsymbol{\theta}$ and $\boldsymbol{\phi}$ that must be specified. Departing from the preceding nomenclature, consider a forward model mapping operator $f (\boldsymbol{\theta}, \boldsymbol{\mathcal{X}})$ to $\boldsymbol{\upsilon} \in \mathbb{R}^\nu$ for which specification of model inputs and combined independent time and space coordinates $\boldsymbol{\mathcal{X}}$ is made to obtain predictions for the response of the system. The observed response of this system $\boldsymbol{y}$ is a realization of a random vector variable $\boldsymbol{Y}$ with multiple sets able to be combined into an overall set $\mathbb{Y}$ of observations. Likewise, departing from the deterministic point of view, inputs of the material response model and the model inadequacy term can also be recognized as corresponding realizations of $\boldsymbol{\Theta}$ and $\boldsymbol{\Phi}$ random vector variables, while the model inadequacy contribution can be considered to be a realization of a statistical random process. Given a formulation for the relation between model output and actual response in Equation \ref{e:ko_extension}, where all sources of uncertainty are now given probabilistic descriptions, the Bayesian inference algorithm provides a suitable framework for a non-deterministic calibration procedure. The solution of the posed statistical inverse problem can subsequently be utilized to solve statistical forward problems that account for total quantified uncertainty \cite{tarantola:2005,kennedyetal:2001,smith:2014}.

Bayesian inference analysis uses Bayes' theorem to update the initial state of information given newly obtained evidence and, through the process, perform model inversion for the unknown parameters given observed data. The terms in the following expression form the basis of the procedure in the present context
\begin{equation}
  \label{e:bayes_theorem}
  \mathcal{P}( \boldsymbol{\theta}, \boldsymbol{\phi} \vert \boldsymbol{y}) = \frac{\mathcal{L}(\boldsymbol{\theta}, \boldsymbol{\phi};\boldsymbol{y} ) \mathcal{P}( \boldsymbol{\theta}, \boldsymbol{\phi} )}{\mathcal{P}(\boldsymbol{y} )},
\end{equation} 
where the prior $\mathcal{P}( \boldsymbol{\theta}, \boldsymbol{\phi} )$ is a joint distribution for inputs under calibration and a statement of current beliefs concerning their respective values. In contrast, the likelihood function $\mathcal{L}(\boldsymbol{\theta}, \boldsymbol{\phi};\boldsymbol{y} )$, also expressed in terms of the likelihood $\mathcal{P}(\boldsymbol{y} \vert \boldsymbol{\theta}, \boldsymbol{\phi} )$, updates the current state of knowledge given new evidence and encodes prior assumptions concerning model response and data error. The present work's likelihood formulation corresponds to the sum of the generalized model inadequacy emulator term and the experimental contribution to the error. The marginal likelihood term $\mathcal{P}(\boldsymbol{y})$ in Equation \ref{e:bayes_theorem} normalizes the expression for the posterior $\mathcal{P}( \boldsymbol{\theta}, \boldsymbol{\phi} \vert \boldsymbol{y})$ to form a proper joint multivariate distribution over uncertain parameters that reflects the updated state of knowledge given new evidence. The obtained posterior distribution is typically utilized as a part of an additional predictive inference step that yields probability 
\begin{equation}
  \label{e:posterior_predictive}
  \mathcal{P}(\widetilde{\boldsymbol{y}} \vert \boldsymbol{y}) = \int \mathcal{L}(\boldsymbol{\theta}, \boldsymbol{\phi};\widetilde{\boldsymbol{y}}) \mathcal{P}( \boldsymbol{\theta}, \boldsymbol{\phi} \vert \boldsymbol{y} ) d\left( \boldsymbol{\theta}, \boldsymbol{\phi} \right)
\end{equation}
corresponding to the probability of new observations $\widetilde{\boldsymbol{y}}$ conditioned on the evidence used in the Bayesian inference procedure. The predictive probability distribution $\mathcal{P}(\widetilde{\boldsymbol{y}} \vert \boldsymbol{y})$ can further be used to make posterior predictive checks to validate the calibration procedure results and the calibrated model's predictive capabilities \cite{gelmanetal:1995}. While as discussed by Sivia and Skilling \cite{siviaskilling:2006} the above formulations are further conditioned on $\mathcal{I}$ corresponding to available background information, the contribution of this term is not made explicit here for brevity and to reduce the complexity of utilized notation throughout the document.

Caution though must be exercised when interpreting the results of the Bayesian inference methodology, and of calibration procedures in general. Kennedy and O'Hagan \cite{kennedyetal:2001} noted that the joint posterior distribution of model quantities obtained through the non-deterministic model inversion process yields the best-fitting values for calibration data. However, these calibrated values are highly dependent on the assumed error structure and do not necessarily correspond to these parameters' actual physical values due to the structural inadequacy of embedded models. These behaviors can lead to a situation where given identical models but different calibration data, the Bayesian inference procedure can yield posterior distributions for the parameters that are not congruent when considering data observed between experiments individually. 

\subsection{Model Averaging For Prediction}

Several data sets may be available to determine uncertain model inputs, including captured data from prediction scenarios. A Bayesian inversion procedure can be carried out with these data sets individually to arrive at $n_e$ posterior distributions for the uncertain parameters. In addition to the parameter prior before any Bayesian updates that comprises the model $\mathcal{M}_{\mathrm{up}}$, each obtained posterior distribution can then be used to define the prior distribution for a new prediction scenario and subsequently a competing candidate model $\mathcal{M}_i$. The corresponding quantity $p \left( \mathcal{M}_j \vert\boldsymbol{y} \right)$ for a particular model $\mathcal{M}_j$ can be obtained by rearranging the Bayesian inference formulation concerning the candidate model
\begin{equation}
  \label{e:model_posterior}
  \mathcal{P}(\mathcal{M}_j \vert \boldsymbol{y}) = \frac{\mathcal{P}(\boldsymbol{y} \vert \mathcal{M}_j)\mathcal{P}(\mathcal{M}_j )}{\sum_{i=1}^{n_e} \mathcal{P}(\boldsymbol{y} \vert \mathcal{M}_i) + \mathcal{P}(\boldsymbol{y} \vert \mathcal{M}_{\mathrm{up}}) } = \frac{\mathcal{P}(\boldsymbol{y} \vert \mathcal{M}_j)\mathcal{P}(\mathcal{M}_j )}{\mathcal{P}(\boldsymbol{y}) }
\end{equation}
where $p \left( \mathcal{M}_j  \right)$ specifies the prior probability for model $\mathcal{M}_j$ considering the complete set $\mathbb{M}$ of total $n_\mathbb{M} = n_e+1$ of candidate models. The dependence of the posterior based on the model $\mathcal{M}_j$ can likewise be made evident by rewriting Equation \ref{e:bayes_theorem} as 
\begin{equation}
  \label{e:model_bayes_theorem}
  \mathcal{P} \left( \boldsymbol{\theta}, \boldsymbol{\alpha} \vert \boldsymbol{y} , \mathcal{M}_j\right) = \frac{\mathcal{P} \left( \boldsymbol{y} \vert \boldsymbol{\theta}, \boldsymbol{\alpha}, \mathcal{M}_j \right) \mathcal{P} \left( \boldsymbol{\theta}, \boldsymbol{\alpha} \vert \mathcal{M}_j \right)}{\mathcal{P} \left( \boldsymbol{y} \vert \mathcal{M}_j \right)} 
\end{equation}
where the likelihood in Equation \ref{e:model_posterior} can be seen to correspond to the evidence term in the inference of model inputs. Given these formulations, model selection in the Bayesian framework is performed with the Bayes factor
\begin{equation}
  BF_{jk} = \frac{\mathcal{P}(\boldsymbol{y} \vert \mathcal{M}_j )}{\mathcal{P}(\boldsymbol{y} \vert \mathcal{M}_k )} = \frac{\mathcal{P}(\mathcal{M}_j \vert \boldsymbol{y} )}{\mathcal{P}(\mathcal{M}_k \vert \boldsymbol{y})} \frac{\mathcal{P}(\mathcal{M}_k )}{\mathcal{P}(\mathcal{M}_j )}
\end{equation}
that indicates the relative probability of observed data considering the two models, $\mathcal{M}_j$ and $\mathcal{M}_k$, involved in the comparison. The expression is further simplified if the same prior probability is assigned to each model and reduces to be equivalent to the posterior odds. However, selecting a single model leads to utilizing updates based on data from a single experiment and overconfidence in a single model, and it is expected that no single model can explain all trends of model residual in comparison to actual responses.

Bayesian Model Averaging (BMA) provides an alternative to model selection whereby all candidate models are taken into account based on their respective weights, whether their priors or data-updated values. As noted by Hoeting et al. \cite{hoetingetal:1999}, averaging all candidate models also provides better average predictive capabilities. A similar approach has been recently used in application of clinical drug trial data obtained with adult sampling population in the pediatrics field by prescribing a two-component prior mixture to address the degree of prior data's applicability to the target underage population \cite{mikolaidisetal:2021,siebers:2017,wadswortgetal:2016,schmidletal:2014,roveretal:2019}. Though in referenced literature, vague priors were used to dilute the impact of the informative component, unlike in the current effort. The approach in the present context consists of conducting a weighted summation of parameter posteriors obtained with individual experimental data along with the prior distribution preceding the Bayesian inference update. This, in turn, yields a BMA prior for the prediction scenario
\begin{equation}
  \label{e:bma_prior}
  \mathcal{P}( \boldsymbol{\theta}, \boldsymbol{\phi} \vert \mathbb{Y}^{\mathrm{exp}}, \mathbb{M} ) = \sum_{i=1}^{n_e} \mathcal{P}(\mathcal{M}_i ) \mathcal{P}(\boldsymbol{\theta}, \boldsymbol{\phi} \vert \boldsymbol{y}^i, \mathcal{M}_i) + \mathcal{P}(\mathcal{M}_{\mathrm{up}} ) \mathcal{P}( \boldsymbol{\theta}, \boldsymbol{\phi} \vert \mathcal{M}_{\mathrm{up}} )
\end{equation}
with weights equated to the prior probability of each model and $\mathbb{Y}^{\mathrm{exp}}$ corresponding to the set of all experimentally captured data. Dependence on the total model collection $\mathbb{M}$ explicitly distinguishes the resulting distribution from the posterior obtained through a single Bayesian inference application with all experimental data simultaneously. A fundamental property of this formulation is in the inclusion of the prior before Bayesian updates for the model parameters as an additional model. While a wealth of data can be obtained through many experiments, these data can show little relevance to the prediction scenario based on scenario parameter differences. The overall prior for the prediction scenario in such cases should significantly be influenced by prior knowledge encoded by the researchers rather than rely on non-compatible data or impose a false state of no information. The degree of compatibility of experimental data and the information extracted from them is indicated by the value of $1-\mathcal{P}(\mathcal{M}_{\mathrm{up}} )$ or equivalently by the sum of $\mathcal{P}(\mathcal{M}_{i} )$ weights over experimentally determined Bayesian models. Predictions with the prior formulation and the general model inadequacy emulator can be applied to the prediction scenario by obtaining the respective prior predictive distribution
\begin{equation}
  \mathcal{P}(\widetilde{\boldsymbol{y}} \vert \mathbb{Y}^{\mathrm{exp}}, \mathbb{M}) = \int \mathcal{P}(\widetilde{\boldsymbol{y}} \vert \boldsymbol{\theta}, \boldsymbol{\phi}) \mathcal{P}( \boldsymbol{\theta}, \boldsymbol{\phi} \vert \mathbb{Y}^{\mathrm{exp}}, \mathbb{M} ) d\left( \boldsymbol{\theta}, \boldsymbol{\phi} \right)
\end{equation}
for the response quantity. A Bayesian inference procedure can again be carried out once captured prediction scenario data are available to obtain the posterior distribution with respect to the BMA prior and obtain updated weights of each model. The posterior is again a weighted sum of posteriors under each considered model as follows
\begin{equation}
  \mathcal{P}( \boldsymbol{\theta}, \boldsymbol{\phi} \vert \boldsymbol{y}, \mathbb{Y}^{\mathrm{exp}}, \mathbb{M} ) = \sum_{i=1}^{n_e} \mathcal{P}(\mathcal{M}_i \vert \boldsymbol{y}) \mathcal{P}(\boldsymbol{\theta}, \boldsymbol{\phi} \vert \boldsymbol{y}, \boldsymbol{y}^i, \mathcal{M}_i ) + \mathcal{P}(\mathcal{M}_{\mathrm{up}} \vert \boldsymbol{y}) \mathcal{P}( \boldsymbol{\theta}, \boldsymbol{\phi} \vert \boldsymbol{y}, \mathcal{M}_{\mathrm{up}} )
\end{equation}
where $\mathcal{P}(\mathcal{M}_i \vert \boldsymbol{y})$ are updated model probabilities, weights enabling computation of the posterior predictive distribution
\begin{equation}
  \mathcal{P}(\widetilde{\boldsymbol{y}} \vert \boldsymbol{y}, \mathbb{Y}^{\mathrm{exp}}, \mathbb{M}) = \int \mathcal{P}(\widetilde{\boldsymbol{y}} \vert \boldsymbol{\theta}, \boldsymbol{\phi}) \mathcal{P}( \boldsymbol{\theta}, \boldsymbol{\phi} \vert \boldsymbol{y}, \mathbb{Y}^{\mathrm{exp}}, \mathbb{M} ) d\left( \boldsymbol{\theta}, \boldsymbol{\phi} \right)
\end{equation}
for the scenario. Expanding the formulations for the prior and the posterior reveals that both prior and posterior predictive distributions are also weighted sums of respective distributions under each model. Additional flight data can refine the model discrepancy term emulator and inform the model probabilities or become included in the BMA prior for other scenarios.

While the model set $\mathbb{M}$ by design requires the presence of the non-updated prior, the remaining models under consideration can be adjusted \cite{hoetingetal:1999}. One possibility is to include Bayesian models based on calibration with data from multiple experiments under similar conditions. Calibration with multiple experiment data simultaneously in a single analysis though can suffer from effects where small residuals for any particular scenario dominate the trends set forth by the likelihood term and reduce the effectiveness of model averaging in comparison. One can otherwise entirely neglect all experiments other than the one most similar to the prediction scenario, albeit this would somewhat hinder the robustness aspect of the BMA result and limit information gained corresponding to embedded models. Alternatively, a prior selection can be made for the members of set $\mathbb{M}$ by considering those most robust when applied to other experiments. This process can be carried out by computing appropriate Bayes factors and retaining models that consistently perform well in comparison. Prior model probabilities can likewise be determined with those results or purely based on a priori knowledge. Different approaches to pruning the model set $\mathbb{M}$ can be found in Hoeting et al. \cite{hoetingetal:1999} whose discussion addresses computational complexity of the procedure when the number of models under consideration is not negligible.

\subsection{Application to a Charring Ablator Scenario}

The framework described herein is applied to the problem of extension of material performance data captured in plasma wind tunnel facilities to flight conditions. In particular, captured thermocouple temperature profiles for an experiment scenario at the HyMETS facility at multiple locations within the material are utilized to calibrate embedded material model parameters and the model inadequacy term hyperparameters. The calibration results are then applied in the non-deterministic prediction of the MSL heat shield temperature response for thermocouple locations within the MISP-4 plug. A brief description of the physical theory and associated closure embedded models implemented as part of NASA's Porous material Analysis Toolbox based on OpenFOAM (PATO) \cite{lachaud:2017} platform that is used to simulate the response of the charring ablator in this work is given in the appendix. 

Formulation of the likelihood in Equation \ref{e:bayes_theorem} is open-ended and implements a priori assumptions concerning model output relationship with observed data. A multiplicative form for the model error is prescribed in the present effort, motivated by the increased complexity of the material response problem and the relevance of embedded models and their inadequacy at elevated temperatures. It is expected at low temperatures that the primary mode of heat transfer is through conduction in the solid medium, but pyrolysis and gas transport phenomena, as well as simplifying assumptions within the model, become more relevant with rising heat loads; hence contribution due to model inadequacy is assumed to scale with material temperature. Likewise, the manufacturer stated thermocouple accuracies also often specify multiplicative error bounds past temperatures commonly encountered during plasma wind tunnel tests and even relatively mild-speed hypersonic entries. The multiplicative error assumption, utilizing the ideas of Kennedy and O'Hagan \cite{kennedyetal:2001} concerning model inadequacy, is formulated as an additive log-error 
\begin{equation}
  \boldsymbol{y}^\ast = f^\ast (\boldsymbol{\theta}, \boldsymbol{\mathcal{X}}) + \boldsymbol{\delta} \left( \boldsymbol{\beta}, \boldsymbol{\phi} \right) + \boldsymbol{\epsilon}
\end{equation}
where in the preceding deterministic statement natural log transformations of system observations and model output realizations are considered \cite{rostkowski:2019,venturietal:2020_a, yodungetal:2013}. Both model inadequacy and observational error terms are here assumed to be distributed according to zero-mean Gaussian formulations defined by white-noise kernels and corresponding $\sigma_{\boldsymbol{\delta}}$ and $\sigma_{\boldsymbol{\epsilon}}$ uncertain hyperparameters. However, the two are not separable in this form, and the contribution due to experimental factors is expected to be negligible compared to the inadequacy of the model. The sum of the contribution of the two sources of uncertainty are thus combined into a single term governed by a single hyperparameter $\sigma_{\mathcal{L}}$ to ameliorate these concerns. It is important to also consider that material response at various depths is subject to differing degrees of modeling complexity. Sections of the material near the heated surface are more likely to be affected by inadequacy of the assumptions concerning surface balance equations and decomposition phenomena than sections of the charring ablator near the bond line. It is prudent for these reasons to consider model inadequacy at different depths by prescribing separate hyperparameters $\sigma_{\mathcal{L}, \mathrm{TC} i}$ per each simulated thermocouple of the total $n_\mathrm{TC}$ thermocouple profiles included in the inference procedure.

These assumptions, in sum, lead to the following compact formulation of the likelihood in statistical inference exercises
\begin{equation}
\label{e:gaussian_likelihood}
\mathcal{L}(\boldsymbol{\theta};\boldsymbol{y},\mathcal{M}) = \prod_{i=1}^{n_\mathrm{TC}} \Big[ (2 \pi \sigma_\mathcal{L}^{2})^{-0.5\nu} \times \exp (- 0.5 (\boldsymbol{y}^\ast-f^\ast (\boldsymbol{\theta}, \boldsymbol{\mathcal{X}}) )^\mathrm{T} \sigma^{-1}_\mathcal{L} (\boldsymbol{y}^\ast-f^\ast (\boldsymbol{\theta}, \boldsymbol{\mathcal{X}}) )) \Big]_{\mathrm{TC} i}
\end{equation}
where the $\mathrm{TC}$ subscript denotes the thermocouple specific operation. A similar formulation based on logarithmic values of the response and data were also utilized by Miki et al. \cite{mikietal:2011,mikietal:2012} and Panesi et al. \cite{panesi:2012} in the realm of shock tube experiments. The values of $\sigma_{\mathcal{L}, \mathrm{TC} i}$ parameters of the likelihood are treated as additional uncertain inputs, the calibration of which enables quantification of uncertainty due to modeling and experimental sources. The complete likelihood function for both MSL and HyMETS calibration exercises is formed by taking a product of likelihood partitions defined per each simulated thermocouple profile in a single code evaluation. The preceding likelihood definition is based on the scenario-specific statistical model due to the scenario and thermocouple location-specific definition of $\sigma_\mathcal{L}$ parameters, and, therefore, calibrated distributions for these parameters cannot be reused in a straightforward manner for different experimental or flight scenarios.

The model inadequacy emulator approach is employed in this work to amend this limitation and allow for extension of quantified uncertainty due to shortcomings of the numerical model for flight scenarios. Existing literature has shown that Gaussian processes provide a suitably versatile framework for statistical treatments of the error model that is in line with the objectives of Bayesian calibration methodology \cite{kennedyetal:2001,xuetal:2015,neufcourtetal:2018,rasmussenetal:2006}. However, unlike past referenced efforts where large sets of data are often available to train the Gaussian process, the current effort is limited to considering a single ground facility test case. The Gaussian process formulation is consequently limited to a zero-mean process using a white noise kernel
\begin{equation}
\Sigma_{ij} = \sigma_\mathrm{em}^2 \delta_{ij}
\end{equation}
considering a multiplicative error assumption. A single $\sigma_\mathrm{em}$ parameter is introduced into the calibration exercise, in place of multiple depth-specific hyperparameters, such that $\sigma_\mathrm{em} = \sigma_{\mathcal{L}, TC i}$ when considering multiple thermocouple profiles. The simplicity of the utilized formulation for the statistical model minimizes overfitting effects during the calibration process. The contribution due to experimental uncertainty commonly prescribed to follow the Gaussian model $\mathrm{N}(0,\sigma_\epsilon)$ is again assumed to be negligible compared to the magnitude of $\sigma_\mathrm{em}$ for simplicity in constructing the general likelihood. Calibration of the stochastic process is also limited to the simultaneous tuning of kernel hyperparameters with material property inputs during the inference procedure, as was done by Miki et al. \cite{mikietal:2011} in the context of Nitrogen ionization reaction rates. The statistical model can be extended in the future to consider additional quantities like environmental conditions, estimated degree of char, and temporal correlation in solution error structure, among others, once more testing campaign data are made available. 

The general BMA prior formulation in Equation \ref{e:bma_prior} is further given a description specific to the current effort. Given that a single experiment is considered here, the BMA prior can be expressed as
\begin{equation}
  \label{e:mixing_formula}
  \mathcal{P} ( \boldsymbol{\theta}, \boldsymbol{\phi} | \boldsymbol{y}_\mathrm{HyMETS}, \mathbb{M}  ) = w \mathcal{P} ( \boldsymbol{\theta}, \boldsymbol{\phi} | \boldsymbol{y}_\mathrm{HyMETS},  \mathcal{M}_\mathrm{HyMETS} ) + ( 1-w ) \mathcal{P} ( \boldsymbol{\theta}, \boldsymbol{\phi} \vert \mathcal{M}_\mathrm{up} )
\end{equation}
where the value of $w$ is prior probability $\mathcal{P} ( \mathcal{M}_\mathrm{HyMETS} )$ of the model corresponding to the posterior obtained through calibration with HyMETS data. Alternatively, the parameter $w$ can be interpreted as the degree of relevance of data captured at the HyMETS facility to the MSL scenario. The choice of $w$ in this work is determined by minimizing divergence between predictive inference results given HyMETS facility and MSL flight data using previously outlined Gaussian process formulation for the model inadequacy term. Namely, the use of both Jeffreys divergence statistic and ubiquitous KL divergence is explored, where the former is computed by summing both asymmetric variants of the latter. Because the statistical and physical models in both scenarios are identical, and given the formulation of the general model inadequacy term, the procedure aims to isolate the effects of the differences in the achieved ground facility and flight conditions on the applicability of captured experimentally determined material performance data. Values of $w$ close to unity indicate the identical state of updated information with ground facility data as flight data. In contrast, values close to zero indicate that the experiment is irrelevant given the Bayesian model, and predictions should be based on prior knowledge alone.

Target posterior distribution of the Bayesian inference exercise is challenging to obtain in practice following a brute force approach and is typically approximated through Markov Chain Monte Carlo (MCMC) sampling strategies. While variational inference approaches provide alternatives for determining the posterior for problems characterized by high-dimensional inputs, they are outside the scope of the present discussion and come with drawbacks \cite{blei:2017}. The posterior in this work is sampled using the Delayed Rejection Adaptive Method (DRAM) MCMC algorithm \cite{haarioetal:06} using 20 independent chains consisting of 500000 samples, with the first 300000 samples per chain discarded as part of the burn-in process. The remaining samples are further filtered by considering every 20th state as part of the ``clean'' chain to reduce the effects of autocorrelation between individual chain states. Following the outlined post-processing procedure of obtained Markov chains, each application of the Bayesian inference methodology in this work yields a 200000 ``clean'' sample aggregate from the posterior distribution collected from 20 independent Markov chains.

\section{Ground Facility and Flight Scenarios}
\label{sec:scenarios}

\subsection{HyMETS Arc-jet}

The arc-jet facility scenario in the present effort corresponds to the HyMETS facility test case reviewed in Diaz et al. \cite{diazetal:2021} for a PICA sample immersed in a Mars-like high-enthalpy flow. Initially constructed in 1968 at NASA Langley Research Center, the HyMETS complex is equipped with a segmented-constrictor-dc-electric-arc-heater supplied by a 400 kW power source that generates a high-voltage electric arc to heat test gas mixture flows. The super-heated gases are subsequently accelerated through a Mach 5 convergent-divergent nozzle and then exhausted into a low-pressure cylindrical test chamber to generate non-equilibrium, high-speed flows that simulate conditions encountered in flight during an atmospheric entry maneuver. The material test sample is mounted on one of the four available pneumatic arms that brings it downstream of the nozzle exit once the target environment is established. The downstream gas mixture flow is decelerated and cooled before being pumped out of the complex after leaving the test chamber \cite{splinteretal:2011}. The facility can conduct experiments with both Earth and Mars-like test gas mixtures, albeit additional Argon is injected into the flow to prevent damage to the electrodes of the arc heater.


\begin{figure}
  \centering
  \begin{subfigure}[b]{0.49\linewidth}
    \centering
    \includegraphics[width=\linewidth, trim=10cm 0cm 12cm 0cm, clip]{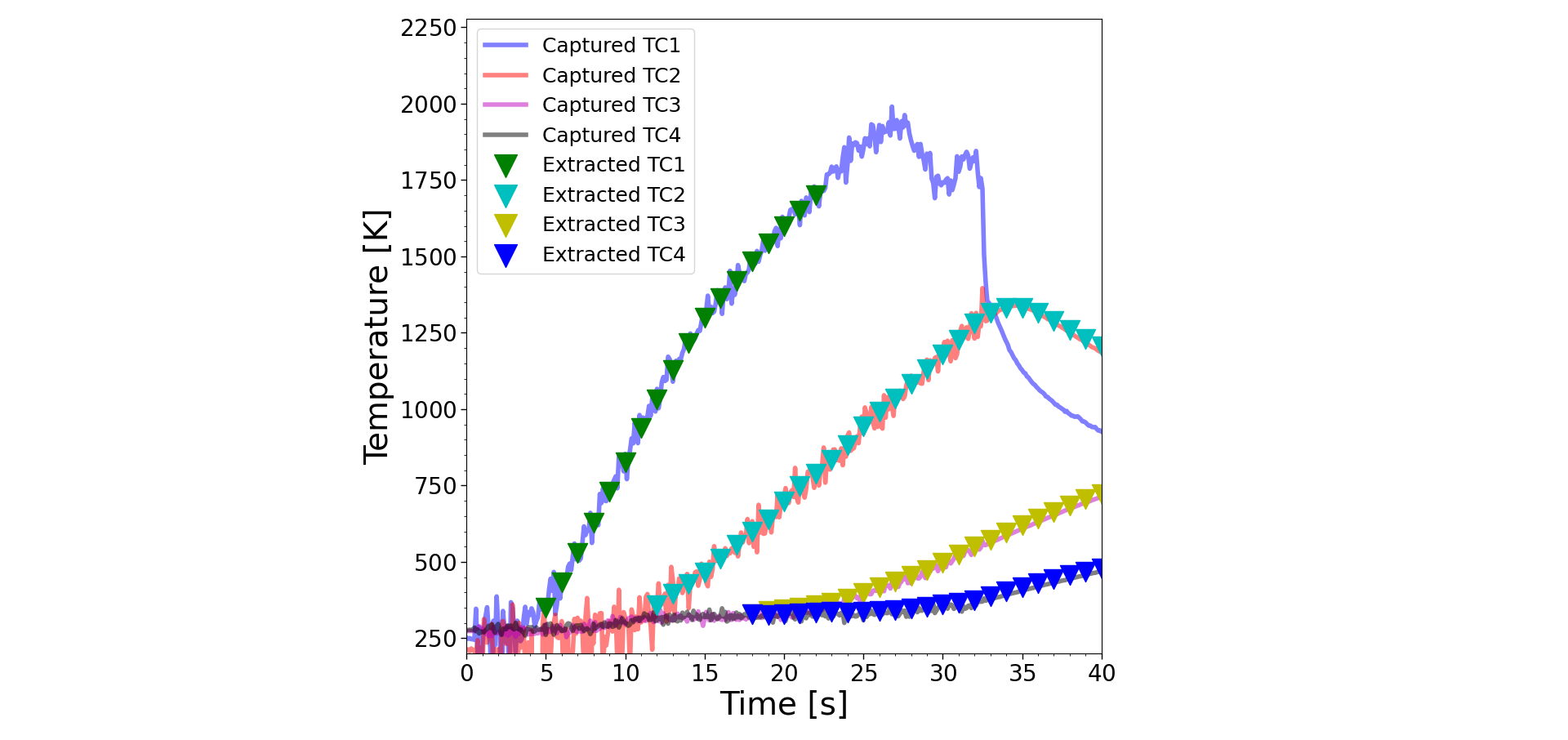}
    \caption{}
    \label{f:digitized_hymets_data}
  \end{subfigure}
  \hfill
  \begin{subfigure}[b]{0.49\linewidth}
    \centering
    \includegraphics[width=0.95\linewidth, trim=19.75cm 0cm 6.5cm 2cm, clip]{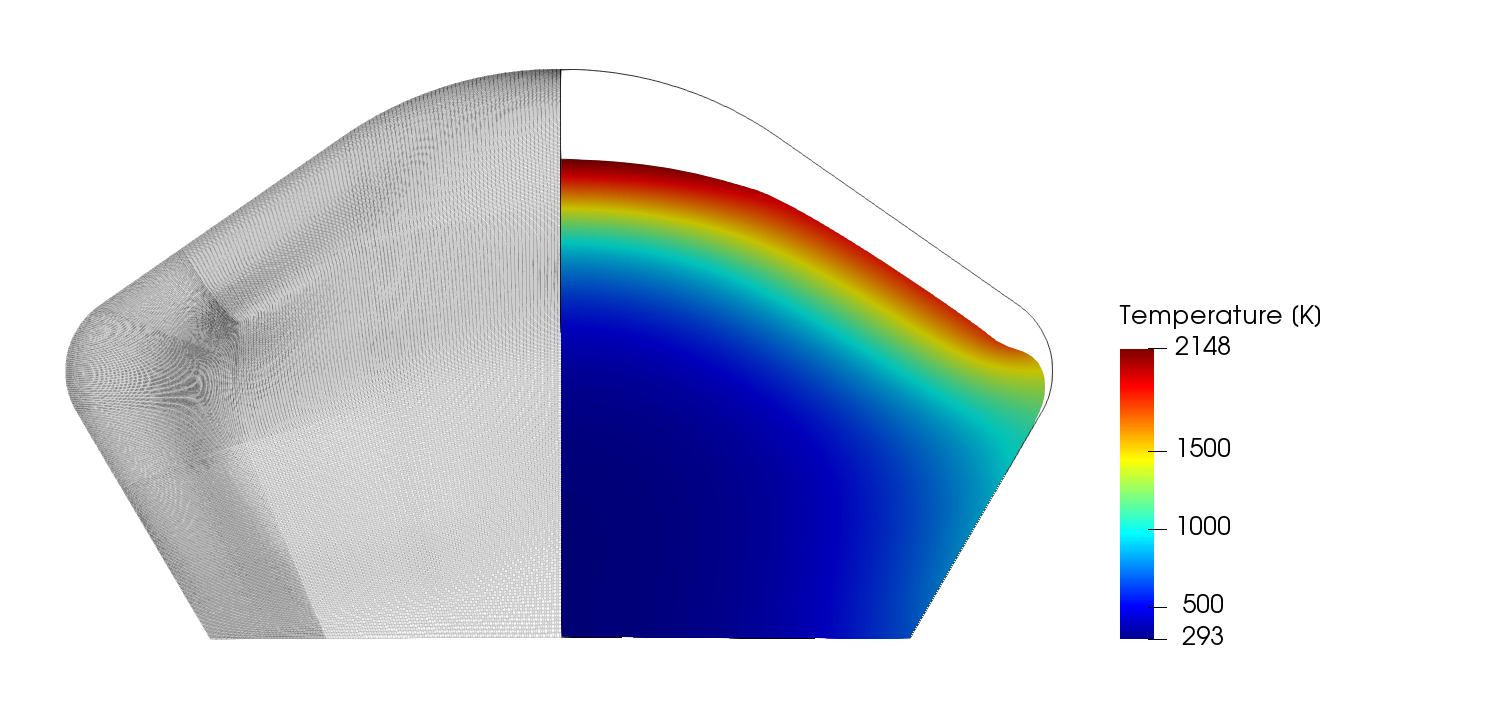}
    \vspace{0.025cm}
    \caption{}
    \label{f:babysprite_assembly}
  \end{subfigure}
  \caption{Baby-sprite captured digitized HyMETS data along with extracted values (a) and solution with nominal parameter values at t=22 s \cite{diazetal:2021}.}
  \label{f:hymets_data}
\end{figure}

The ground facility scenario under study was performed as part of the PICA-N test campaign conducted in March 2019 at the HyMETS facility. Both NuSIL-coated and non-coated PICA material samples were subjected to various heating conditions at the complex alongside numerous data acquisition approaches \cite{bessireetal:2019}. The scenario studied herein consisted of a baby-SPRITE geometry uncoated PICA material sample subjected to a heated Mars-like gas mixture with a target heat flux of 126 $\mathrm{W} / \mathrm{cm}^2$ and 5.3 kPA stagnation pressure for a 31 s heating duration. The test specimen design included four thermocouples embedded at various material depths along the axis of symmetry installed at depths indicated in the first half of Table \ref{t:tc_depths}. The raw captured temperature profiles of the material are plotted in Figure \ref{f:hymets_data}, along with the axisymmetric solution for sample temperature at time t=22s. Further information on the utilized computational grid can be found in Diaz et al. \cite{diazetal:2021}.

\begin{table}[t]
 \begin{center}
    \caption{Thermocouple Locations in MSL and HyMETS Scenarios \cite{meurisse:2018,whiteetal:2013,diazetal:2021}.}
    \label{t:tc_depths}
    \begin{tabular}{lcc}
      \toprule
      \toprule
      Scenario & Thermocouple & Distance From Substructure (mm)\\
      \midrule
      \multirow[c]{4}{*}{HyMETS}     & TC-1 & 22.02 \\
                                     & TC-2 & 17.53 \\
                                     & TC-3 & 12.13 \\
                                     & TC-4 & 7.224 \\\hline
      \multirow[c]{4}{*}{MSL MISP-4} & TC-1 & 29.28 \\
                                     & TC-2 & 26.36 \\
                                     & TC-3 & 20.43 \\
                                     & TC-4 & 13.81 \\
    \end{tabular}
  \end{center}
\end{table}

\subsection{MSL Flight}

The MSL craft successfully executed a hypersonic entry in March 2021 through the Martian atmosphere. The vehicle design utilized a 4.5 m diameter heat shield to protect the Curiosity rover payload during the entry maneuver, where speeds of up to 5.9 km/s were achieved. The vehicle heat shield was constructed of 113 individual tiles manufactured out of PICA material of uniform 31.75 mm thickness with a silicone elastomer bonding agent used to fill the remaining intermediate gaps \cite{wrightetal:2014, meurisse:2018, boseetal:2014, whiteetal:2013}. The TPS structure was designed to protect the vehicle's integrity from heating rates and total loads of up to 226 $\mathrm{W} / \mathrm{cm}^2$ and 6400 $\mathrm{J} / \mathrm{cm}^2$ throughout the entire maneuver \cite{boseetal:2014}.

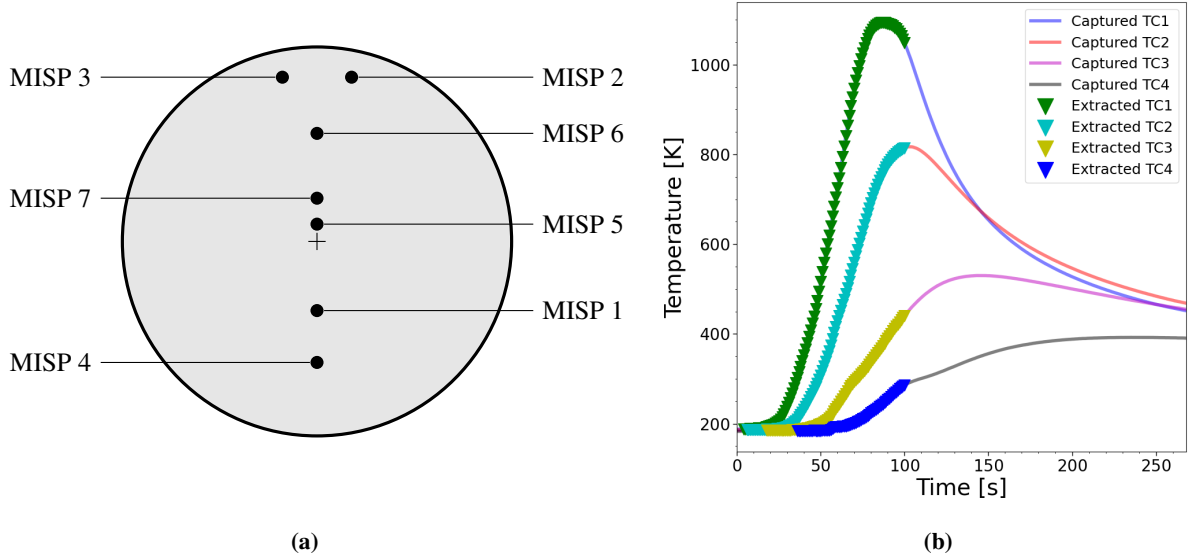
\begin{figure}
  \begin{subfigure}[b]{0.49\linewidth}
  \centering
  \begin{tikzpicture}[scale=1.15]

    \filldraw[fill=none, very thick, fill=gray!20!white] (0,0) circle [radius=2.25cm];
    
    \draw (0,-0.1) -- ++ (0,0.2);
    \draw (-0.1,0) -- ++ (0.2,0);
    
    \node[anchor=west] (M2) at (2.5, 1.9) {MISP 2};
    \node[circle, fill=black, inner sep=1.75pt] (M2tc) at (0.4,1.9) {};
    \draw (M2) -- (M2tc);
    
    \node[anchor=east] (M3) at (-2.5, 1.9) {MISP 3};
    \node[circle, fill=black, inner sep=1.75pt] (M3tc) at (-0.4,1.9) {};
    \draw (M3) -- (M3tc);
    
    \node[anchor=west] (M6) at (2.5, 1.25) {MISP 6};
    \node[circle, fill=black, inner sep=1.75pt] (M6tc) at (0,1.25) {};
    \draw (M6) -- (M6tc);
    
    \node[anchor=east] (M7) at (-2.5, 0.5) {MISP 7};
    \node[circle, fill=black, inner sep=1.75pt] (M7tc) at (0,0.5) {};
    \draw (M7) -- (M7tc);
    
    \node[anchor=west] (M5) at (2.5, 0.2) {MISP 5};
    \node[circle, fill=black, inner sep=1.75pt] (M5tc) at (0,0.2) {};
    \draw (M5) -- (M5tc);
    
    \node[anchor=east] (M4) at (-2.5, -1.4) {MISP 4};
    \node[circle, fill=black, inner sep=1.75pt] (M4tc) at (0,-1.4) {};
    \draw (M4) -- (M4tc);
    
    \node[anchor=west] (M1) at (2.5, -0.8) {MISP 1};
    \node[circle, fill=black, inner sep=1.75pt] (M1tc) at (0,-0.8) {};
    \draw (M1) -- (M1tc);
    
  \end{tikzpicture}

    \vspace{1cm}
    \caption{}
    \label{f:msl_misp_locations}
  \end{subfigure}
  \hfill
  \begin{subfigure}[b]{0.49\linewidth}
    \centering
    \includegraphics[width=\linewidth, trim=10cm 0cm 12cm 0cm, clip]{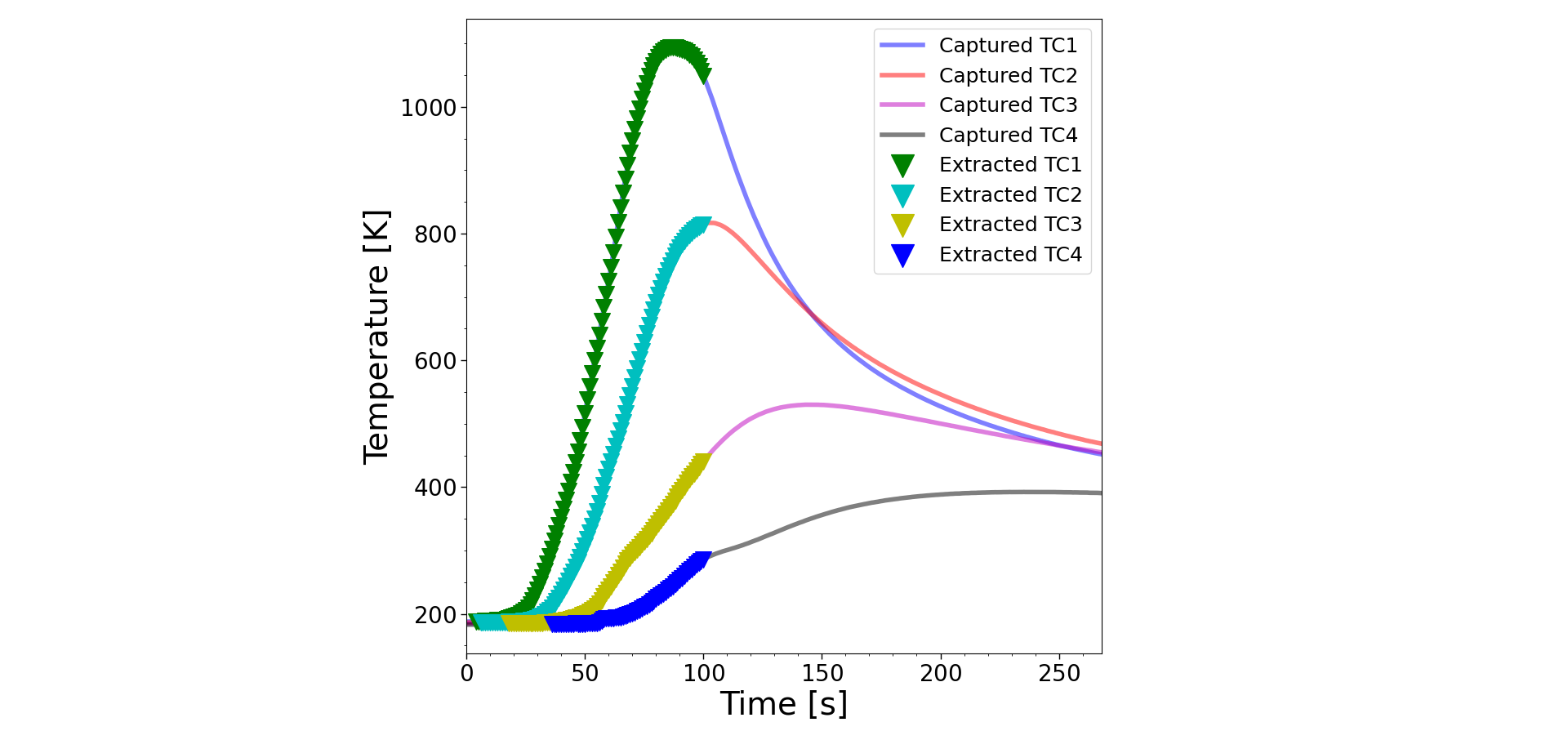}
    \caption{}
    \label{f:msl_tc_data}
  \end{subfigure}
  \caption{MISP locations on the MSL heat shield (a) and captured MISP-4 TC profiles (b) along with extracted values \cite{whiteetal:2013}.}
  \label{f:msl_data}
\end{figure}
Instrumentation collection included in the MSL heat shield design implemented the extensive Mars Science Laboratory Entry, Descent, and Landing Instrumentation (MEDLI) suite that captured critical temperature and pressure measurements during the hypersonic entry. The MEDLI Integrated Sensor Plug (MISP) system sensor array included 24 type-K thermocouples distributed between seven individual PICA plug assemblies installed at various points of interest. Measured response by foremost TCs was sensitive to changes in surface heating conditions, whereas deeper TCs were aimed at capturing the in-depth response of the charring ablator \cite{boseetal:2014}. The current discussion focuses on the measured material thermal response by MISP-4 assembly TCs located in the stagnation region of the forebody. Measured depths of each TC installation are given in the latter half of Table \ref{t:tc_depths}, along with captured material temperature profiles in Figure \ref{f:msl_data} and the location of each MISP assembly on the heat shield. Material response reconstruction at the MISP-4 location in the MSL scenario is reduced to a 1-D problem based on the findings in Rostkowski et al. \cite{rostkowski:2022} that investigated sensitivity of the response between different configurations of the MSL numerical scenario. In contrast, the HyMETS facility scenario is simulated using a 2-D axisymmetric setup due to the non-negligible influence of 3-D phenomena on material response along the stagnation line.

\section{Sensitivity Analysis and Surrogate Modeling}
\label{sec:sa_sm}

\subsection{Morris Method}

Sensitivity analysis is carried out with respect to defined uncertain parameters and material response framework prior to the application of planned calibration and uncertainty quantification treatments. In the employment of the sensitivity treatments, variation in model output quantities is apportioned qualitatively or quantitatively to varied inputs \cite{saltelli:2000b, smith:2014}. Obtained insights into how the model depends on supplied information can then be used to prune the set of uncertain parameters, which then decreases the computational complexity of planned statistical procedures and their computational cost. In the present context of charring ablators under high heating conditions, the model under study approximates numerous non-linear and interacting phenomena requiring sensitivity analysis methodologies that do not impose restrictive, simplifying assumptions concerning model complexity.

Non-influential parameters are identified here through the application of the Morris screening procedure. The methodology provides qualitative sensitivity information at a relatively low computational expense while being sufficiently versatile for use with models characterized by non-monotonic and discontinuous complexity \cite{morris:1991,campolongo:2011,rostkowski:2022}. Consider a forward model operator $f ( \boldsymbol{\theta})$ that yields scalar realizations $\upsilon \in \mathbb{R}^1$ of the random variable $\Upsilon$ that is obtained when large number of realizations of $\Theta$ are fed through the model. If it is also assumed that values of $\boldsymbol{\theta}$ fall on a $d$-level discretized hypercube $[0,1]^p_d$ with equal spacing, and components of the random variable vector follow uniform distributions, an Elementary Effect ($EE$) with respect to individual component $\theta_i$ is defined as 
\begin{equation}
\label{e:elementary_effect}
EE_i = \frac{f(\theta_1,\theta_2,\ldots,\theta_i+\Delta,\ldots,\theta_k)-f(\theta_1,\theta_2,\ldots,\theta_{i},\ldots,\theta_k)}{\Delta}
\end{equation}
considering a corresponding perturbation $\Delta$ in its value. Value of the perturbation component is furthermore obtained with
\begin{equation}
\Delta = \frac{c}{d-1}    
\label{e:delta_def}
\end{equation}
where $d$ is again related to the discretization of the unit hypercube spacing such that $\theta_i + \Delta$ rests on the discretized grid and integer $c$ controls the number of level jumps for the value of the perturbation. 

The Morris procedure is based on analyzing elementary effects distributions obtained by a sampling of individual elementary effects. Morris statistics $\mu_i$ and $\sigma_i$ are subsequently based on mean and standard deviation given $r$ number of $EE_i$ samples. The magnitude of the value of $\mu_i$ implies the importance of linear trends due to variation of the input $\theta_i$, while the value of $\sigma_i$ signifies the importance of non-linear and higher-order effects over output quantities, including interaction effects with other varied inputs. However, the importance of each statistic, and by extension of the input being varied, can only be ascertained by simultaneous analysis of the whole ensemble of obtained Morris statistic values. A graphical approach is typically taken in the post-processing stage, where points $(\mu_i, \sigma_i)$ for each input $\theta_i$ are plotted on a plane spanned by $\mu$ and $\sigma$ axes. The relative distance of plotted points from the origin is then used to determine the importance of each input where non-influential parameters are clustered about the origin. Moreover, the methodology can be utilized with general non-uniformly distributed inputs through perturbation of values in the joint quantile space. Campolongo et al.\cite{campolongo:2007} noted that analysis based on values of $\mu$ of obtained $EE$ distributions could lead to Type \textrm{II} classification errors due to cancellation effects and is argued by the authors that the analysis should instead focus on the $\mu^\ast$ statistic values based on the distribution of the absolute values of sampled elementary effects. Results corresponding to the analysis of $\left| EE_i \right|$ can be further interpreted to signify the relative importance of linear and non-linear trends based on the location of the point $(\mu^\ast_i, \sigma_i)$ away from the $\mu^\ast = \sigma$ line on the Morris plot.

The sampling of elementary effects in this work is performed using the trajectory approach detailed by Morris in his seminal work \cite{morris:1991}. While other approaches have been utilized in existing literature, the trajectory approach, in conjunction with small perturbation steps, can be applied in cases where correlation in inputs is known to exist a priori following an unknown correlation structure. The use of small perturbation steps in combination with numerous samples also aids in reducing the likelihood of Type \textrm{I} and Type \textrm{II} classification errors in the pruning of non-influential parameters. The sensitivity analysis procedure in this work is based on 500 trajectories for the MSL scenario and 100 trajectories for the HyMETS case with perturbation magnitudes $\Delta=0.01$ on the discretized unit hypercube corresponding to the joint quantile space, where the initial coordinates of each trajectory are populated with the Latin Hypercube Sampling (LHS) space-filling design.

\begin{figure}
  \centering
  \begin{subfigure}[b]{0.49\linewidth}
    \centering
    \includegraphics[width=\linewidth, trim=10cm 0cm 13cm 0cm, clip]{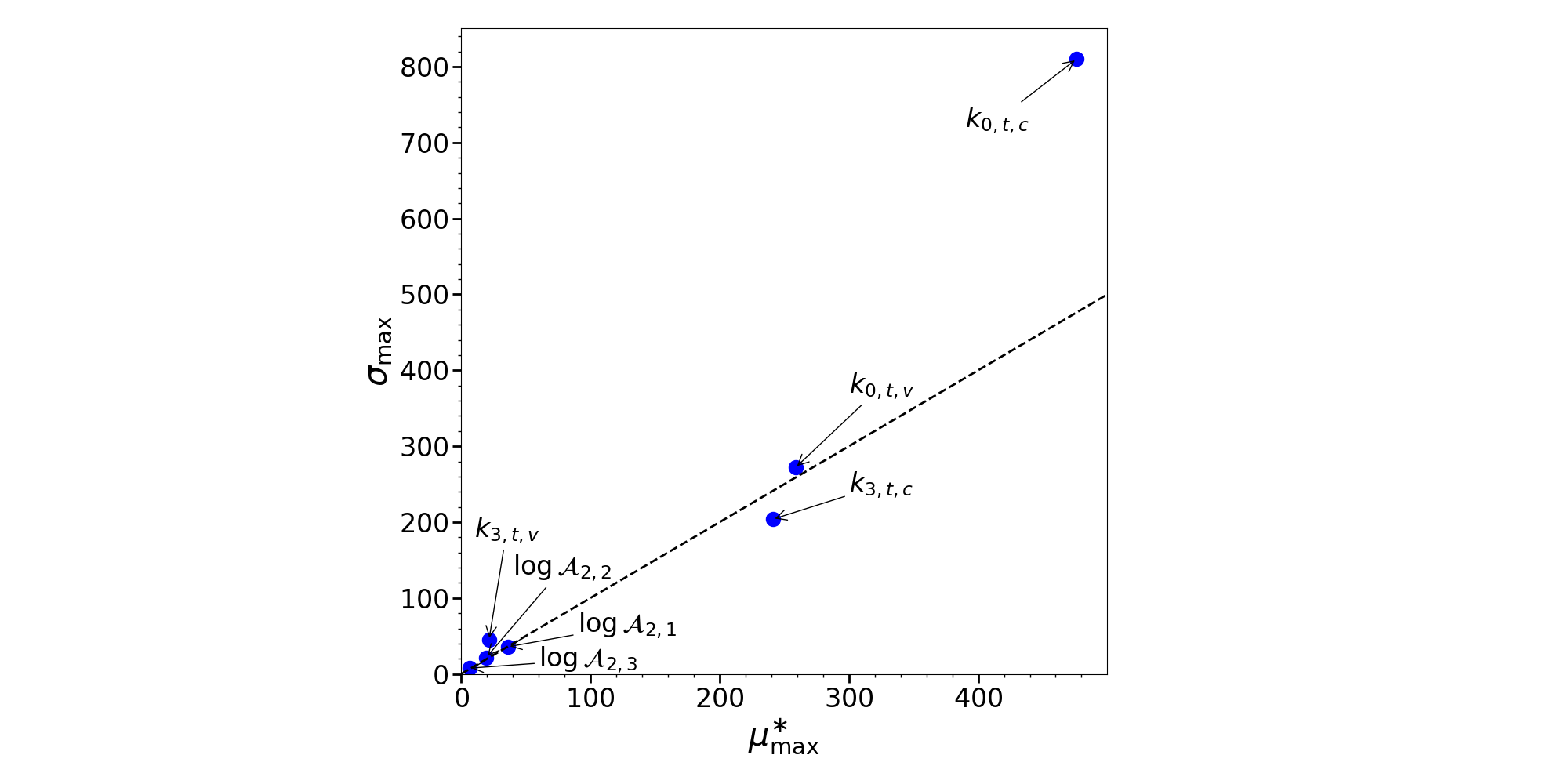}
    \caption{}
    \label{f:morris_hymets}
  \end{subfigure}
  \hfill
  \begin{subfigure}[b]{0.49\linewidth}
    \centering
    \includegraphics[width=\linewidth, trim=10cm 0cm 13cm 0cm, clip]{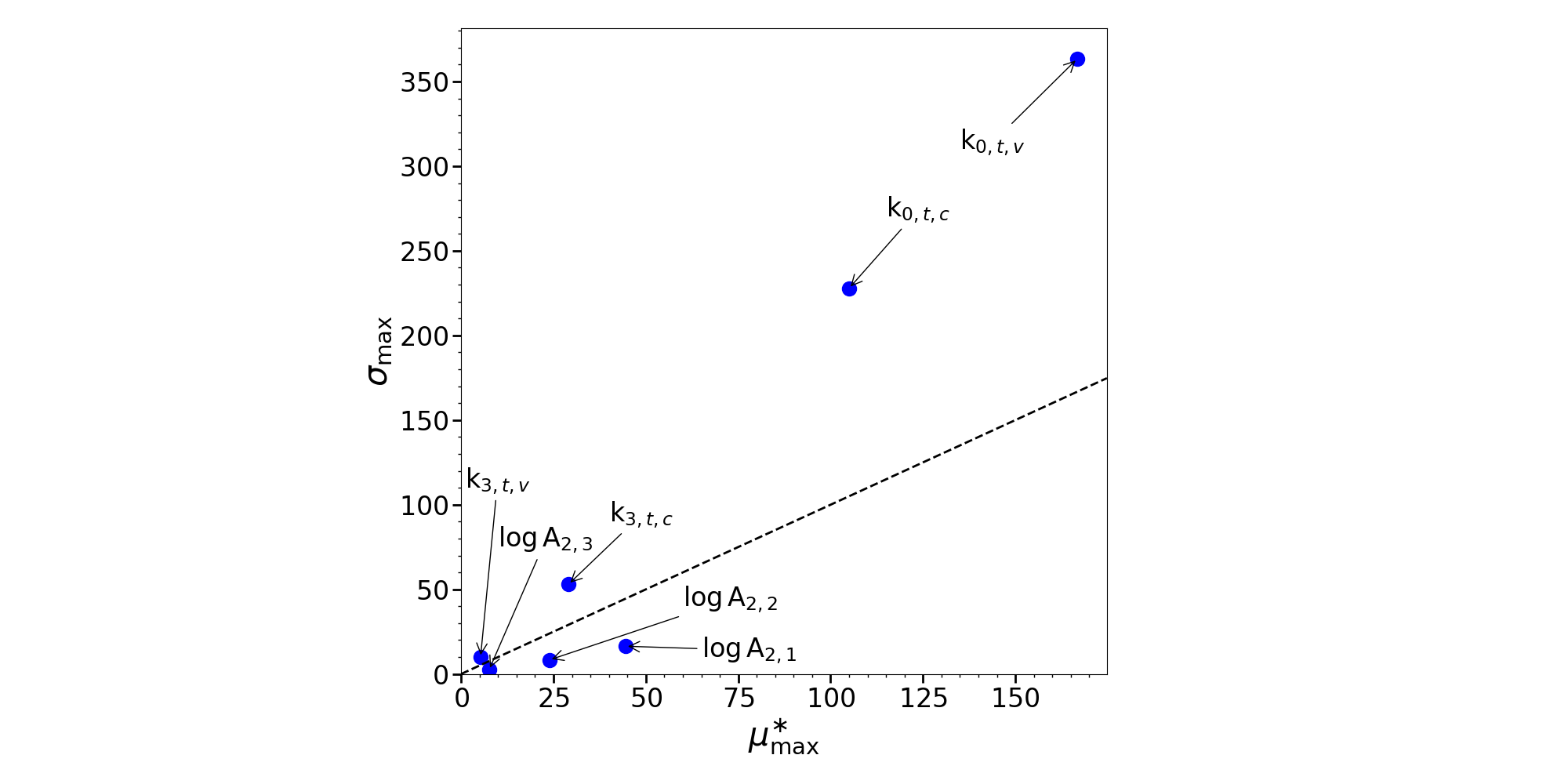}
    \caption{}
    \label{f:morris_msl}
  \end{subfigure}
  \caption{Morris method aggregate results for HyMETS (a) and MSL MISP-4 (b) scenarios.}
  \label{f:morris}
\end{figure}

\begin{figure}
  \centering
  \begin{subfigure}[b]{0.49\linewidth}
    \centering
    \includegraphics[width=\linewidth, trim=10cm 0cm 12cm 0cm, clip]{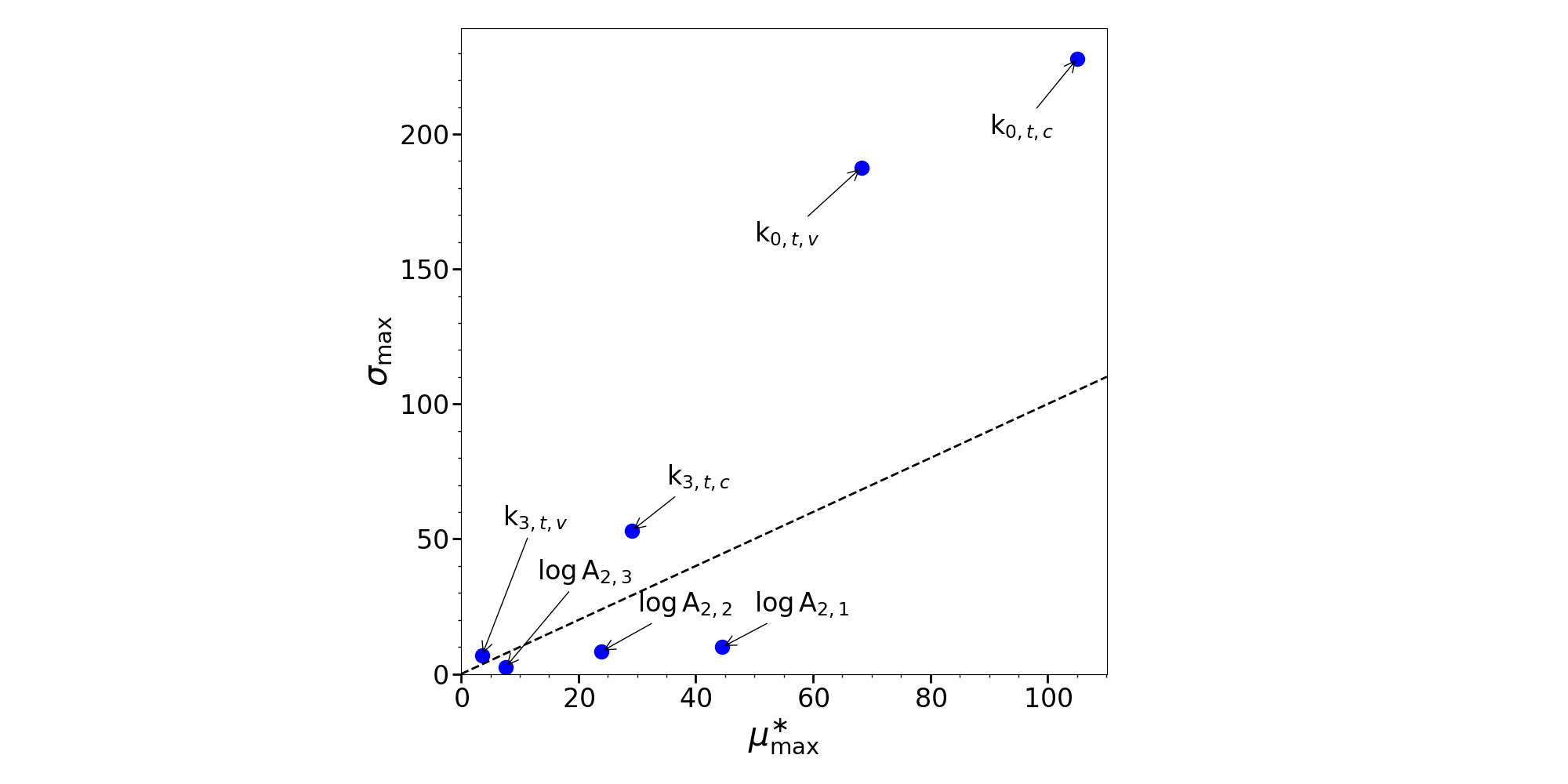}
    \caption{}
    \label{f:morris_misp4_tc1}
  \end{subfigure}
  \hfill
  \begin{subfigure}[b]{0.49\linewidth}
    \centering
    \includegraphics[width=\linewidth, trim=10cm 0cm 12cm 0cm, clip]{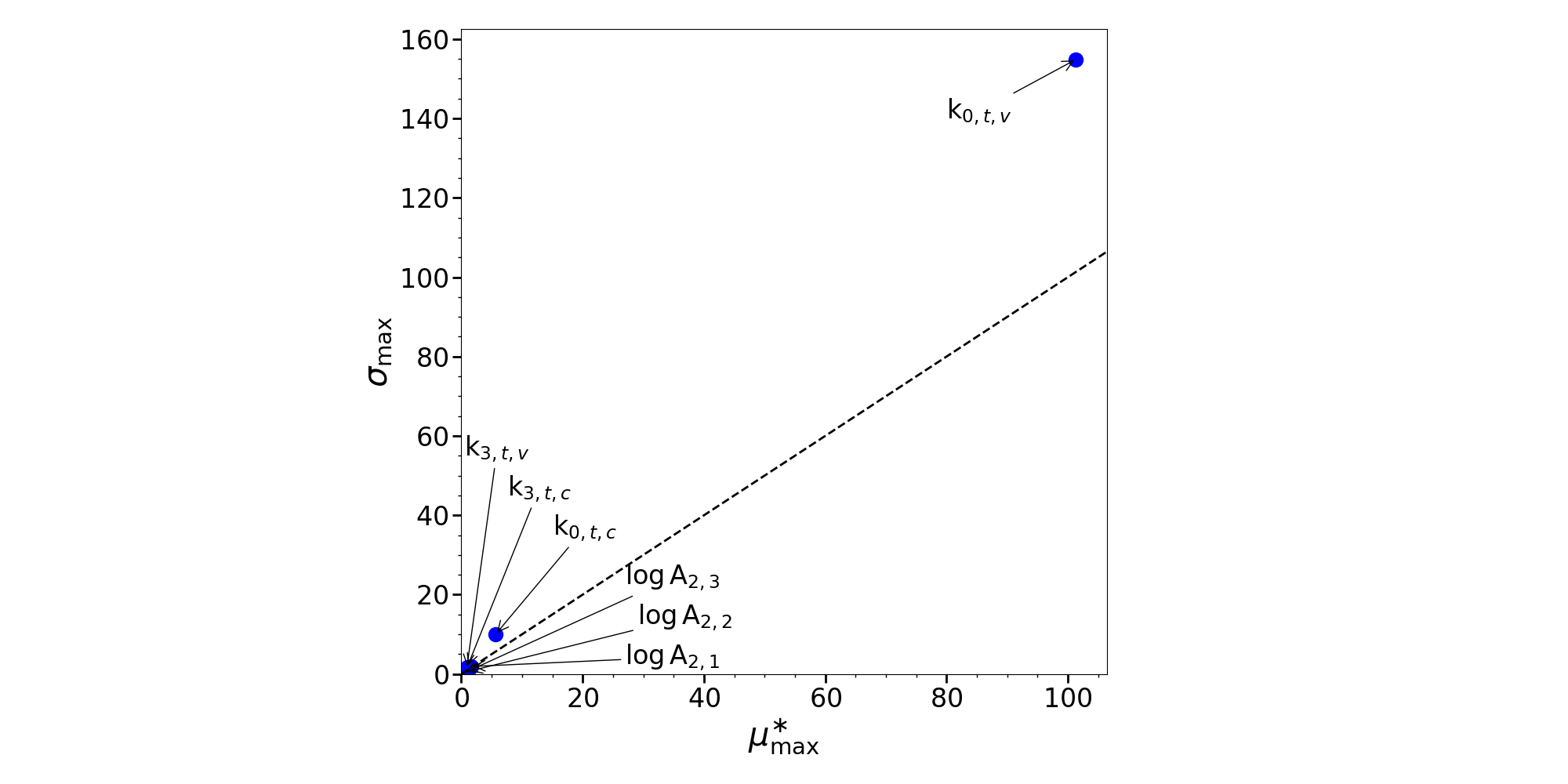}
    \caption{}
    \label{f:morris_misp4_tc4}
  \end{subfigure}
  \caption{Morris method results for MSL MISP-4 scenario with focus on TC1 (a) and TC4 (b) locations.}
  \label{f:morris_tcs}
\end{figure}

Applying the Morris screening methodology leads to results shown in Figure \ref{f:morris} and the identification of different subsets of influential parameters between the two considered scenarios. The discussion in this work is necessarily condensed to the analysis of maximum sampled values $\mu^\ast_\mathrm{max}$ and $\sigma_\mathrm{max}$ instead of the complete set of 433 Morris plots corresponding to each discretized point in space and time of the field response; conclusions were verified against the complete set of results to ensure their validity. It is observed that thermal energy transfer phenomena are significant in both scenarios, and variation of thermal conductivity parameters imposes mostly non-linear trends in predicted material temperature. However, uncertainty in the $\mathrm{k}_{3,t,v}$ tuning input's value can be ignored and the input set to its nominal value. Sampled $\mu^\ast$ and $\sigma$ for material decomposition Arrhenius pre-exponential parameters are also less significant, and variation in the values of these parameters in unison with others imparts negligible variance in model output in the ground facility case. Though the classification of material decomposition parameters and the $\mathrm{k}_{3,t,c}$ input for the MSL scenario is ambiguous based on Figure \ref{f:morris_msl} alone and is instead based on values of $\mu^\ast_\mathrm{max}$ and $\sigma_\mathrm{max}$ determined independently per each thermocouple location. A subset of corresponding Morris plots is shown in Figure \ref{f:morris_tcs} for TC1 and TC4 locations in the MISP-4 assembly, where significant trends for the sensitivity of the solution with material depth are observed. Solution near the bondline, for instance, is primarily affected only due to the variation of the $\mathrm{k}_{0,t,v}$ input. Near the surface, on the other hand, computed material temperature values are also sensitive to variation of $\mathrm{k}_{0,t,c}$ and $\mathrm{k}_{3,t,c}$ inputs and the Arrhenius pre-exponential parameters corresponding to $j=1,2$ sub-phase decomposition reactions of the matrix component. The differences in identified subsets of influential inputs per scenario are resolved by including their union in subsequent statistical treatments and are identified in Table \ref{t:uncertain_parameters} as those being included in calibration exercises. Moreover, while a parameter may be less influential, any departure from prior uncertainty PDFs can significantly alter forward propagation results in a scenario where respective parameter uncertainty is not dismissible.

\subsection{Polynomial Chaos}

The statistical approach employed in this work imposes large computational requirements that are alleviated through the generalized Polynomial Chaos (gPC) surrogate modeling technique. In applying the methodology, a scalar model response is projected onto a space spanned by an appropriately chosen orthogonal polynomial basis based on the work of Wiener \cite{wiener:1938}. The spectral representation of the non-deterministic, scalar response in the seminal work took the form of the infinite sum 
\begin{equation} 
\label{e:polychaos}
\Upsilon = \sum_{i=0}^{\infty} a_{\boldsymbol{\eta}_i}\Psi_{\boldsymbol{\eta}_i}(\boldsymbol{\Theta}) \ , \ a_{\boldsymbol{\eta}_i} = \frac{\left\langle f(\boldsymbol{\Theta}), \Psi_{\boldsymbol{\eta}_i}(\boldsymbol{\Theta}) \right\rangle_w}{\lVert\Psi_{\boldsymbol{\eta}_i}(\boldsymbol{\Theta})\rVert^2_w}
\end{equation}
consisting of multivariate, orthogonal polynomials $\Psi_{\boldsymbol{\eta}_i}(\boldsymbol{\Theta}) : \mathbb{R}^p \to \mathbb{R}^1$, basis coefficients $a_{\boldsymbol{\eta}_i} \in \mathbb{R}^1$, and multi-indices $\boldsymbol{\eta}_i \in \mathbb{R}^p$ of form $\boldsymbol{\eta}_i = [\eta_1, \eta_2, \ldots , \eta_p]_i$ that are equal in dimension to the random vector input of the model \cite{sudret:2008}. In practice, the infinite sum in Equation \ref{e:polychaos} is truncated a priori, and the basis is chosen according to the Wiener-Askey scheme for non-normally distributed model inputs \cite{xiukarn:2002}. The integration involved in the formula for coefficient values can be carried out by applying quadrature schemes in conjunction with Smolyak sparse grids that impart relatively lower computational costs than brute force integration. An alternative procedure for computing gPC basis coefficients consists of minimization of mean-square approximation error through regression simulation approaches. 

The gPC surrogate modeling methodology is applied with MSL and HyMETS scenarios concerning the identified pruned set of influential inputs. The model's time and space-dependent responses are approximated through the frozen-time gPC approach, where a metamodel is obtained at 1 s intervals of the response domain. The polynomial chaos series is truncated following a hyperbolic scheme, and corresponding basis coefficients are obtained using a sparse, adaptive procedure based on the Least Angle Regression (LAR) solver application given 10000 LHS-based input realizations and evaluations of the model for the MSl scenario and 1000 samples for the HyMETS case due to computational constraints \cite{blatman:2009}. The truncation order of the series is incrementally increased until a variance normalized cross-validation target of 0.001 is achieved while taking steps to prevent overfitting phenomena; the obtained models often perform well beyond this threshold. 

\section{Discussion}
\label{sec:discussion}

\subsection{Forward Propagation of Ground Facility Results}
\label{sec:propagation}

\begin{figure}
    \centering
    \includegraphics[width=\linewidth, trim=0cm 1.5cm 0cm 0cm, clip]{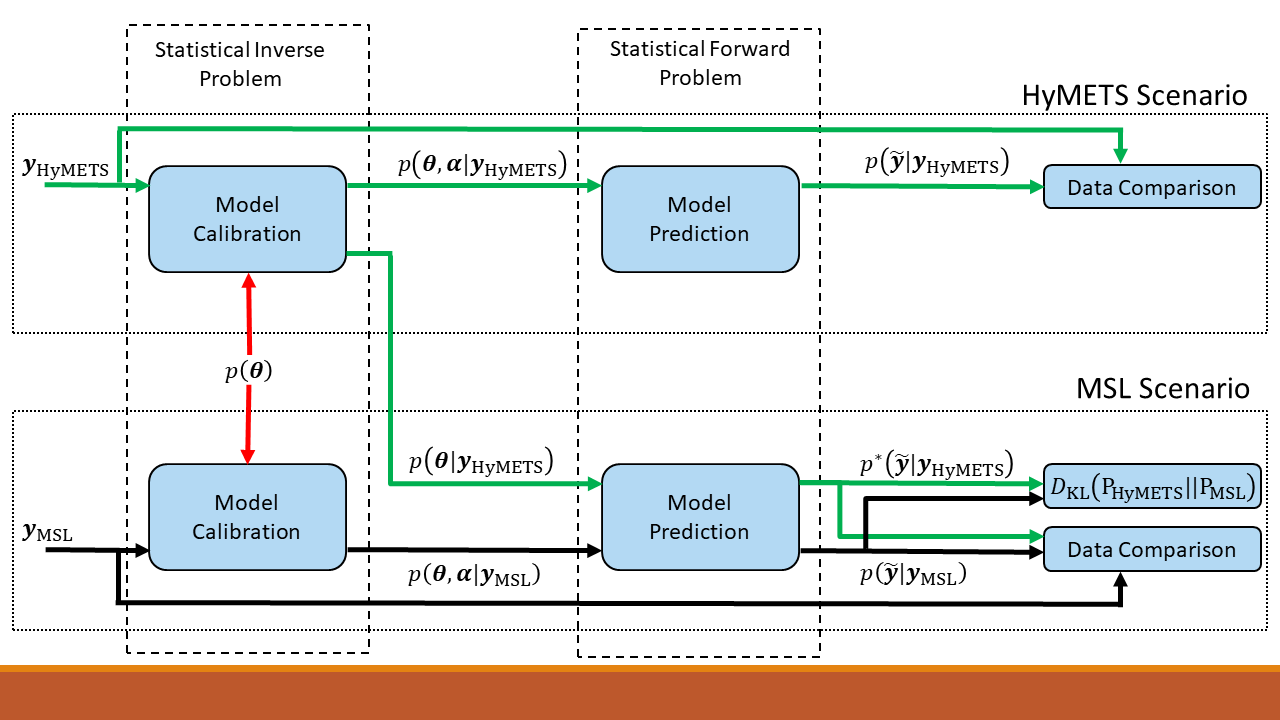}
    \caption{Process diagram for evaluating validity of the straightforward propagation procedure of results obtained with HyMETS data to the MSL scenario.}
    \label{f:extrapolation_process}
\end{figure}

The objective of the formulated non-deterministic extension framework is methodological quantification of uncertainty due to modeling and parametric sources for flight predictions based on quantified uncertainty and calibrated results with ground facility data. However, it is first worthwhile to explore the effects of straightforward propagation of statistical inference exercise results obtained with plasma wind tunnel material performance data for the flight scenario. A comparison can then be made with the quantified degree of uncertainty when the same exercise is performed with measurements captured in flight. The process flow of this exercise is illustrated in  Figure \ref{f:extrapolation_process}, where Bayesian inference is carried out for both HyMETS and MSL scenarios independently with an identical prior $\mathcal{P}( \boldsymbol{\theta} )$ for the model input parameters. Predictive inference and subsequent posterior predictive checks in the data comparison stage of the process are performed to assess the quality of calibration results and the capacity of the model to capture material performance trends. The MSL scenario statistical forward propagation problem results using the posterior obtained from the application of Bayesian inference for the HyMETS scenario would match predictive inference results from the MSL scenario calibration procedure if ground facility test campaigns yielded duplicate evidence as data captured in flight. The magnitude of differences between the two non-deterministic solutions of the forward problem following this logic can be used to assess the efficacy of current ground testing procedures based on graphical comparisons of obtained PDFs and the KL divergence $D_\mathrm{KL} \left( p || q \right)$ value computation. 

Yet as previously ascertained, forward propagation of quantified uncertainty in the HyMETS calibration exercise through the MSL scenario is not intuitive in the present context. The quantified degree of uncertainty apportioned to modeling errors is specific to the calibration scenario, and no general formulation for the estimation of uncertainty due to inadequacy of the model has been formulated in literature for charring ablators. While predictive inference exercises are straightforward in the HyMETS facility case, calibrated posterior distributions for the $\sigma_\mathcal{L}$ parameters assigned per thermocouple location are scenario specific and depend on location within the material and scenario conditions. The application of Bayesian inference results obtained with the ground facility scenario is thus limited to forward propagation of the quantified parametric uncertainty associated with model inputs obtained through marginalization of the posterior over the likelihood parameters. The non-deterministic solution of the limited uncertainty forward propagation problem $\mathcal{P}^\ast ( \widetilde{\boldsymbol{y}} | \boldsymbol{y}_\mathrm{HyMETS} )$ is made distinct in the process flow diagram to highlight this difference. Often approaches like the addition of noise to response values based on experience are utilized to account for the missing contribution to uncertainty due to model inadequacy, but these ad-hoc approaches are undesirable in the present work due to their limited use and subjectivity.

\begin{figure}
  \begin{subfigure}{0.49\linewidth}
    \centering
    \includegraphics[width=\linewidth, trim=10cm 0cm 12cm 0cm, clip]{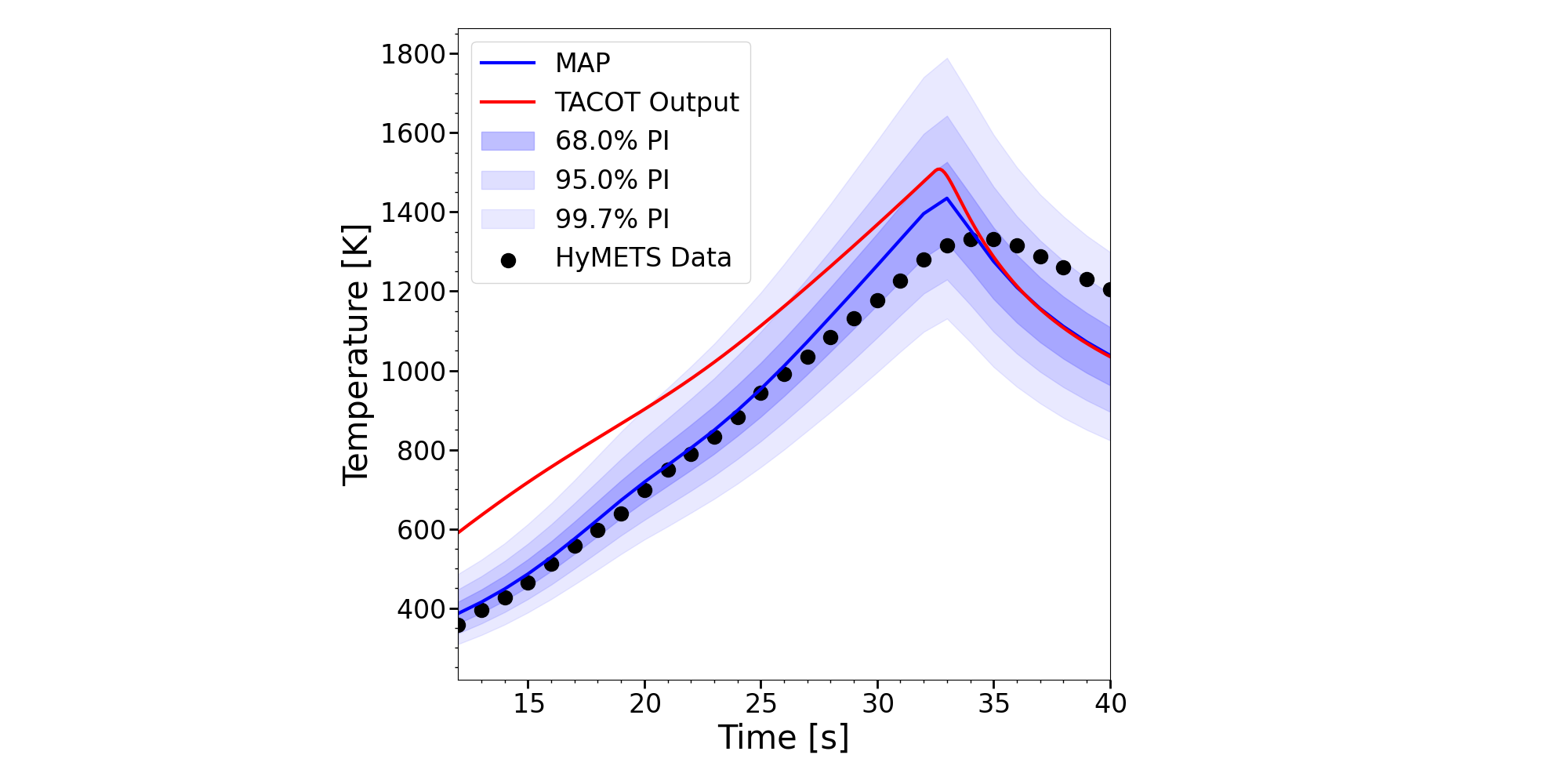}
    \caption{}
    \label{f:calibration_hymets_tc2}
  \end{subfigure}
  \hfill
  \begin{subfigure}{0.49\linewidth}
    \centering
    \includegraphics[width=\linewidth, trim=10cm 0cm 12cm 0cm, clip]{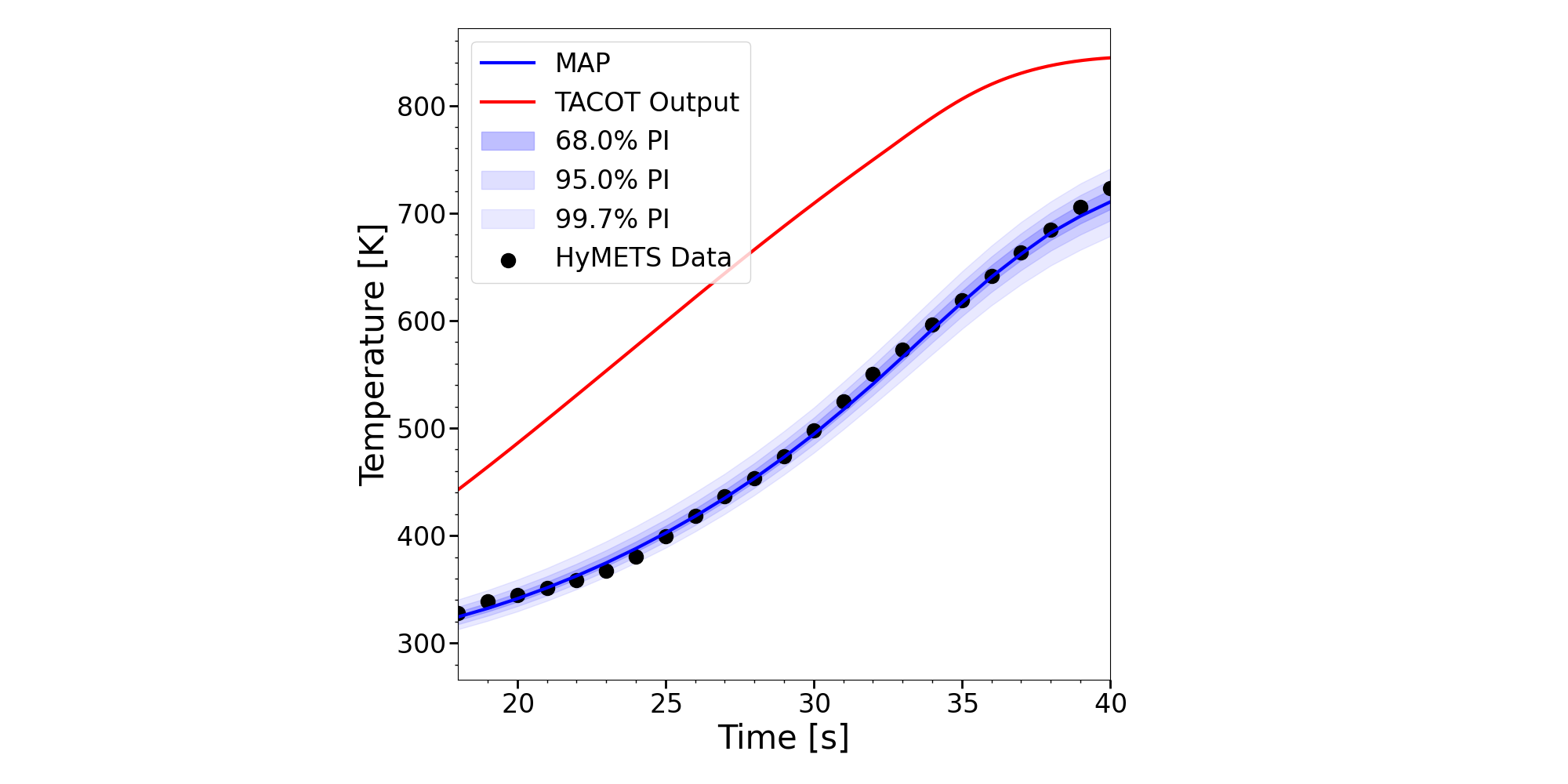}
    \caption{}
    \label{f:calibration_hymets_tc3}
  \end{subfigure}
  \caption{Predictive posteriors for HyMETS scenario material temperature at probed TC2 (a) and TC3 (b) locations.}
  \label{f:calibration_hymets}
\end{figure}
\begin{figure}
  \begin{subfigure}{0.49\linewidth}
    \centering
    \includegraphics[width=\linewidth, trim=10cm 0cm 12cm 0cm, clip]{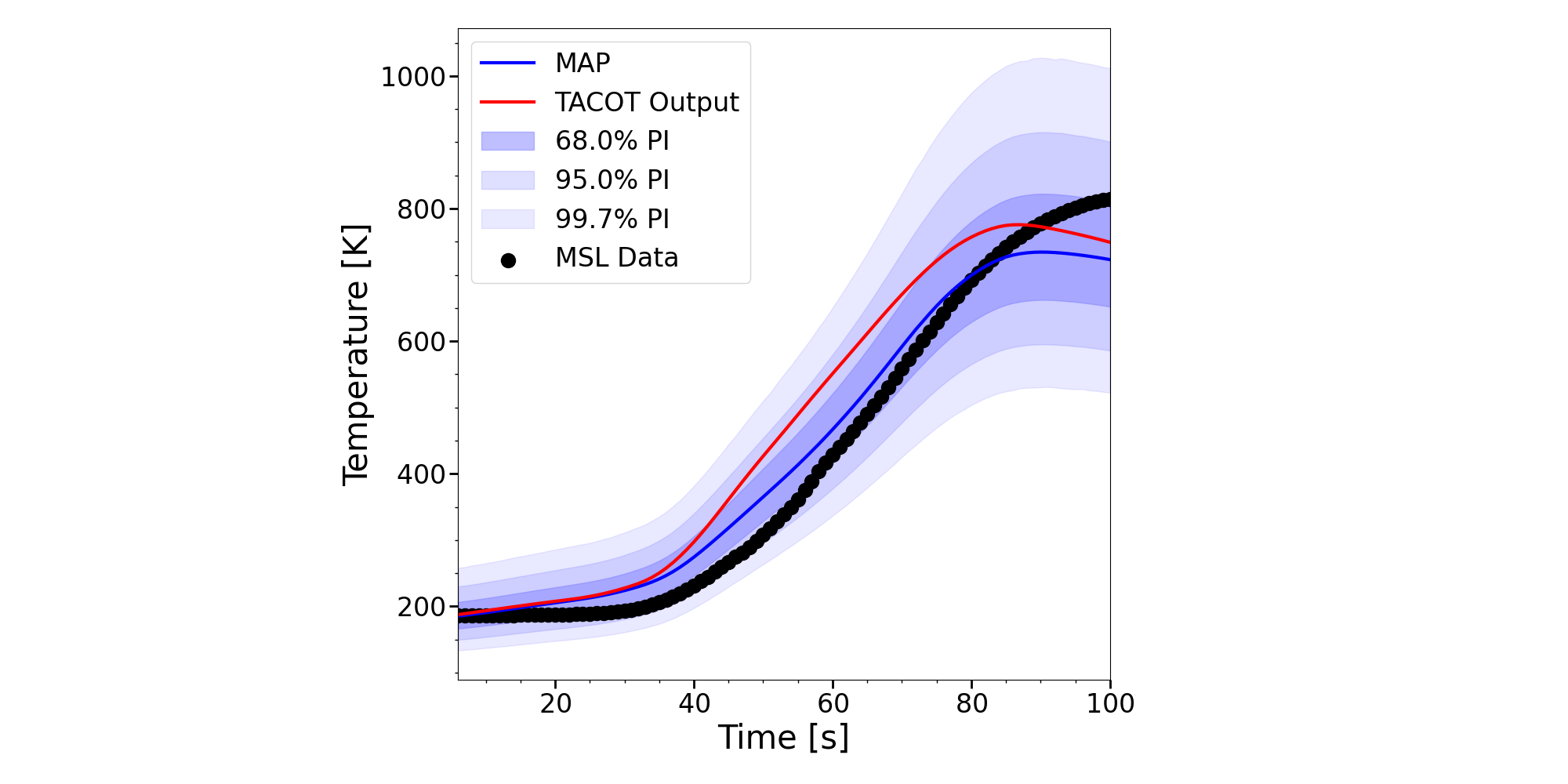}
    \caption{}
    \label{f:calibration_msl_tc2}
  \end{subfigure}
  \hfill
  \begin{subfigure}{0.49\linewidth}
    \centering
    \includegraphics[width=\linewidth, trim=10cm 0cm 12cm 0cm, clip]{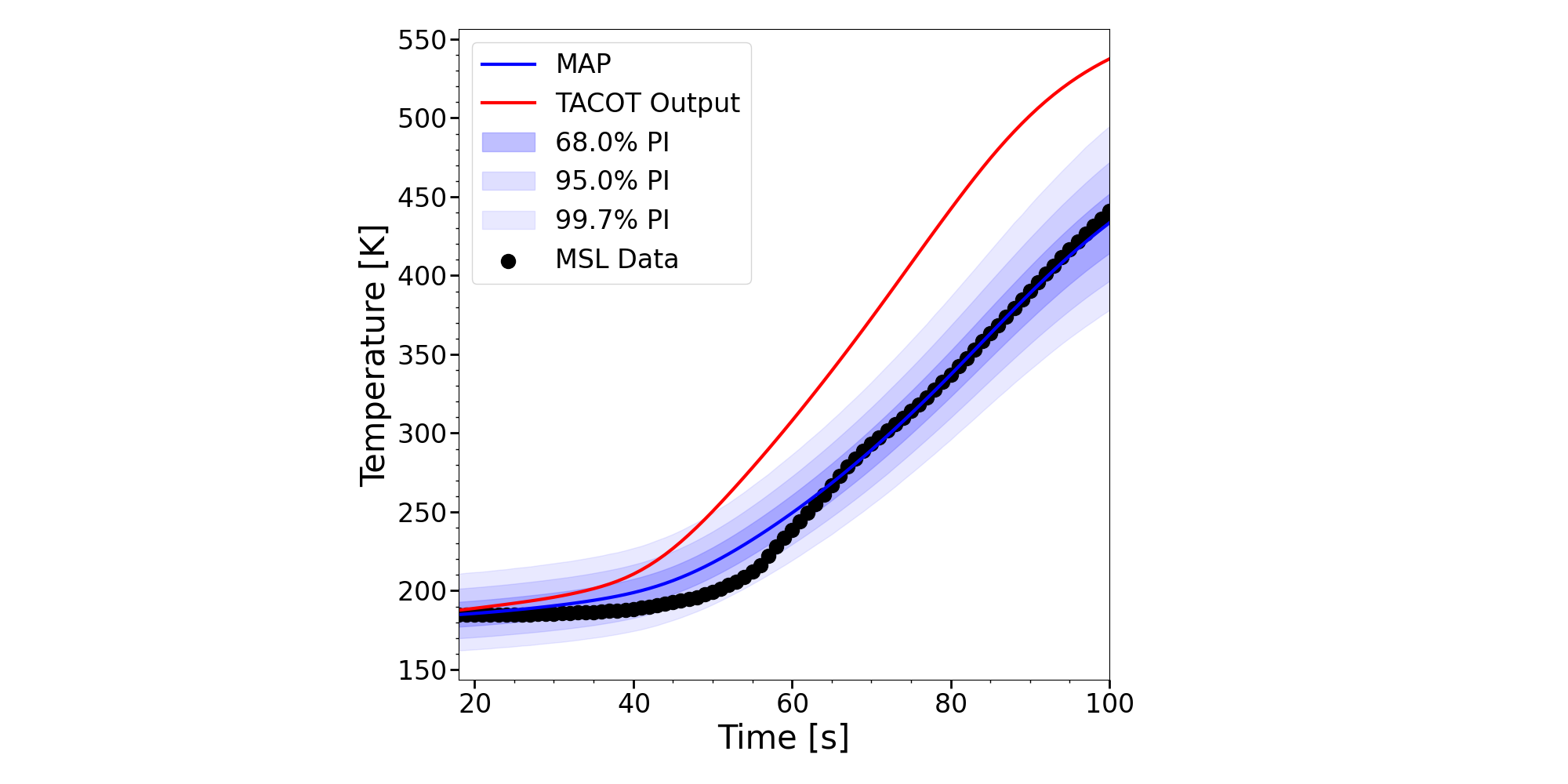}
    \caption{}
    \label{f:calibration_msl_tc3}
  \end{subfigure}
  \caption{Predictive posteriors for MSL scenario material temperature at probed TC2 (a) and TC3 (b) locations.}
  \label{f:calibration_msl}
\end{figure}
Application of the Bayesian inference procedure with MSL or HyMETS data yields improved predictive model capabilities considering their respective scenarios, albeit with a non-negligible degree of uncertainty. A subset of the results for the HyMETS scenario is shown in Figure \ref{f:calibration_hymets} for TC2 and TC3 locations in the Baby-sprite geometry sample. The Maximum a Posteriori (MAP) prediction can be seen to more closely align with facility data than model output with nominal input values, and the majority of data points are also captured within the plotted 95\% prediction intervals. Nonetheless, despite these improvements, MAP temperature predictions still exhibit some deviations from the captured response. The calibrated material response framework predicts significantly faster cooling of the material once removed from the high-speed, high-enthalpy flow jet combined with broad 99.7\% prediction intervals apportioned primarily to model inadequacy. Similar findings are derived from the MSL data-based statistical inference exercise with a subset of corresponding predictive inference results in Figure \ref{f:calibration_msl} for TC2 and TC3 thermocouple locations in the MISP-4 assembly. General improvements in MAP predictions are again observed, and the 95 \% prediction intervals again adequately capture flight data. However, the MAP solution underestimates peak material temperatures, and calibrated non-deterministic response prediction is associated with non-negligible 99.7\% uncertainty bounds.

\begin{figure}
  \begin{subfigure}{0.49\linewidth}
    \centering
    \includegraphics[width=\linewidth, trim=10cm 0cm 12cm 0cm, clip]{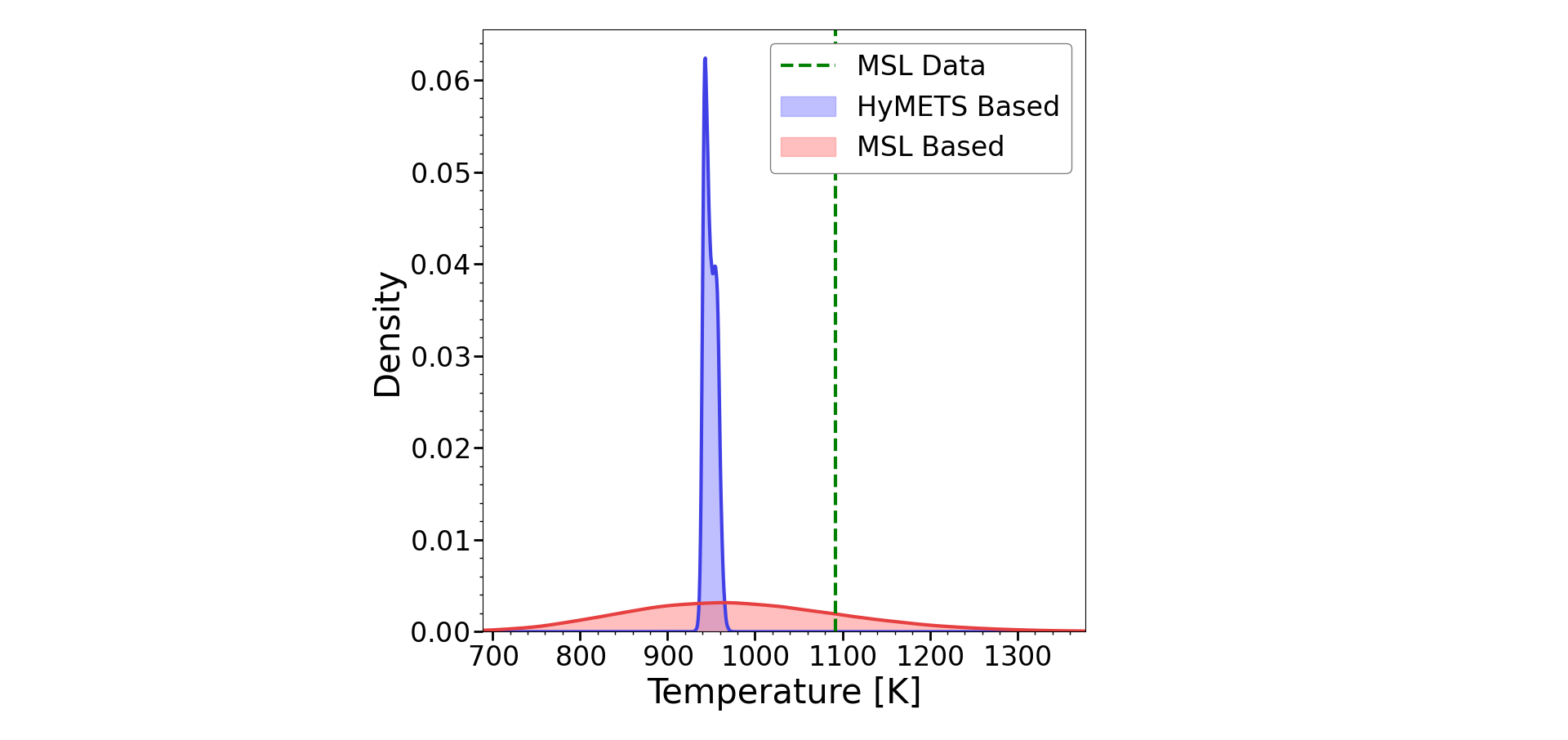}
    \caption{}
    \label{f:msl_tc1_t85s_extrapolation}
  \end{subfigure}
  \hfill
  \begin{subfigure}{0.49\linewidth}
    \centering
    \includegraphics[width=0.9\linewidth, trim=10cm 0cm 12cm 0cm, clip]{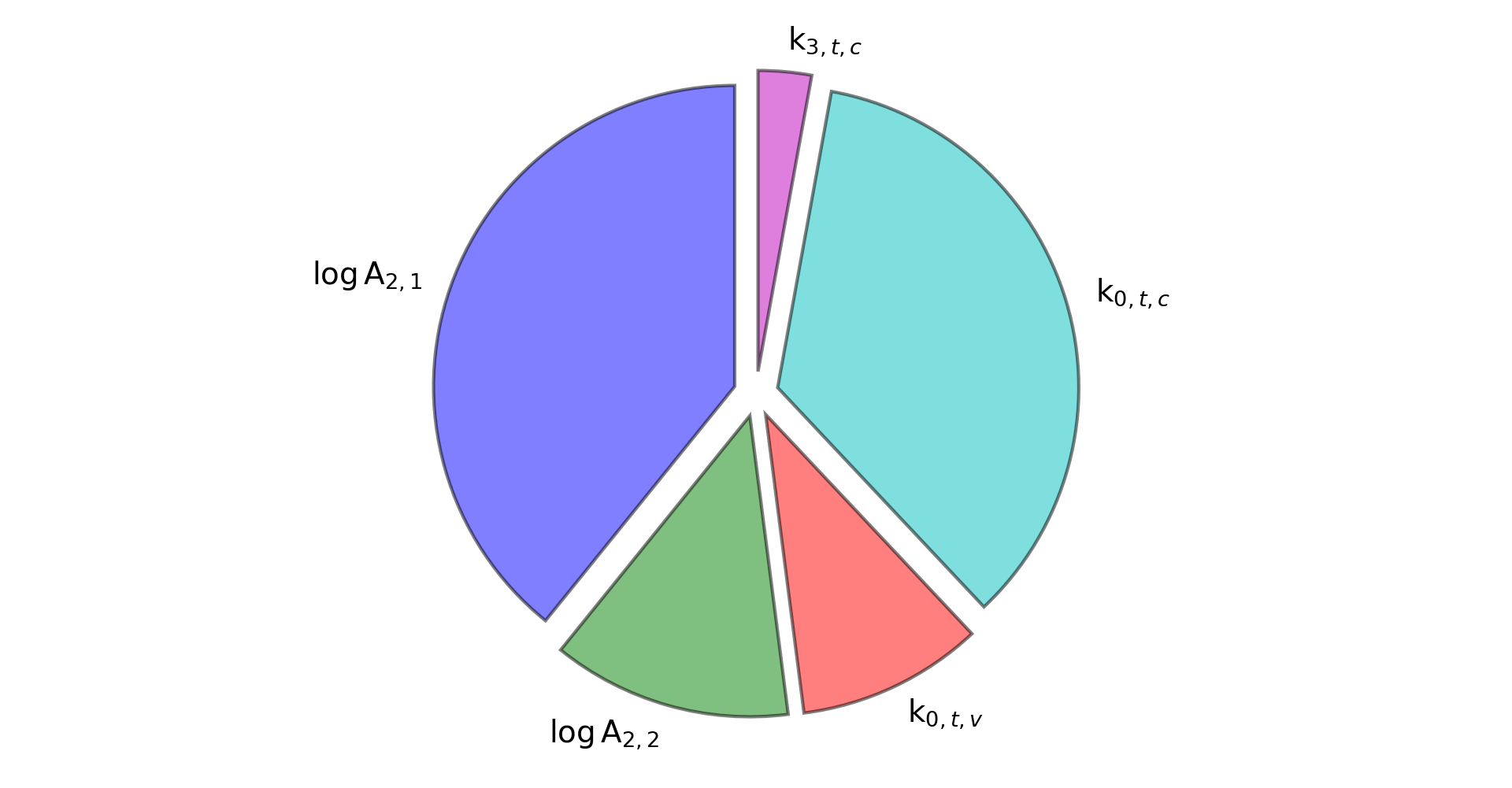}
    \caption{}
    \label{f:msl_tc1_t85s_variance_decomposition}
  \end{subfigure}
  \caption{Comparison of propagated HyMETS results with MSL-based target distribution (a) along with variance decomposition of the response (b) with respect to the prior at t=85s for the TC1 location.}
  \label{f:representative}
\end{figure}
\begin{figure}
  \begin{subfigure}{0.49\linewidth}
    \centering
    \includegraphics[width=\linewidth, trim=9cm 0cm 12cm 0cm, clip]{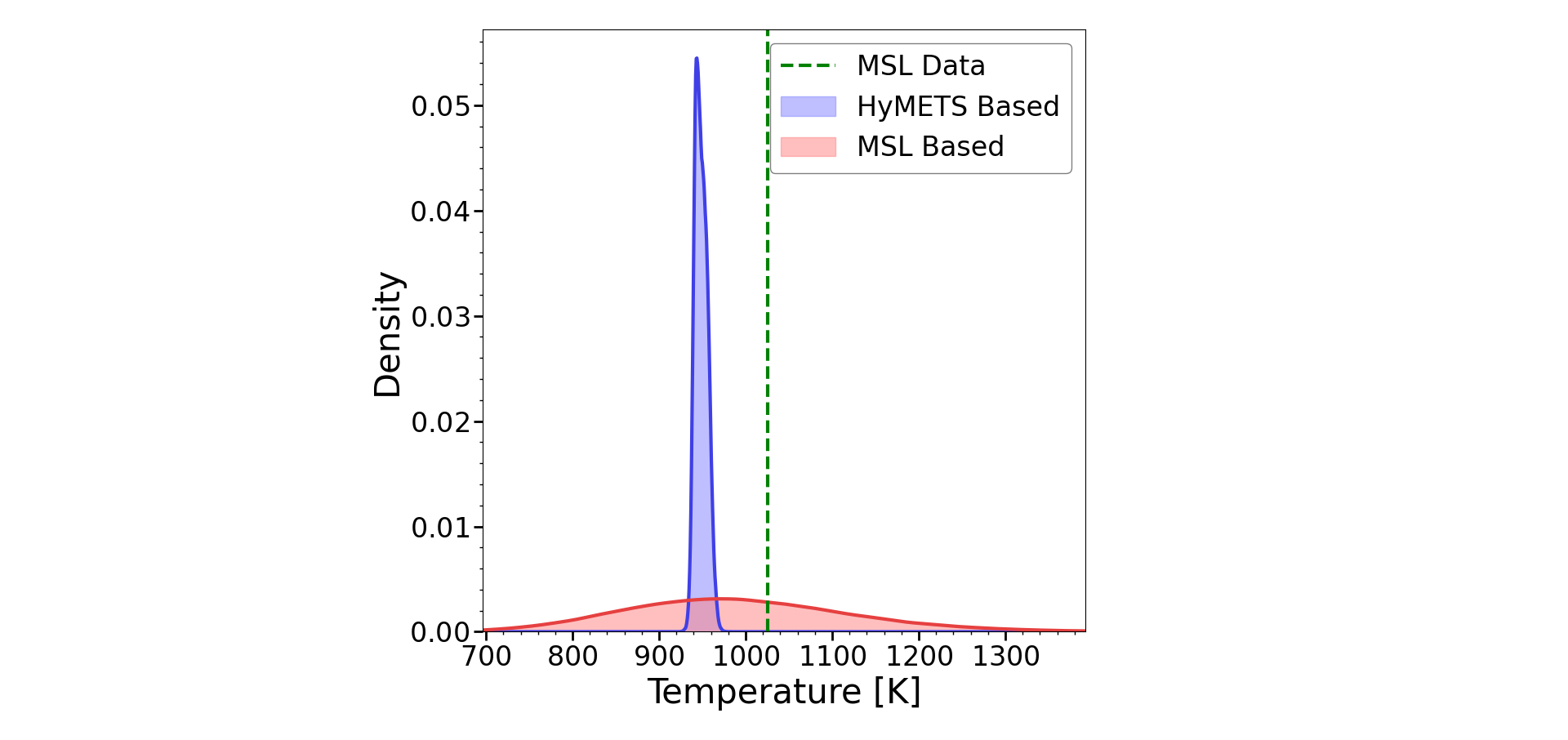}
    \caption{}
    \label{f:extrapolation_hymets_msl_tc1}
  \end{subfigure}
  \hfill
  \begin{subfigure}{0.49\linewidth}
    \centering
    \includegraphics[width=\linewidth, trim=9cm 0cm 12cm 0cm, clip]{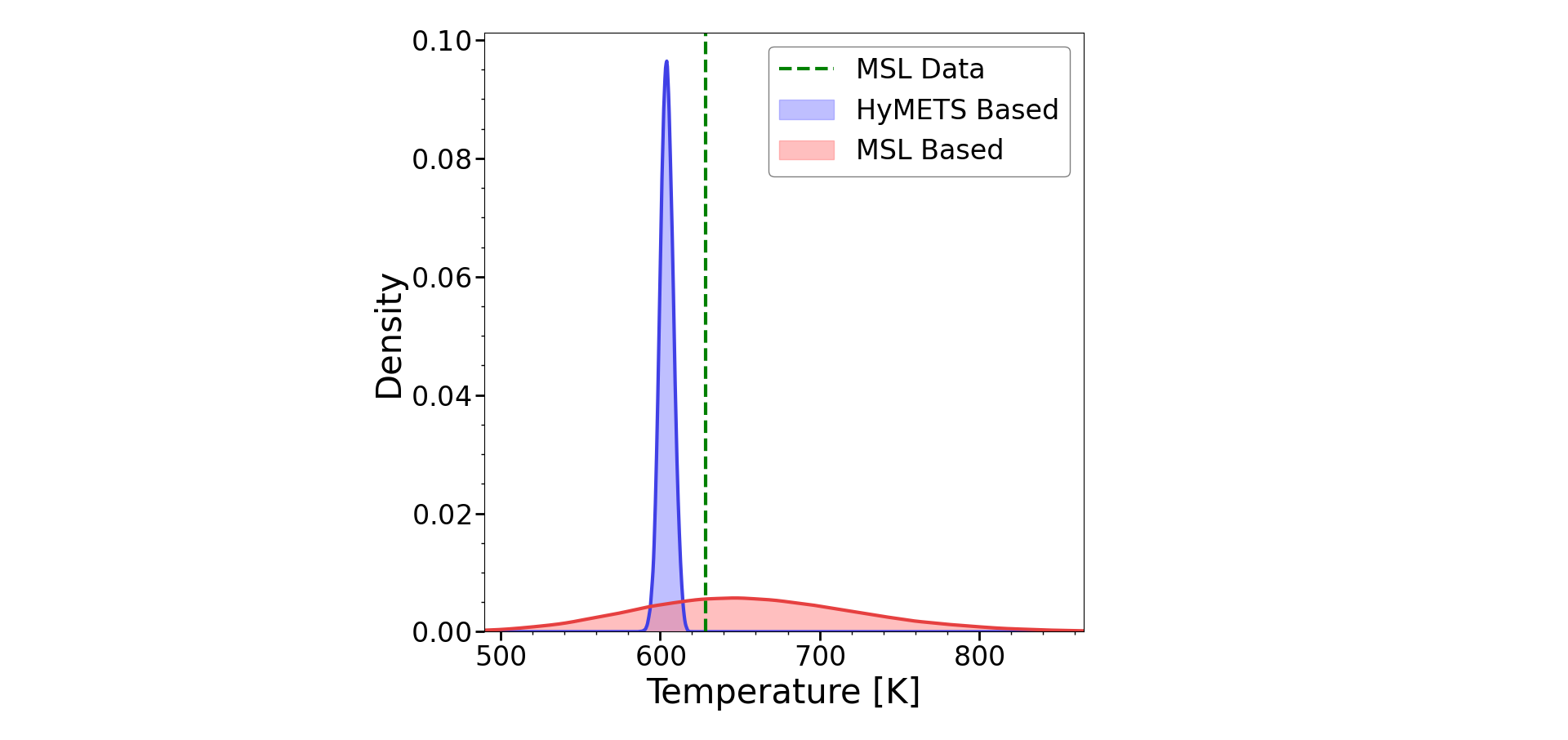}
    \caption{}
    \label{f:extrapolation_hymets_msl_tc2}
  \end{subfigure}
  \medskip
  \begin{subfigure}{0.49\linewidth}
    \centering
    \includegraphics[width=\linewidth, trim=9cm 0cm 12cm 0cm, clip]{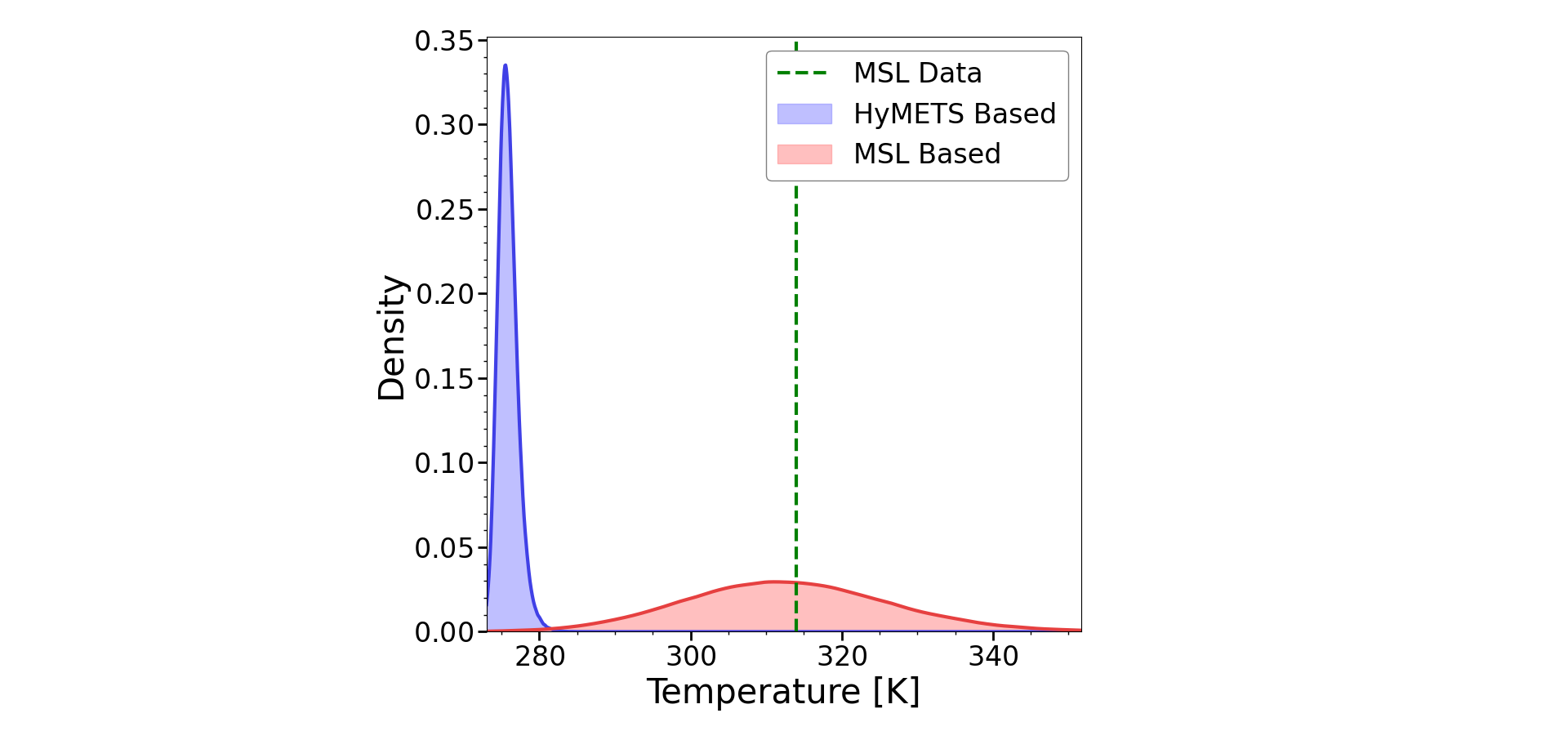}
    \caption{}
    \label{f:extrapolation_hymets_msl_tc3}
  \end{subfigure}
  \hfill
  \begin{subfigure}{0.49\linewidth}
    \centering
    \includegraphics[width=\linewidth, trim=9cm 0cm 12cm 0cm, clip]{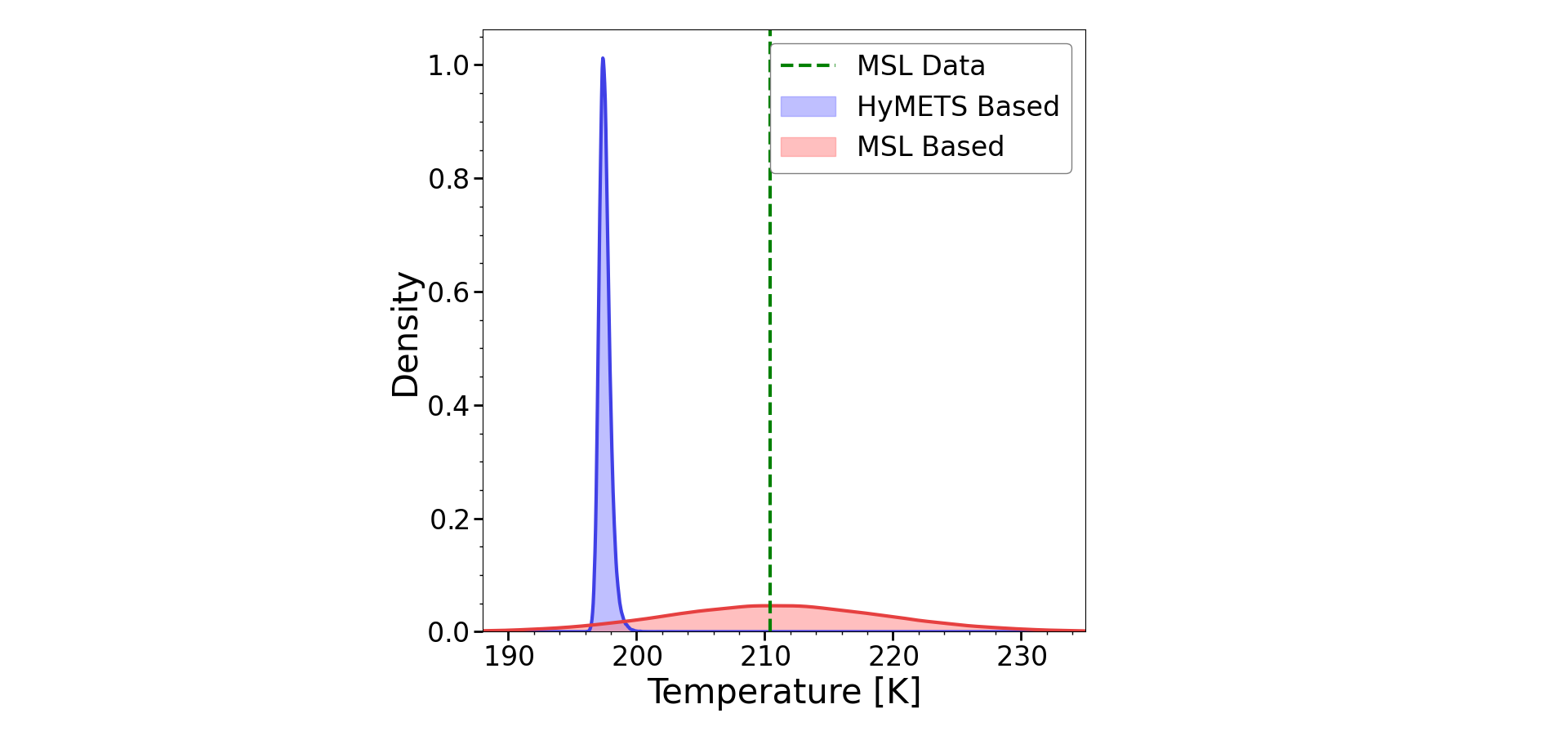}
    \caption{}
    \label{f:extrapolation_hymets_msl_tc4}
  \end{subfigure}
  \caption{Comparison of propagated HyMETS results with MSL-based target distribution at t=75s for TC1-TC4 (a-d) locations.}
  \label{f:extrapolation_hymets_msl}
\end{figure}
Forward propagation of the statistical inference solution given HyMETS facility data through the MSL scenario, more importantly, yields results at odds with predictive inference output obtained with data captured in flight. Discrepancies between solutions of the two forward propagation exercises stem from limited applicability of the ground facility-based result described in the preceding discussion and disparities between the two material property posterior distributions despite identical likelihood structures. A sample of obtained results showing encountered general trends is shown in Figure \ref{f:representative} for the solution at time t=85s for the TC1 location of the MISP-4 assembly. Compared to predictive inference results given MSL data, the forward propagated HyMETS based result significantly underestimates total uncertainty in response values. Predicted uncertainty also fails to capture measured thermocouple values adequately, whereas those values fall well within reasonable bounds of the MSL-based non-deterministic prediction. This solution feature is again shown in Figure \ref{f:extrapolation_hymets_msl} for all four thermocouple locations at a different point in the domain where the degree of discrepancy varies with material depth. Parametric uncertainty alone can be seen to not be sufficient in estimating prediction intervals when applied to flight conditions beyond the testing envelope of ground facilities, and a general model for modeling inadequacy must be employed.

\subsection{Application of the Novel Prediction Framework}

Straightforward propagation of plasma wind tunnel statistical inference results at flight conditions, withholding ad-hoc approaches, is inadequate in estimating uncertainty in material response predictions. The propagated uncertainty estimate based on Bayesian inference results given HyMETS data is limited to considering only the parametric uncertainty, and other differences are also present in posterior distributions of material model parameters. The following discussion is dedicated to applying a model error emulator approach to extend modeling uncertainty beyond the ground testing envelope as well as to resolve the disagreement between obtained posteriors by a posterior-prior mixing strategy with subsequent coarse optimization of involved hyperparameters of the approach.

\begin{figure}
  \begin{subfigure}{0.49\linewidth}
    \centering
    \includegraphics[width=\linewidth, trim=10cm 0cm 12cm 0cm, clip]{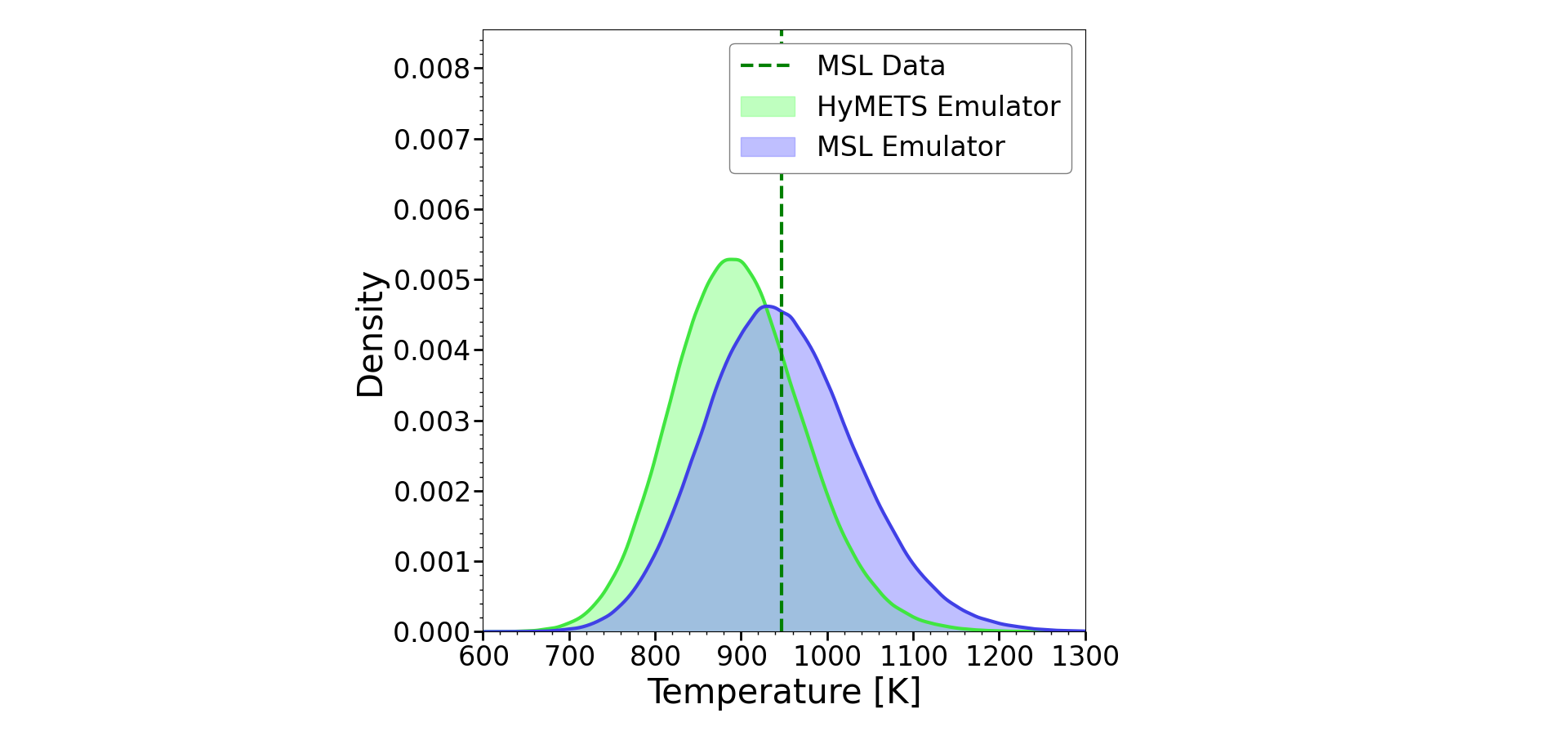}
    \caption{}
    \label{f:emulator_comparison_tc1_t70s}
  \end{subfigure}
  \hfill
  \begin{subfigure}{0.49\linewidth}
    \centering
    \includegraphics[width=\linewidth, trim=10cm 0cm 12cm 0cm, clip]{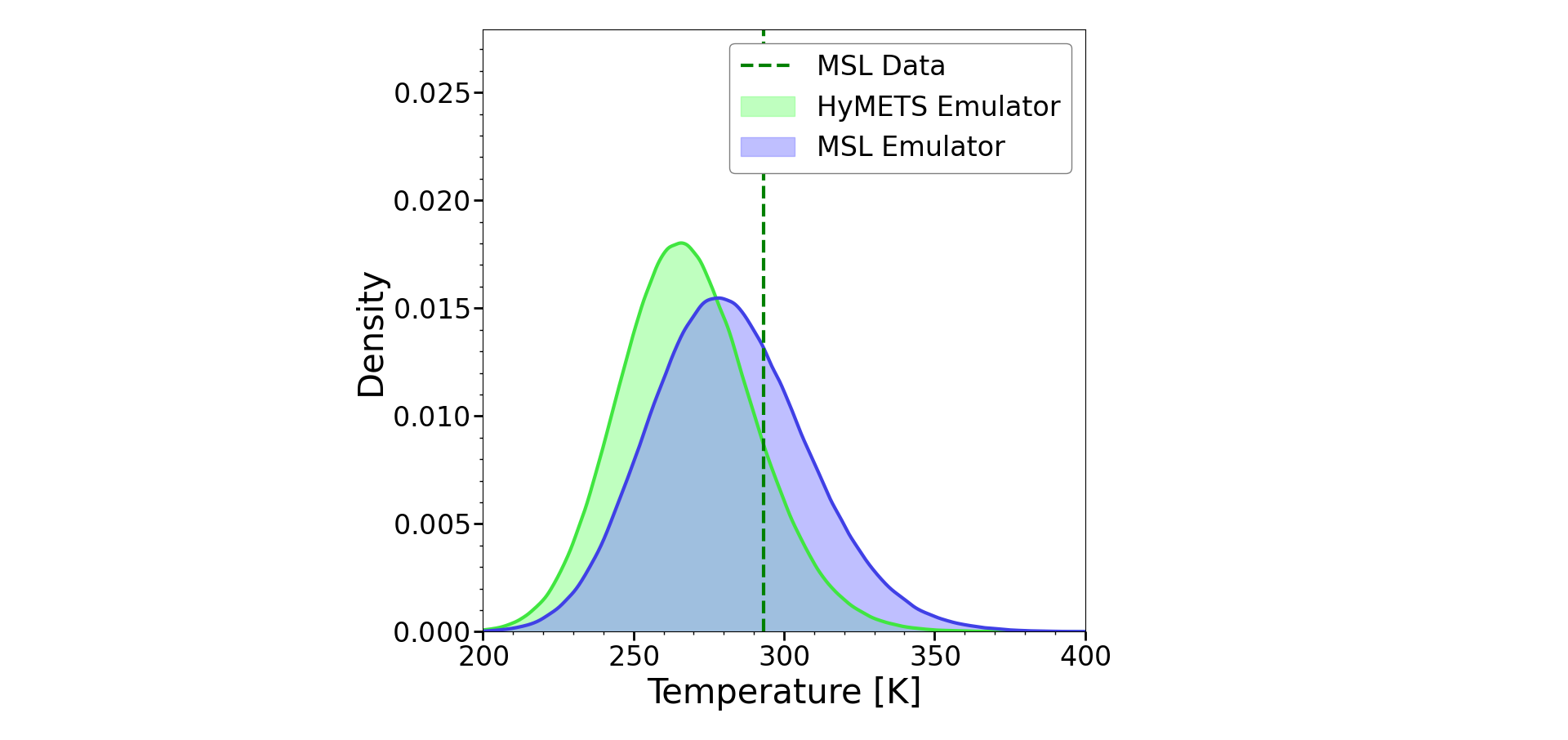}
    \caption{}
    \label{f:emulator_comparison_tc3_t70s}
  \end{subfigure}
  \caption{Comparison of predictive inference done with emulator term calibrated with HyMETS and MSL data in comparison to the target MSL predictive inference result for TC1 (a) and TC3 (b) locations at time t=70s in the MISP-4 assembly.}
  \label{f:emulator_comparison}
\end{figure}

\subsubsection*{Error Emulator}

\begin{figure}
  \begin{subfigure}{0.49\linewidth}
    \centering
    \includegraphics[width=\linewidth, trim=11cm 0cm 13cm 0cm, clip]{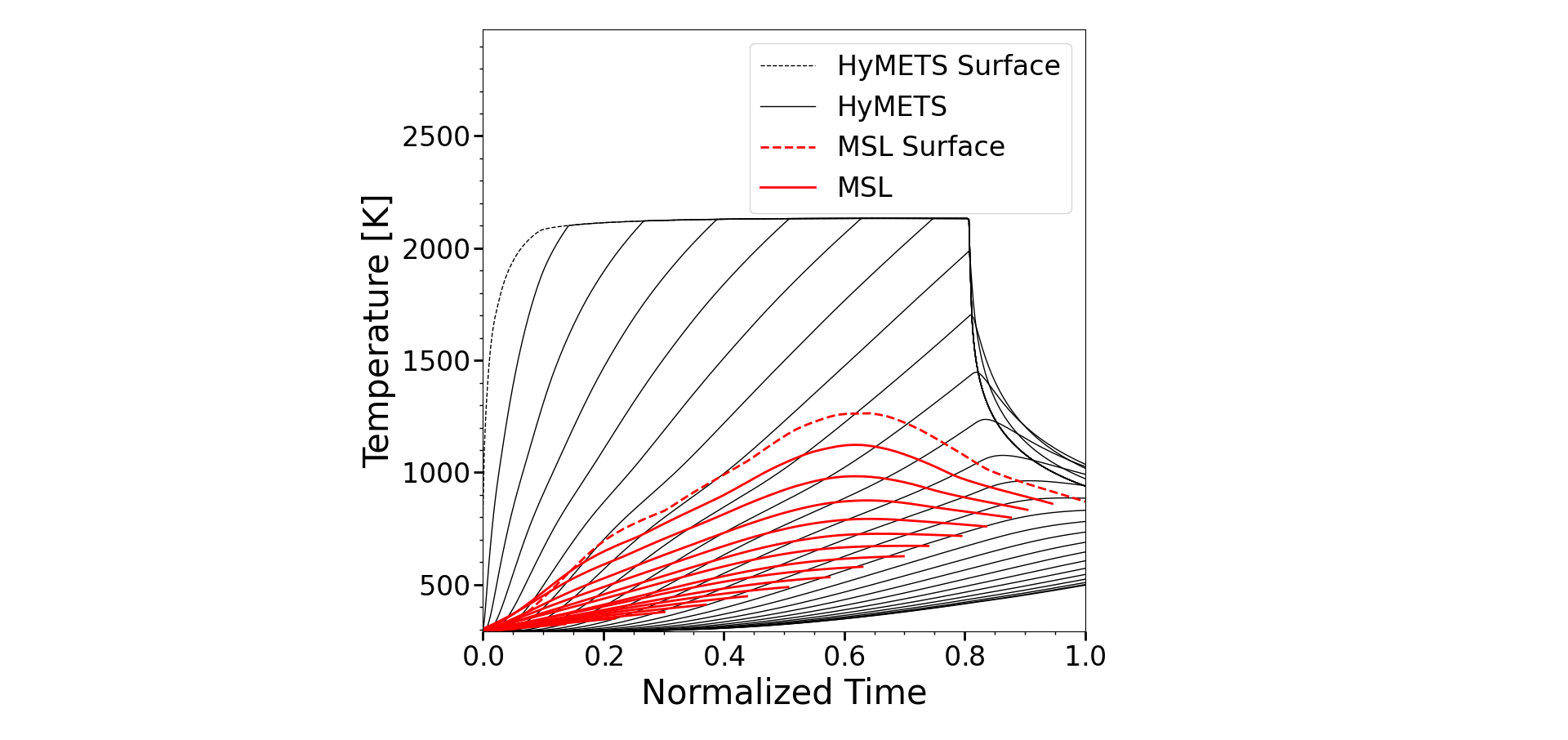}
    \caption{}
    \label{f:arcjet_msl_simulation_comparison}
  \end{subfigure}
  \hfill
  \begin{subfigure}{0.49\linewidth}
    \centering
    \includegraphics[width=\linewidth, trim=11cm 0cm 13cm 0cm, clip]{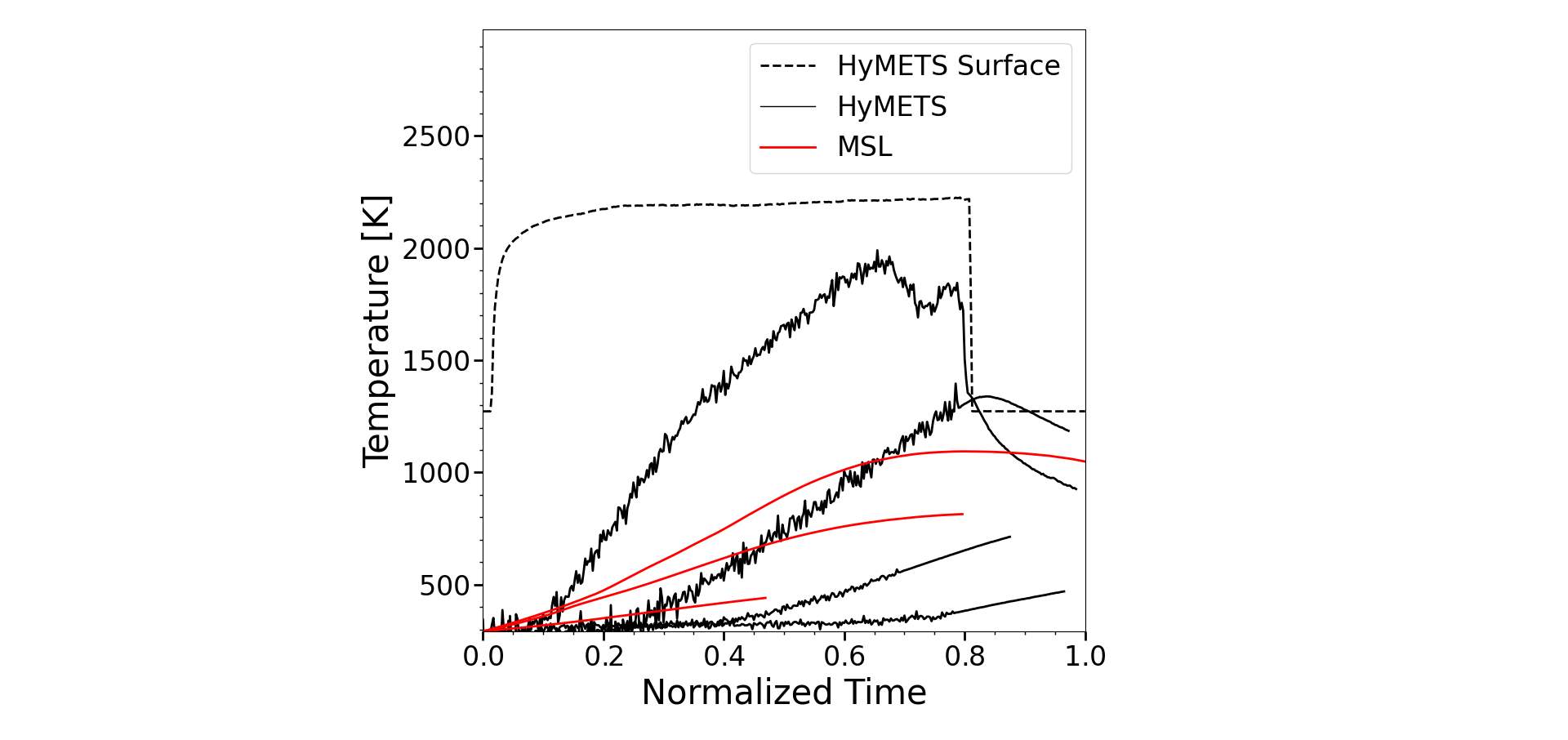}
    \caption{}
    \label{f:arcjet_msl_data_comparison}
  \end{subfigure}
  \caption{Comparison of simulated material temperature for MSL and HyMETS scenarios (a) and captured data (b) considering normalized time.}
  \label{f:msl_hymets_comparison}
\end{figure}

Before utilizing the model inadequacy emulator approach, it is imperative to ensure that the model is interpolative and does not attempt to extrapolate beyond material behaviors observed in the ground facility. To this end, the material sample in the ground facility can be observed to reach at least the temperature simulated and captured in the MSL flight case such that the phenomena present in flight inside the material are adequately represented in the ground facility. This conclusion is made more apparent in Figure \ref{f:msl_hymets_comparison}, where the simulated temperature response of the material using default TACOT property values and at incremental depths starting from the surface is plotted against the simulated response for the PICA sample in the plasma wind tunnel in the same manner. To further aid in this comparison, the temperature profiles for the various depths are shifted to start at the origin after surpassing 290 K, where influence modeling assumptions can be said to become more influential, albeit negligibly at this point. These profiles are shown on a normalized time axis obtained by normalizing the duration of time at which the surface of the material surpasses the temperature threshold. The same procedure is performed when comparing the captured thermocouple profiles between flight and ground facility experiments. It is immediately evident that the temperature of the material during the MSL entry maneuver does not exceed that recorded or simulated in the ground facility. The response of the material in flight thus does not encompass phenomena that were not reproduced in the HyMETS facility case. Because the model inadequacy emulator is concerned with the material temperature response in this work, the statistical model is here believed to be interpolative based on the preceding observations and is not expected to attempt to extrapolate beyond the trends simulated and observed in the HyMETS test case utilized in this work. The interpolative domain of the model, if additional aspects are considered in further augmentations of the model, also significantly increases once flight data is formally introduced into the prediction process of other flight scenarios.

\begin{figure}
  \centering
  \includegraphics[width=0.5\linewidth, trim=11cm 0cm 13cm 0cm, clip]{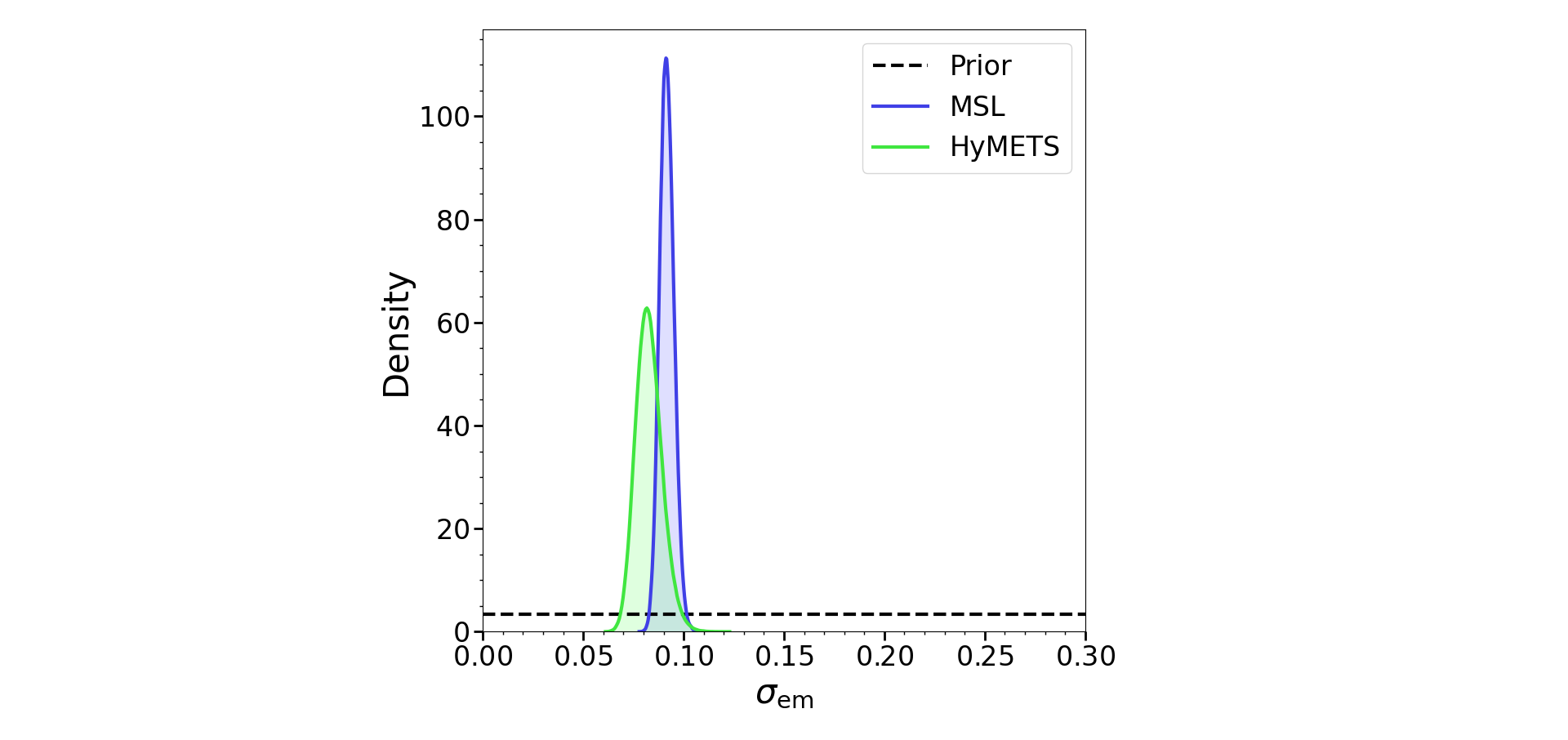}
  \caption{Comparison of posteriors for $\sigma_\mathrm{em}$ for the inadequacy model based on inference with HyMETS and MSL data separately.}
  \label{f:arcjet_msl_sigma_comparison}
\end{figure}

The emulator model, together with material model parameters, is subsequently here calibrated with the likelihood relationship characterized by the white noise kernel given HyMETS ground facility and MSL flight performance data in separate exercises. A subset of respective forward propagation and predictive inference results is shown in Figure \ref{f:emulator_comparison} for TC1 and TC3 locations within the MISP-4, assembly where a significant improvement in the estimation of uncertainty is observed using ground facility-based predictions. Unlike corresponding results of the limited forward propagation exercise in Figure \ref{f:extrapolation_hymets_msl}, data are consistently captured by obtained non-deterministic output based on the HyMETS calibration procedure results combined with the emulator approach. Arguments for the interpolative nature of the statistical model are further supported when comparing the posterior distributions for the $\sigma_\mathrm{em}$ between the two calibrations that as seen in Figure \ref{f:arcjet_msl_sigma_comparison} are in relatively close agreement. The HyMETS based posterior distribution encompasses the one obtained with MSL data, and the most probable values are also in proximity to each other.

A series of steps until this point were taken to improve the predictive uncertainty quantification capabilities based on material response data obtained within plasma wind tunnels. It was recognized that most literature concerning charring ablators and the design process for spacecraft involves deterministic methodologies with which material response tools and databases are updated. These are then used to make deterministic predictions concerning material performance in flight. Deterministic predictions and data are typically compared in the validation of the performance of the computational model and, at most, prior uncertainty concerning aerothermal and material response aspects is forward propagated onto the problem. The obtained uncertainty bounds are then utilized to make sizing decisions. In this work, Bayesian inference has thus far been employed to update the prior state of knowledge concerning various aspects of the problem with ground facility data, which are then forward propagated onto the prediction scenario. Although the quantified degree of uncertainty due to modeling inadequacy can not be propagated in the naive approach, the conducted procedure already extends beyond state-of-the-art approaches in the field. The procedure employed in this section further elaborates on the limitations of the naive approach by employing an inadequacy term model, which is shown to be interpolative given available data, that vastly improves upon the predictive uncertainty quantification capabilities of the naive forward propagation procedure. These improved results are achieved with a relatively simple statistical model for model inadequacy that allows for the utilization of the same likelihood formulation between scenarios.

Still, sampled marginal posterior distributions for material model inputs show non-negligible differences. It is also observed that obtained posterior predictive distributions for model inputs using the emulator approximation for the model error, respectively trained with MSL and HyMETS data, are not identical despite the same statistical and material response models. The HyMETS based prediction for the MSL scenario underestimates solution uncertainty in comparison, and the means of the distributions also somewhat diverge with greater material depths. The emulator-based prediction trained with MSL data also tends to overestimate uncertainty for deeper material depths due to the prescribed homoscedastic behavior imposed by a lack of training data. Ground facility and flight data in the present work update the prior state of knowledge to a different extent despite utilizing the same priors and discrepancy modeling approaches in the emulator-based formulation.

\begin{figure}
  \begin{subfigure}{0.49\linewidth}
    \centering
    \includegraphics[width=\linewidth, trim=10cm 0cm 12cm 0cm, clip]{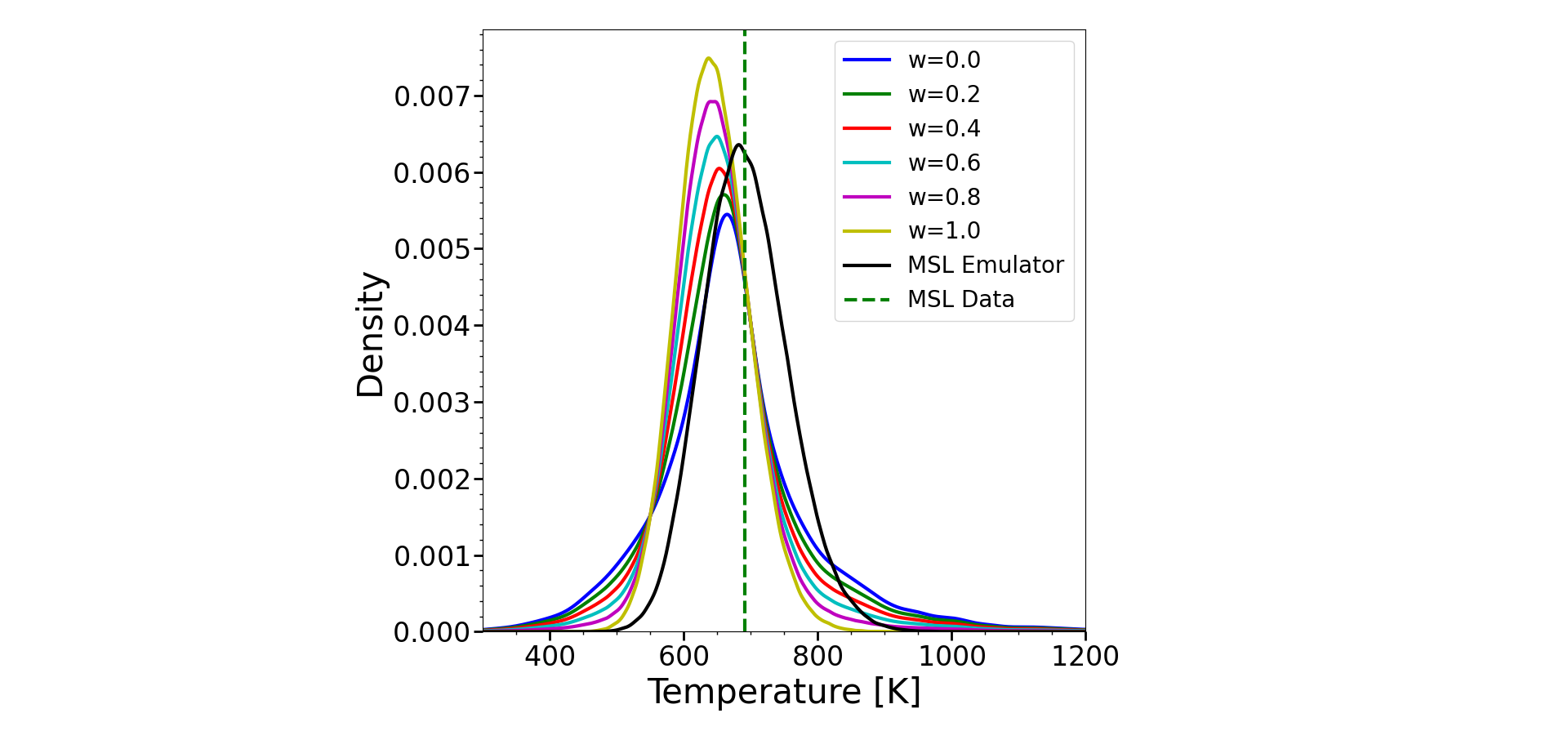}
    \caption{}
    \label{f:w_comparison_tc2_t80s}
  \end{subfigure}
  \hfill
  \begin{subfigure}{0.49\linewidth}
    \centering
    \includegraphics[width=\linewidth, trim=10cm 0cm 12cm 0cm, clip]{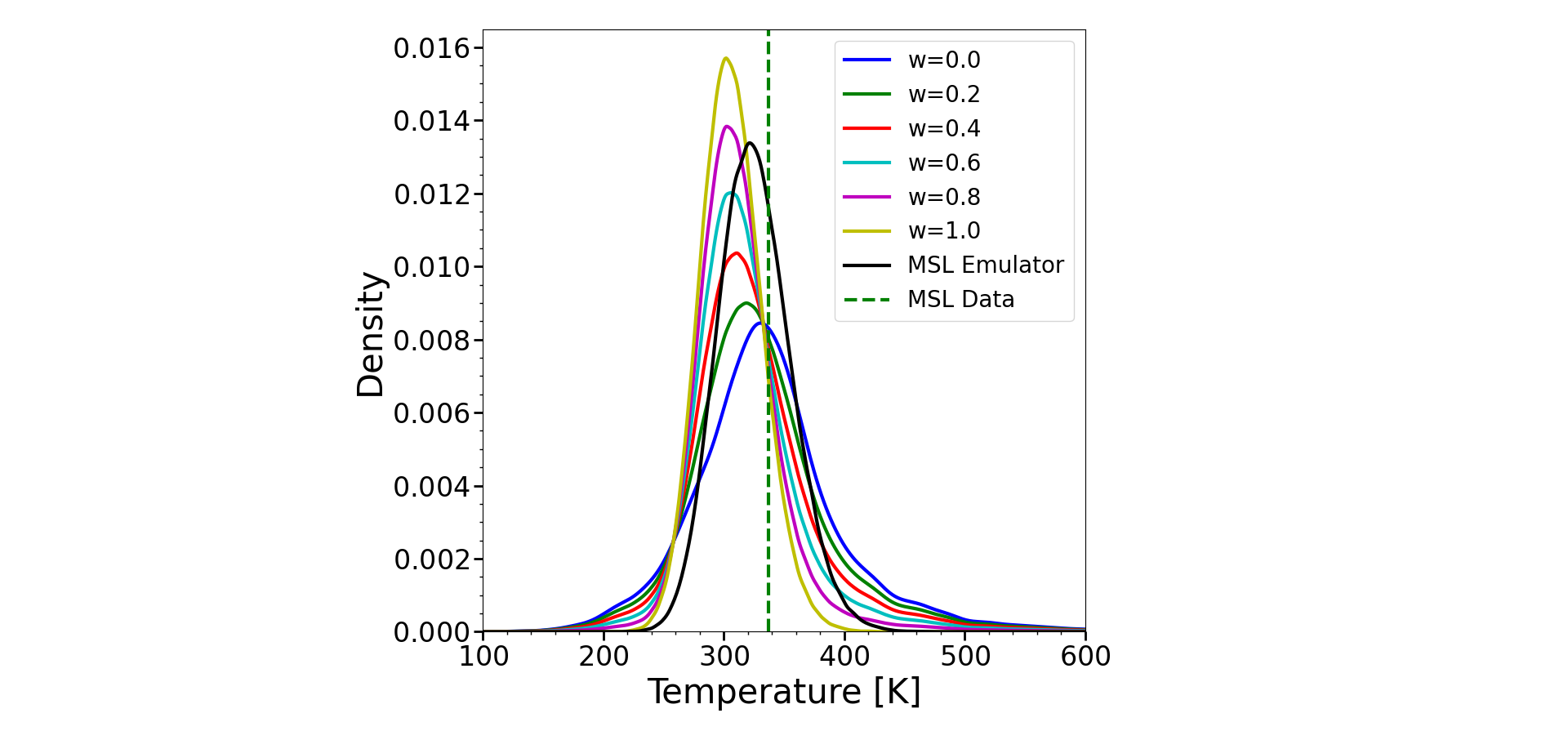}
    \caption{}
    \label{f:w_comparison_tc3_t80s}
  \end{subfigure}
  \caption{Effects of varying value of w parameter for MISP-4 TC2 (a) and TC3 (b) at t=80s on the predictive posterior.}
  \label{f:w_comparison}
\end{figure}

\subsubsection*{Model Averaging for Mixture Priors}

Ground facility and flight data conflict in informing material and emulator model inputs is resolved by applying the prior-posterior mixture approach detailed in Section \ref{sec:nondeterministic_framework}. Variation of the $w$ parameter of the mixture prior significantly impacts the predictive distribution compounded by complex behaviors of the response prediction. Effects of various values of $w$ are shown in Figure \ref{f:w_comparison}, where the predictive prior based on the mixture approach described in the present work is plotted at a single point in time corresponding to TC2 and TC3 locations within the MISP-4 assembly. Predictive responses are also compared to the posterior predictive distribution corresponding to the Bayesian inference exercise associated with the model inadequacy emulator addition and MSL flight data. A higher kurtosis measure generally characterizes distributions that rely more heavily on prior knowledge. It is also seen that the value of $w$ affects both the tail behavior and the location of the mean of the distribution, with more stark behaviors expected when calibration exercises result in more significant deviations of the posterior from the prior.

\begin{table}
 \begin{center}
    \caption{Total Integrated Divergence Values for Increments of $w$.}
    \label{t:total_divergence}
    \begin{tabular}{lccc}
      \toprule
      \toprule
      w & $D_\mathrm{KL} \left( \mathrm{P}_\mathrm{Mixture} || \mathrm{P}_\mathrm{MSL} \right)$ & $D_\mathrm{KL} \left( \mathrm{P}_\mathrm{MSL} || \mathrm{P}_\mathrm{Mixture} \right)$ & $D_\mathrm{J} \left( \mathrm{P}_\mathrm{MSL}, \mathrm{P}_\mathrm{Mixture} \right)$ \\
      \midrule
      
      0.0 & 425.3 & 64.32 & 489.6 \\
      0.1 & 298.1 & 55.82 & 353.9 \\
      0.2 & 203.4 & 48.37 & 251.8 \\
      0.3 & 138.9 & 41.88 & 180.8 \\
      0.4 & 99.67 & 36.33 & 136.0 \\
      0.5 & 74.80 & 31.72 & 106.5 \\ 
      0.6 & 54.84 & 28.08 & 82.92 \\
      0.7 & 39.14 & 25.56 & 64.70 \\
      0.8 & 28.18 & 24.47 & 52.65 \\
      0.9 & 24.79 & 26.11 & 50.90 \\
      1.0 & 29.07 & 34.17 & 63.23

    \end{tabular}
  \end{center}
\end{table}

\begin{figure}
  \begin{subfigure}{0.49\linewidth}
    \centering
    \includegraphics[width=\linewidth, trim=15cm 0cm 5cm 0cm, clip]{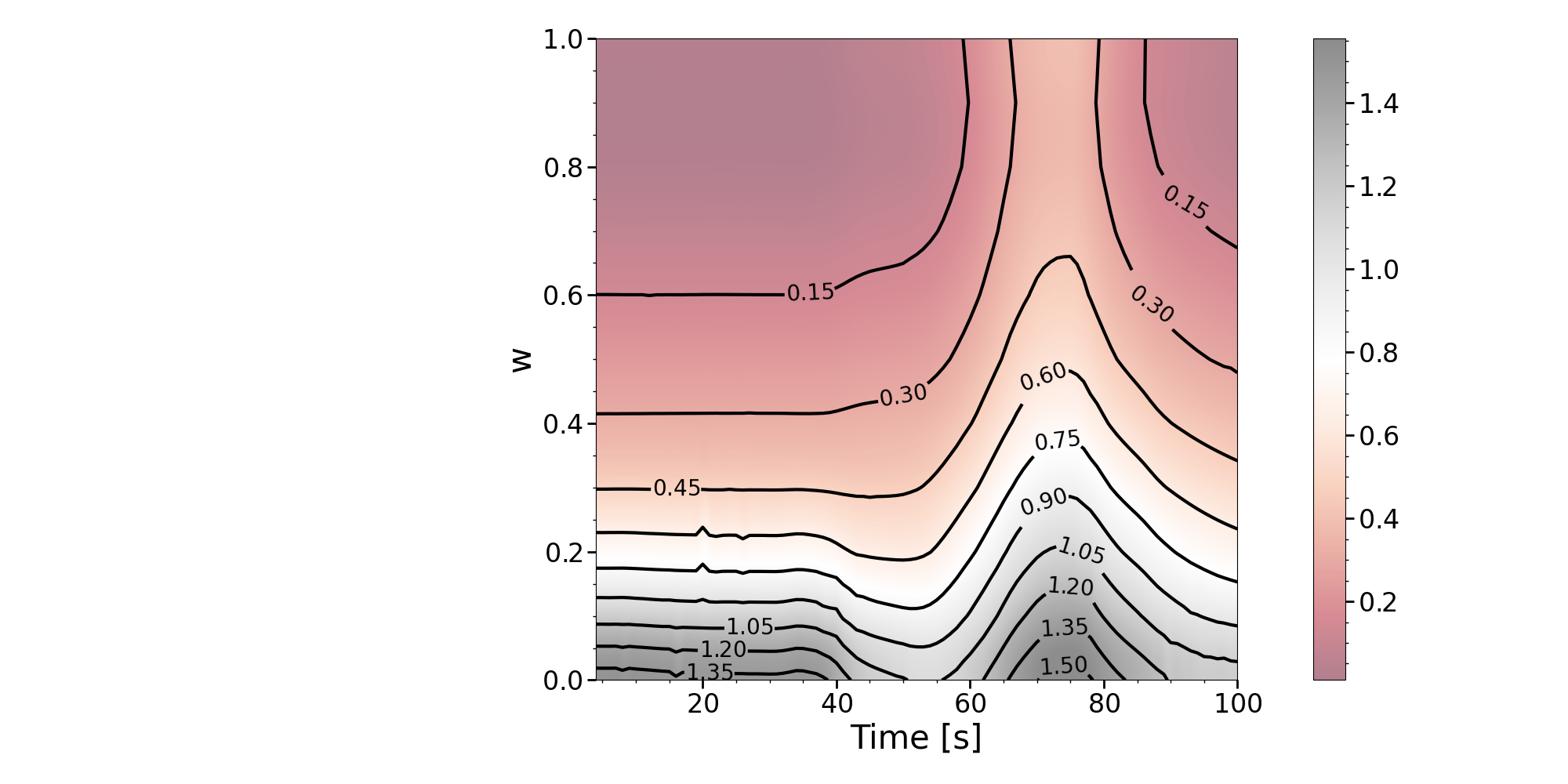}
    \caption{}
    \label{f:kl_contour_tc1}
  \end{subfigure}
  \hfill
  \begin{subfigure}{0.49\linewidth}
    \centering
    \includegraphics[width=\linewidth, trim=15cm 0cm 5cm 0cm, clip]{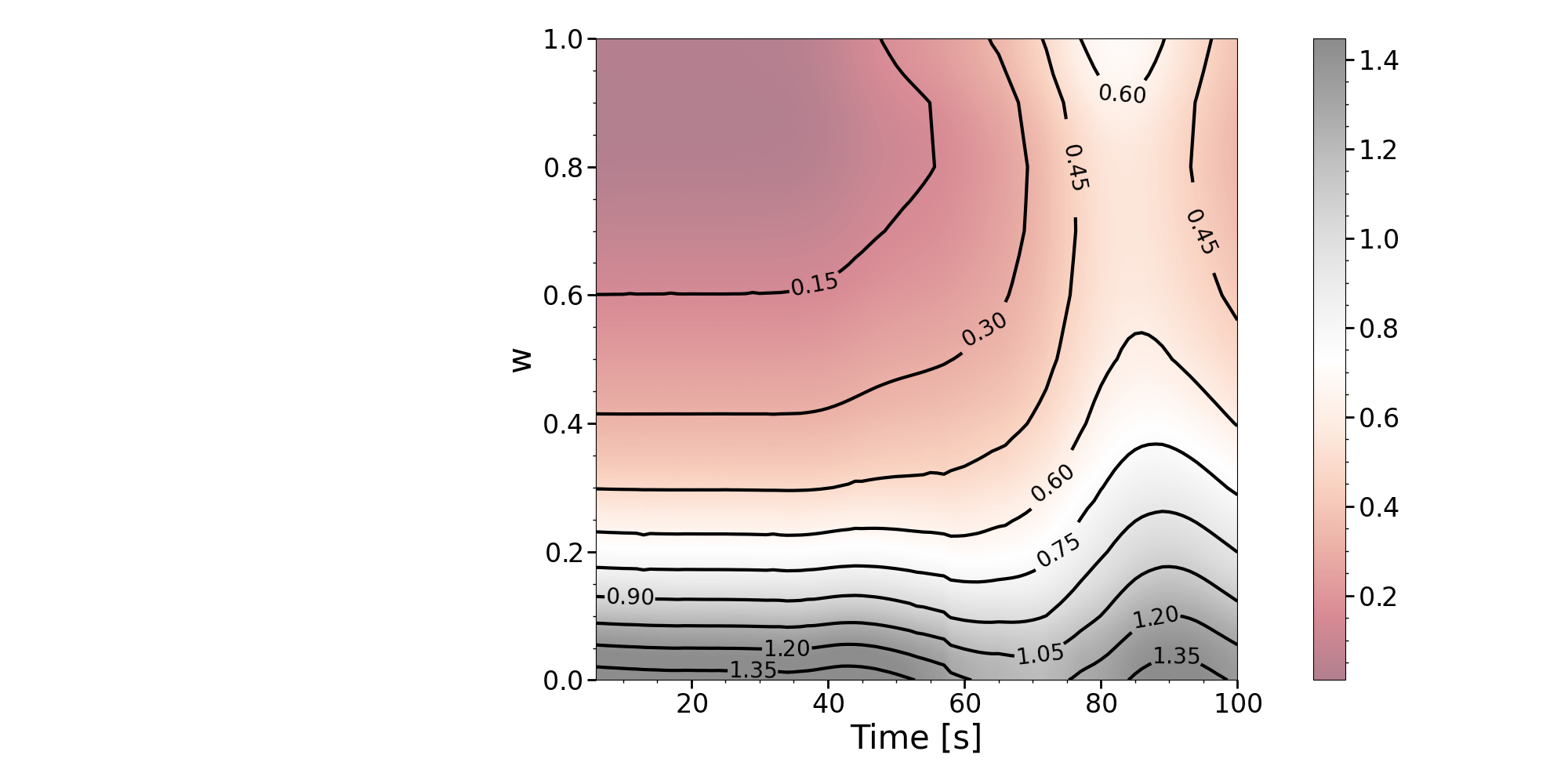}
    \caption{}
    \label{f:kl_contour_tc2}
  \end{subfigure}
  \medskip
  \begin{subfigure}{0.49\linewidth}
    \centering
    \includegraphics[width=\linewidth, trim=15cm 0cm 5cm 0cm, clip]{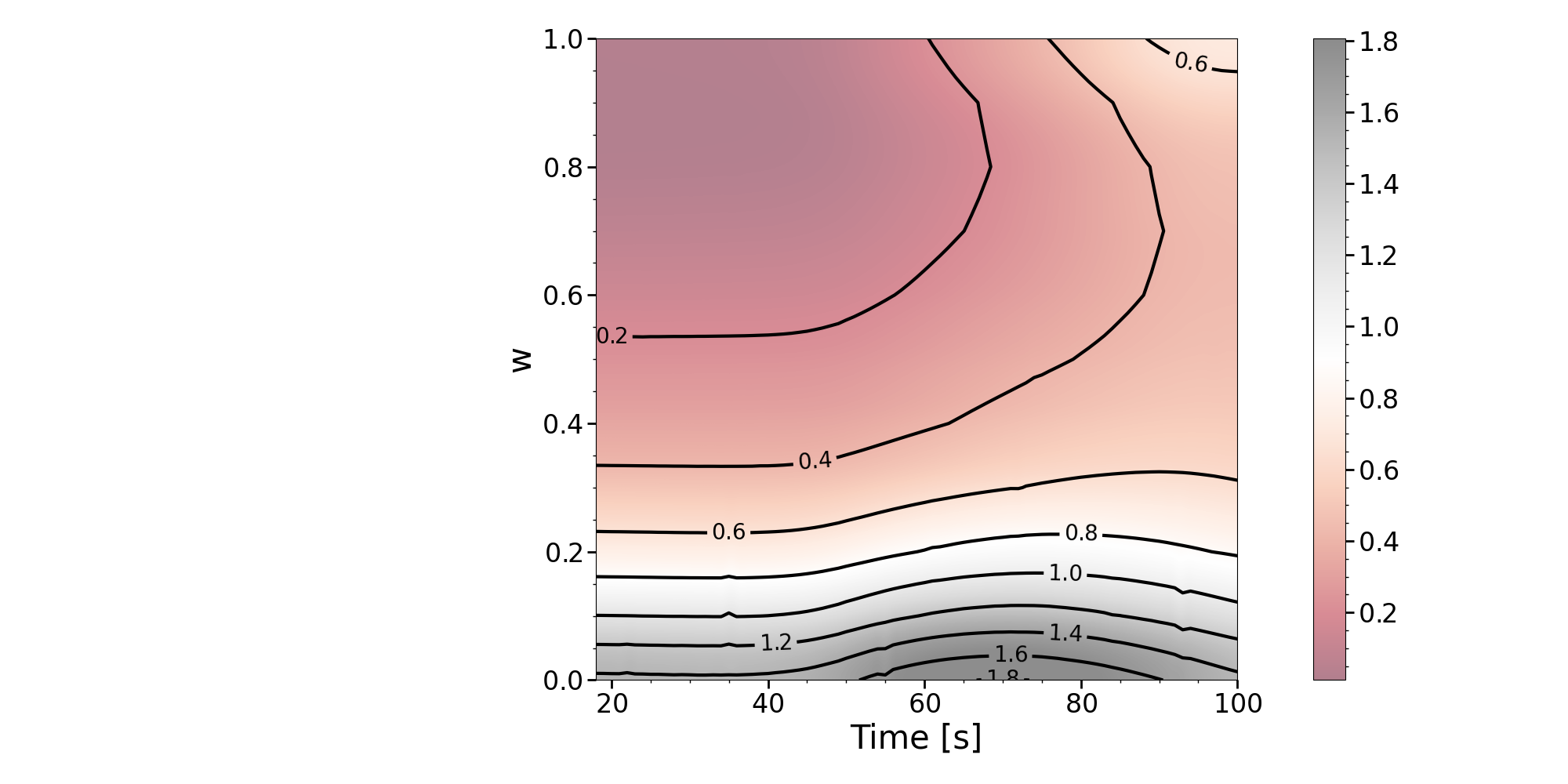}
    \caption{}
    \label{f:kl_contour_tc3}
  \end{subfigure}
  \hfill
  \begin{subfigure}{0.49\linewidth}
    \centering
    \includegraphics[width=\linewidth, trim=15cm 0cm 5cm 0cm, clip]{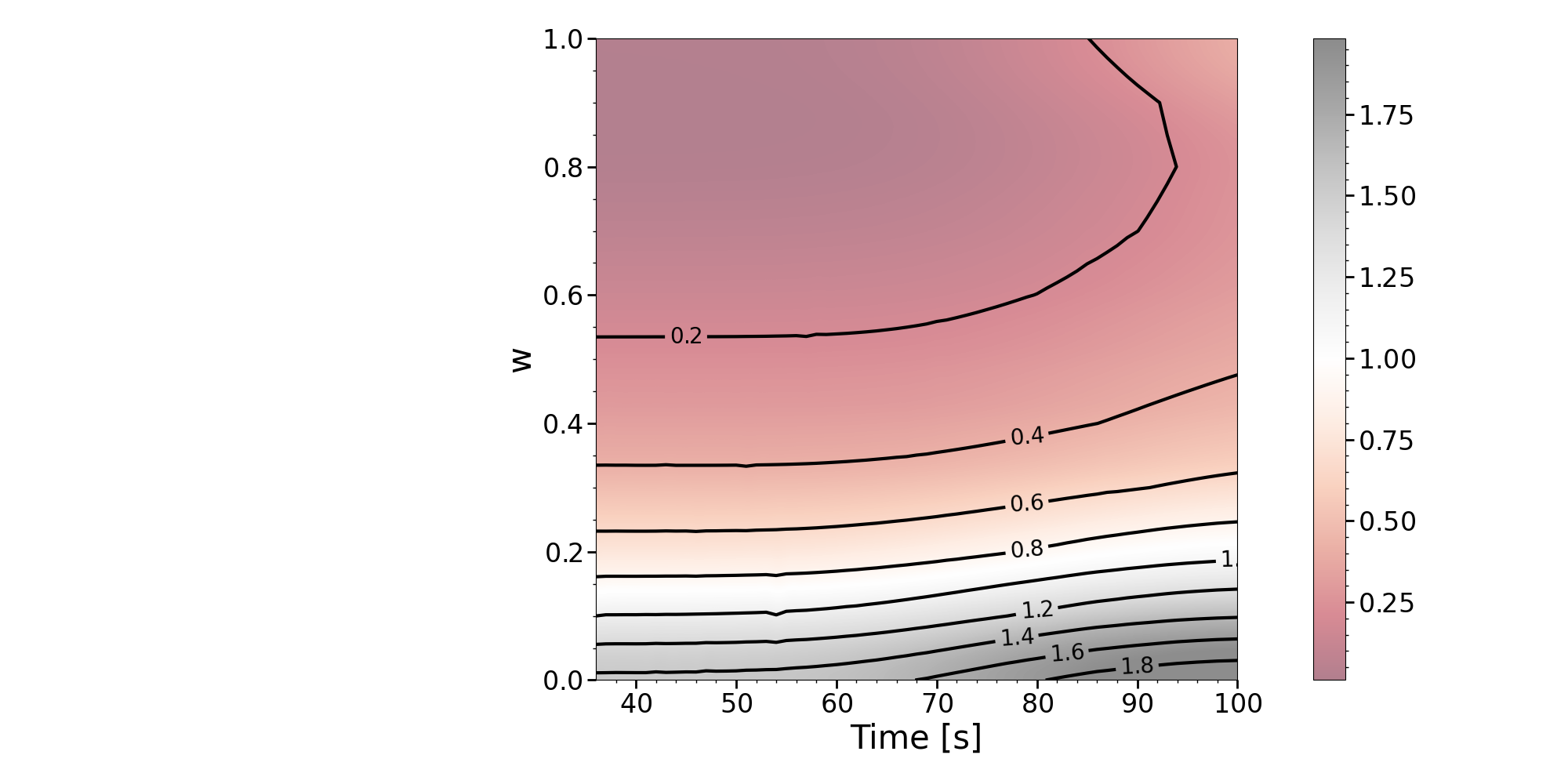}
    \caption{}
    \label{f:kl_contour_tc4}
  \end{subfigure}
  \caption{Jeffreys divergence contours as a function of time and mixing parameter $w$ value.}
  \label{f:kl_contour}
\end{figure}

Nevertheless, the value of $w$ that minimizes the difference between the predictive distribution associated with the mixture approach and calibration with MSL data of the emulator methodology is not immediately evident. Analysis of computed Jeffreys divergence also reveals non-trivial trends across the domain of material temperature response as a function of the mixture relevance parameter. The divergence measure was computed in increments of $\Delta w = 0.1$ in the $[0,1]$ range, and corresponding contour images are shown Figure \ref{f:kl_contour} of the results. It is seen that Jeffreys divergence behavior as a function of time contains complex features but attains the highest values for the predicted response of deeper thermocouples, and the transition between regions of small and large divergence values is generally observed to be somewhat gradual. Plots like those in Figure \ref{f:kl_contour} can be used in future efforts to identify component submodels that impose significant ground facility and flight data discrepancies in calibration exercises. The symmetric Jeffreys divergence balances the underestimation and overestimation of the spread. On the other hand, optimization based on the backward KL divergence value designates the predictive distribution obtained with the mixture prior as the reference distribution and yields conservative results that penalize heavy-tailed behavior to a lesser extent. Both of these options for determining relative entropy between two distributions are utilized in the procedure to explore best options for the value of the mixture hyperparameter.

\begin{figure}
  \begin{subfigure}{0.49\linewidth}
    \centering
    \includegraphics[width=\linewidth, trim=10cm 0cm 12cm 0cm, clip]{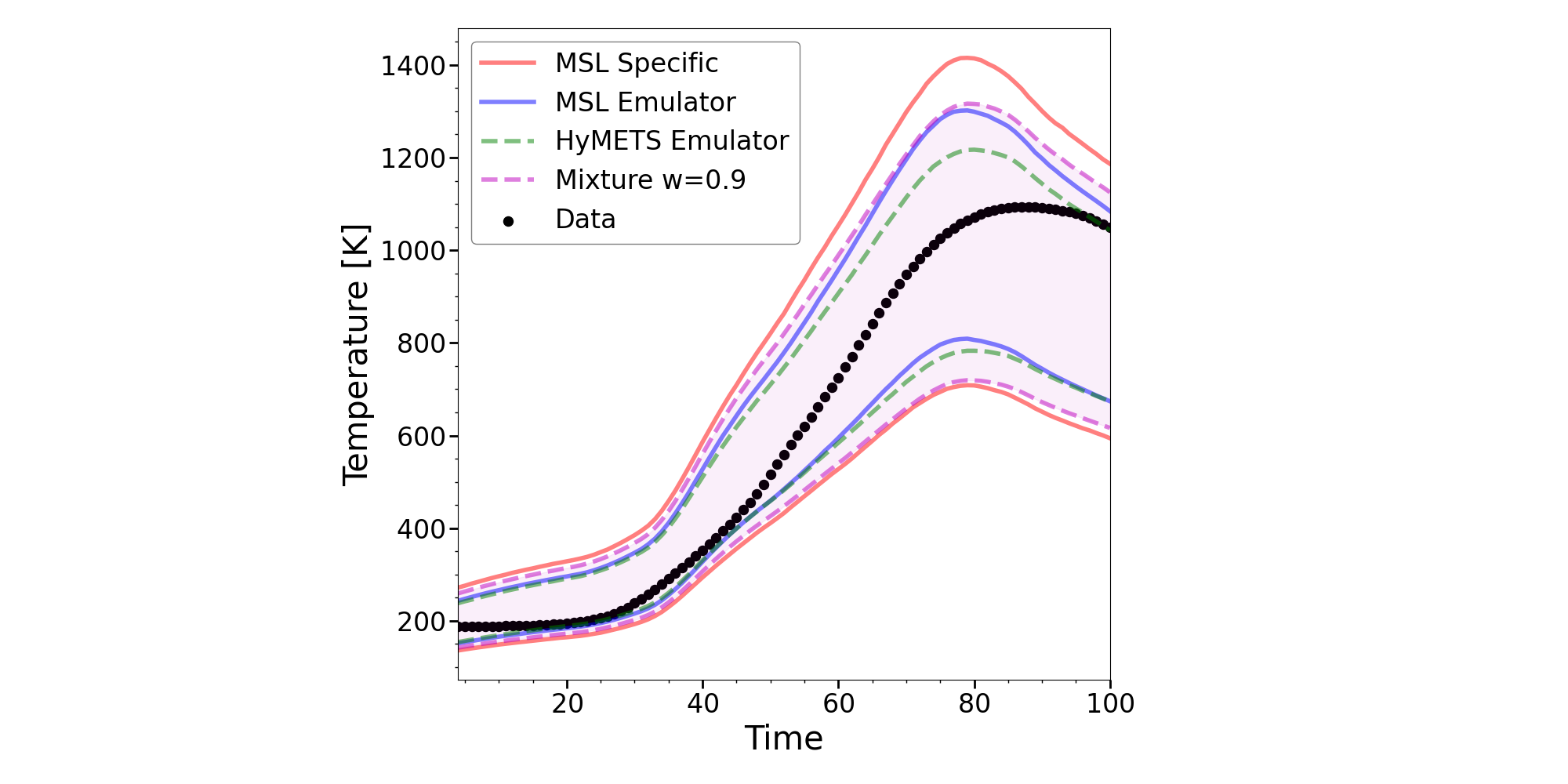}
    \caption{}
    \label{f:optimal_mixture_a0.99_w0.9_tc1}
  \end{subfigure}
  \hfill
  \begin{subfigure}{0.49\linewidth}
    \centering
    \includegraphics[width=\linewidth, trim=10cm 0cm 12cm 0cm, clip]{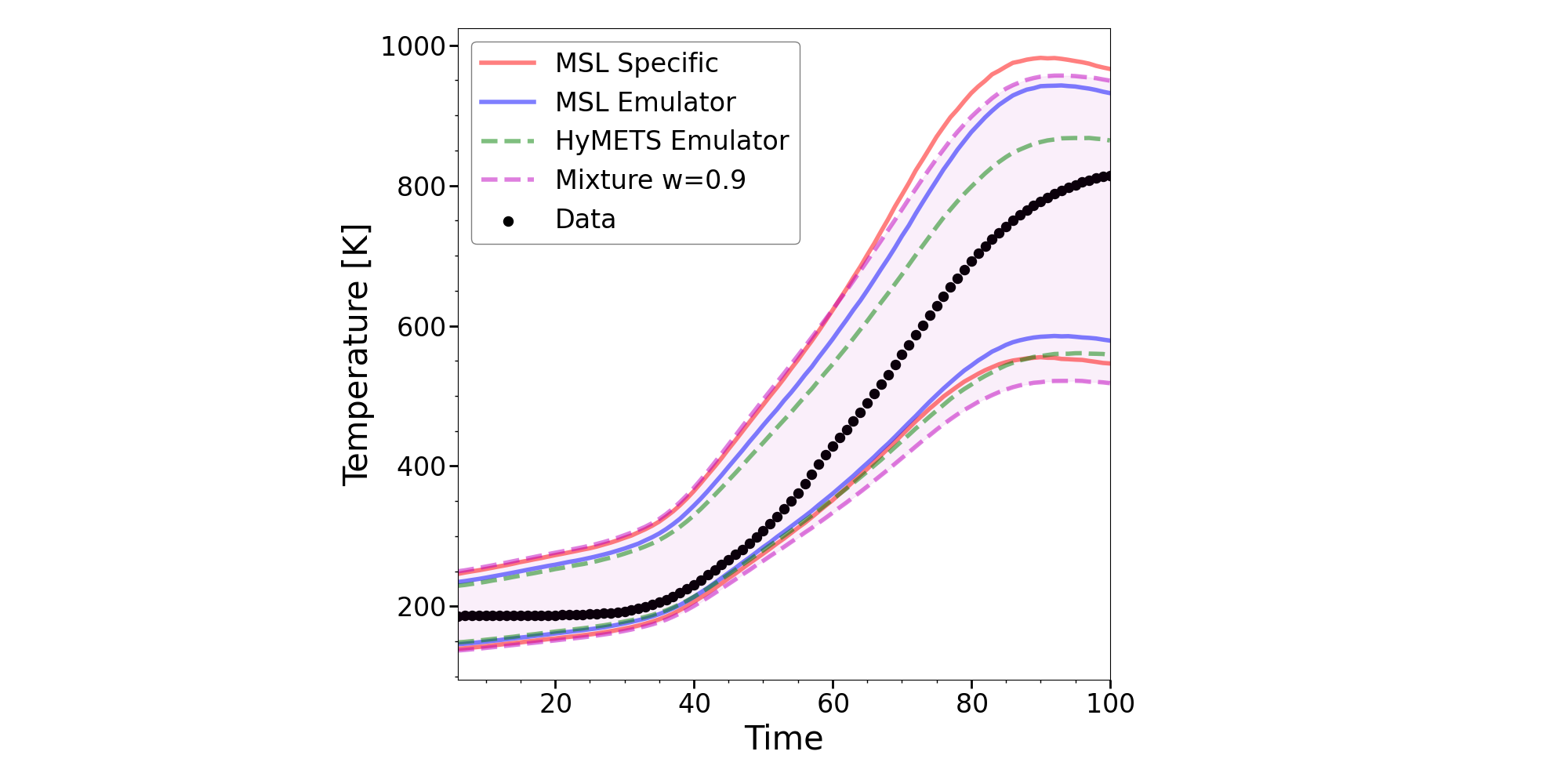}
    \caption{}
    \label{f:optimal_mixture_a0.99_w0.9_tc2}
  \end{subfigure}
  \medskip
  \begin{subfigure}{0.49\linewidth}
    \centering
    \includegraphics[width=\linewidth, trim=10cm 0cm 12cm 0cm, clip]{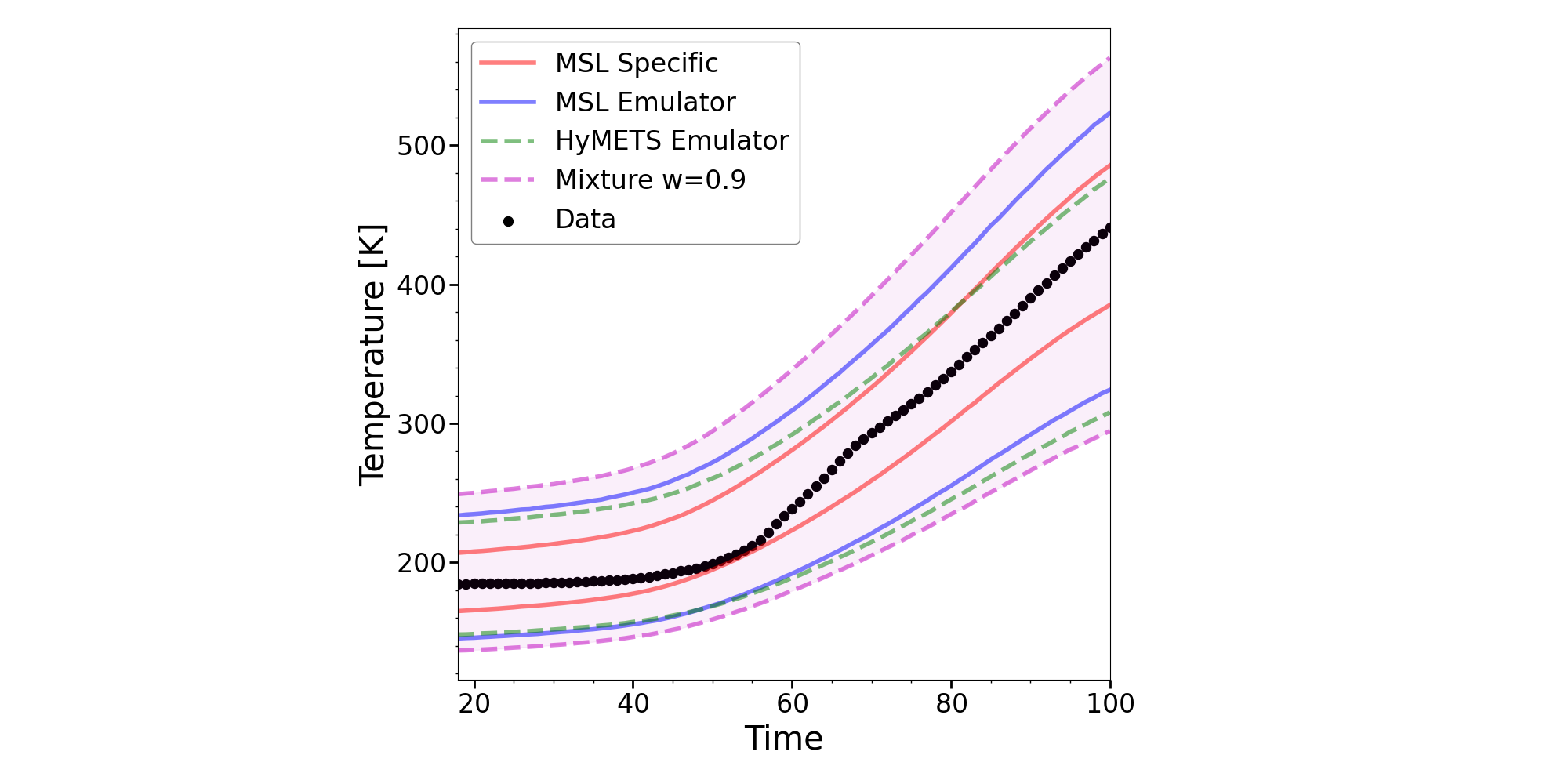}
    \caption{}
    \label{f:optimal_mixture_a0.99_w0.9_tc3}
  \end{subfigure}
  \hfill
  \begin{subfigure}{0.49\linewidth}
    \centering
    \includegraphics[width=\linewidth, trim=10cm 0cm 12cm 0cm, clip]{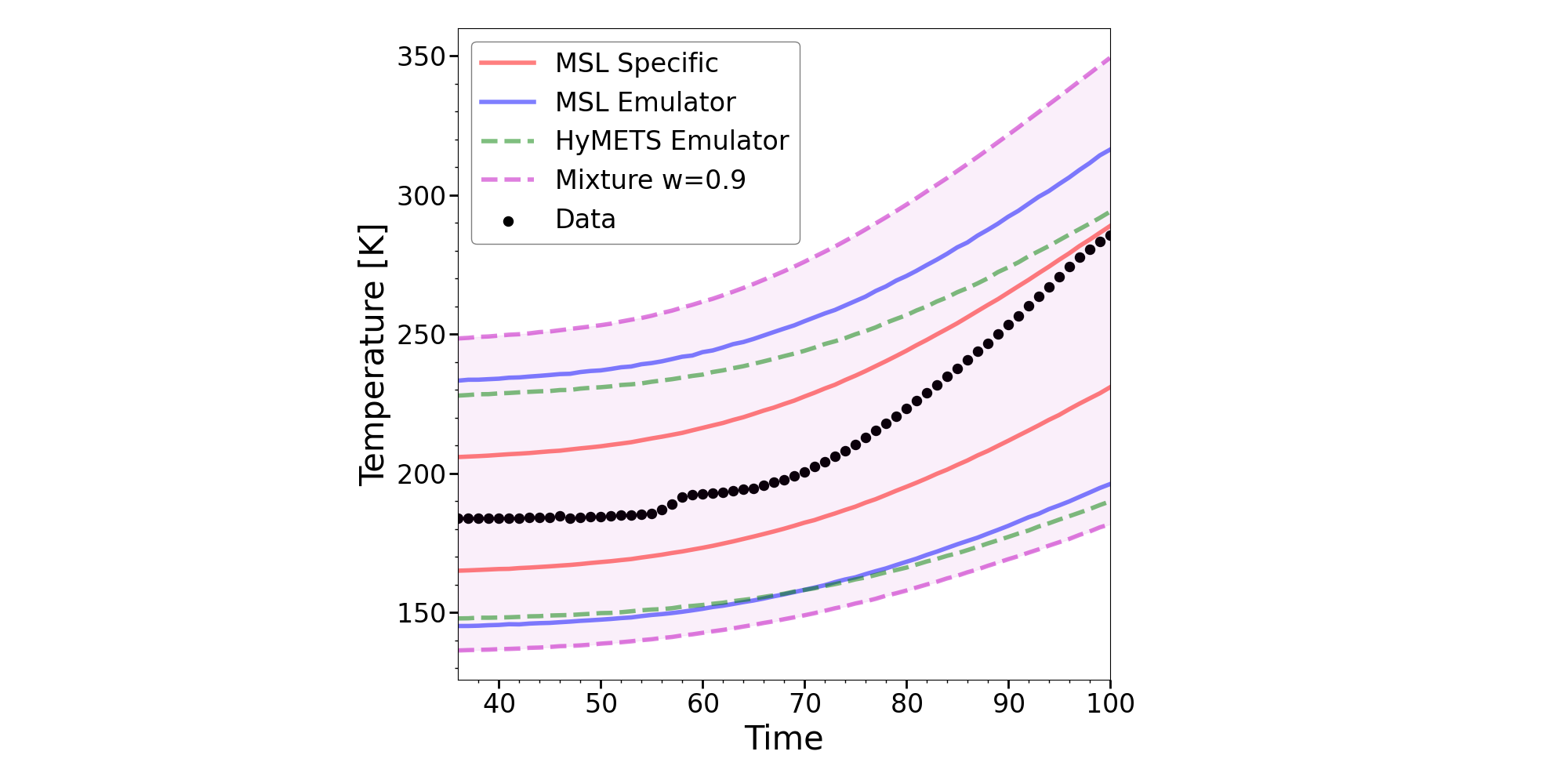}
    \caption{}
    \label{f:optimal_mixture_a0.99_w0.9_tc4}
  \end{subfigure}
  \caption{Comparison of 99\% prediction intervals corresponding to the predictive inference result with scenario specific likelihood, MSL emulator trained with MSL data, unmodified HyMETS data based emulator distribution forward propagated for the MSL scenario, and the prior predictive distribution given the mixture approach with w=0.9 for TC1-TC4 locations (a-d) within the MISP-4 assembly.}
  \label{f:optimal_mixture_a0.99_w0.9}
\end{figure}

Total divergence values for the field material temperature response are obtained by summing integrated values across the response domain of each thermocouple and are reported in Table \ref{t:total_divergence}. Predictive distributions $\mathrm{P}_\mathrm{MSL}$ and $\mathrm{P}_\mathrm{Mixture}$ in the table correspond to predictive inference results with the MSL emulator approach and HyMETS data-based mixture formulation for the MSL scenario. In addition to Jeffrey's divergence, both asymmetric options for KL divergence values following the same integration and summation schemes are included, where Jeffreys divergence measure balances the effects of both $D_\mathrm{KL}$ options. Consequently, the value of $w=0.9$ is seen to provide an overall best matching non-deterministic prediction based on the tabulated values compared to the MSL data-based emulator result. The mixture parameter value of $w=0.8$ alternatively yields the optimal conservative prediction and neither purely prior nor HyMETS calibration-based posterior options constitute the best choice. Mild obtained values of the optimal mixture input are partly due to the considerable reduction in variance in the posterior compared to the prior when model input quantities are calibrated with HyMETS calibration data. This behavior necessitates some degree of prior knowledge influence to attenuate effects of the ground facility and flight data discrepancies in calibration procedure outcomes as obtained prediction intervals are otherwise inadequate in estimating those derived with captured MSL data.

\begin{figure}
  \begin{subfigure}{0.49\linewidth}
    \centering
    \includegraphics[width=\linewidth, trim=10cm 0cm 12cm 0cm, clip]{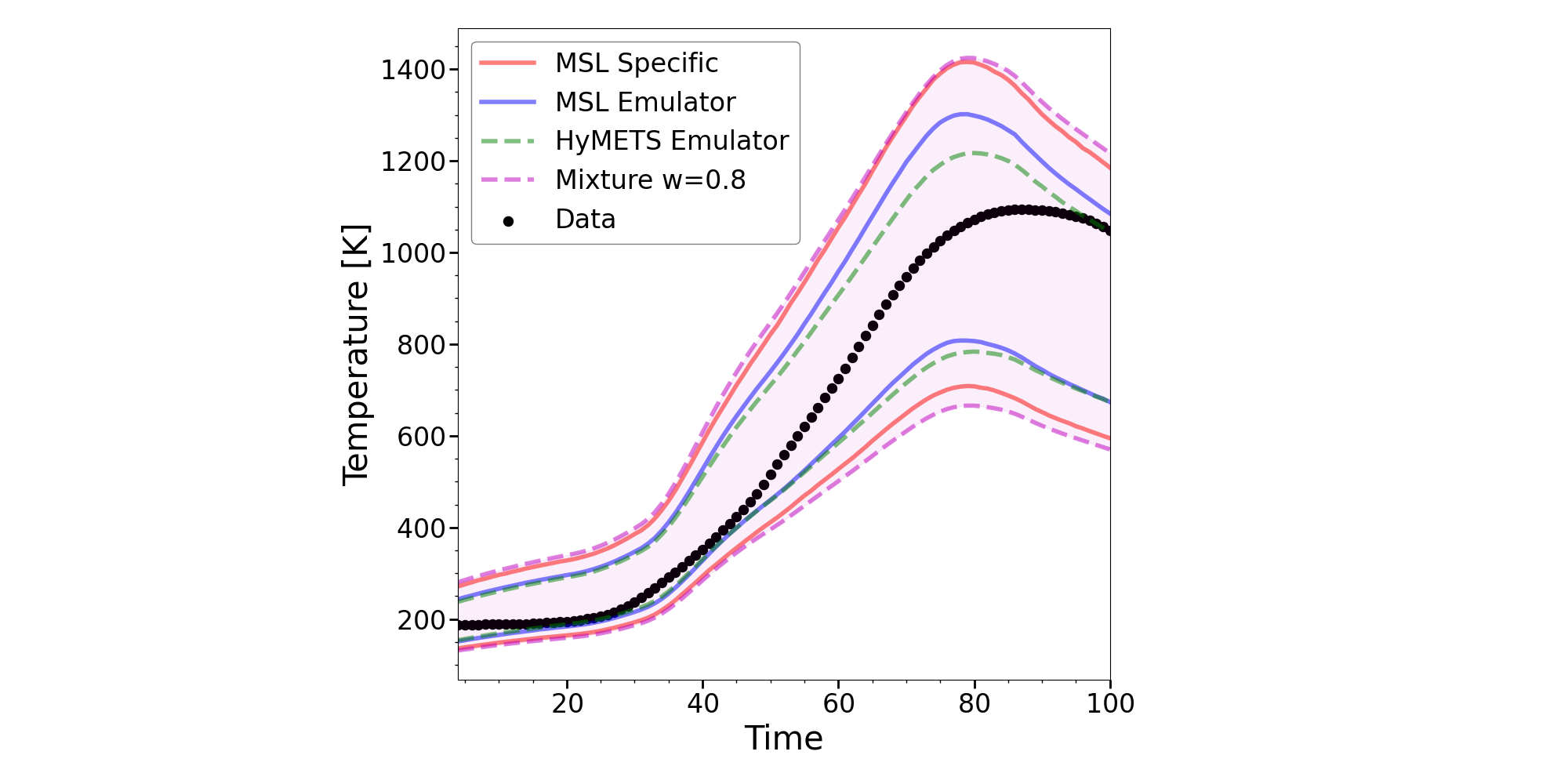}
    \caption{}
  \end{subfigure}
  \hfill
  \begin{subfigure}{0.49\linewidth}
    \centering
    \includegraphics[width=\linewidth, trim=10cm 0cm 12cm 0cm, clip]{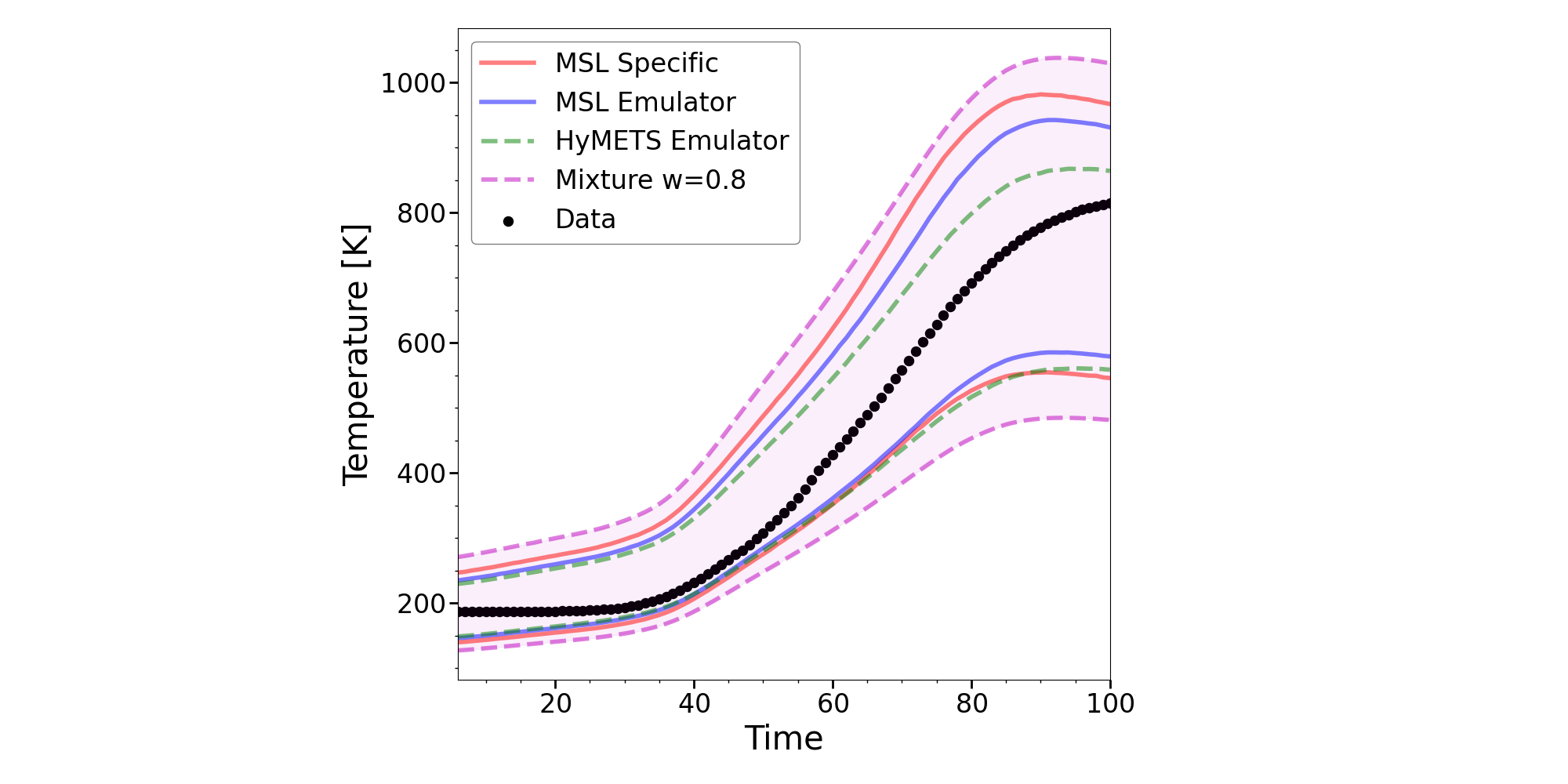}
    \caption{}
  \end{subfigure}
  \caption{Comparison of 99\% prediction intervals corresponding to the predictive inference result with scenario specific likelihood, MSL emulator trained with MSL data, unmodified HyMETS data based emulator distribution forward propagated for the MSL scenario, and the prior predictive distribution given the mixture approach with w=0.8 for TC1 (a) and TC2 (b) locations within the MISP-4 assembly.}
  \label{f:optimal_mixture_a0.99_w0.8}
\end{figure}

Optimal mixture results based on Jeffrey's divergence values are shown in Figure \ref{f:optimal_mixture_a0.99_w0.9}, where 99\% prediction intervals are plotted for all considered thermocouple location responses. These intervals are also compared to ones obtained previously from the posterior predictive distributions of Bayesian inferences exercises conducted with MSL data, using scenario-specific likelihood and the emulator approach, and those obtained from the forward propagated emulator-based solution trained with HyMETS data. It is immediately evident that the bounds obtained with the emulator approach and HyMETS data are insufficient in estimating uncertainty quantified with MSL data, where they are underestimated to a significant degree. The corresponding bounds also do not adequately encapsulate captured flight data nor the 99\% bounds obtained from the posterior predictive distribution associated with the Bayesian inference exercise conducted with the scenario-specific likelihood formulation. The performance of the emulator approach trained with HyMETS data improves upon previous results obtained by propagating solely parametric uncertainty quantified with ground facility data, but alone it is not sufficient in the present context for scenarios outside the testing envelope of ground facilities. The mixture approach furthermore yields bounds that not only capture the prediction intervals obtained when the inadequacy emulator model is calibrated with MSL data but also comes close to capturing the intervals obtained with the scenario specific likelihood characterized by model inadequacy that is allowed to vary between thermocouple locations. The more conservative option of $w=0.8$ can be seen in Figure \ref{f:optimal_mixture_a0.99_w0.8} to close the remaining gap between the two sets of prediction intervals for the foremost thermocouples.

\begin{figure}
  \begin{subfigure}{0.49\linewidth}
    \centering
    \includegraphics[width=\linewidth, trim=10cm 0cm 12cm 0cm, clip]{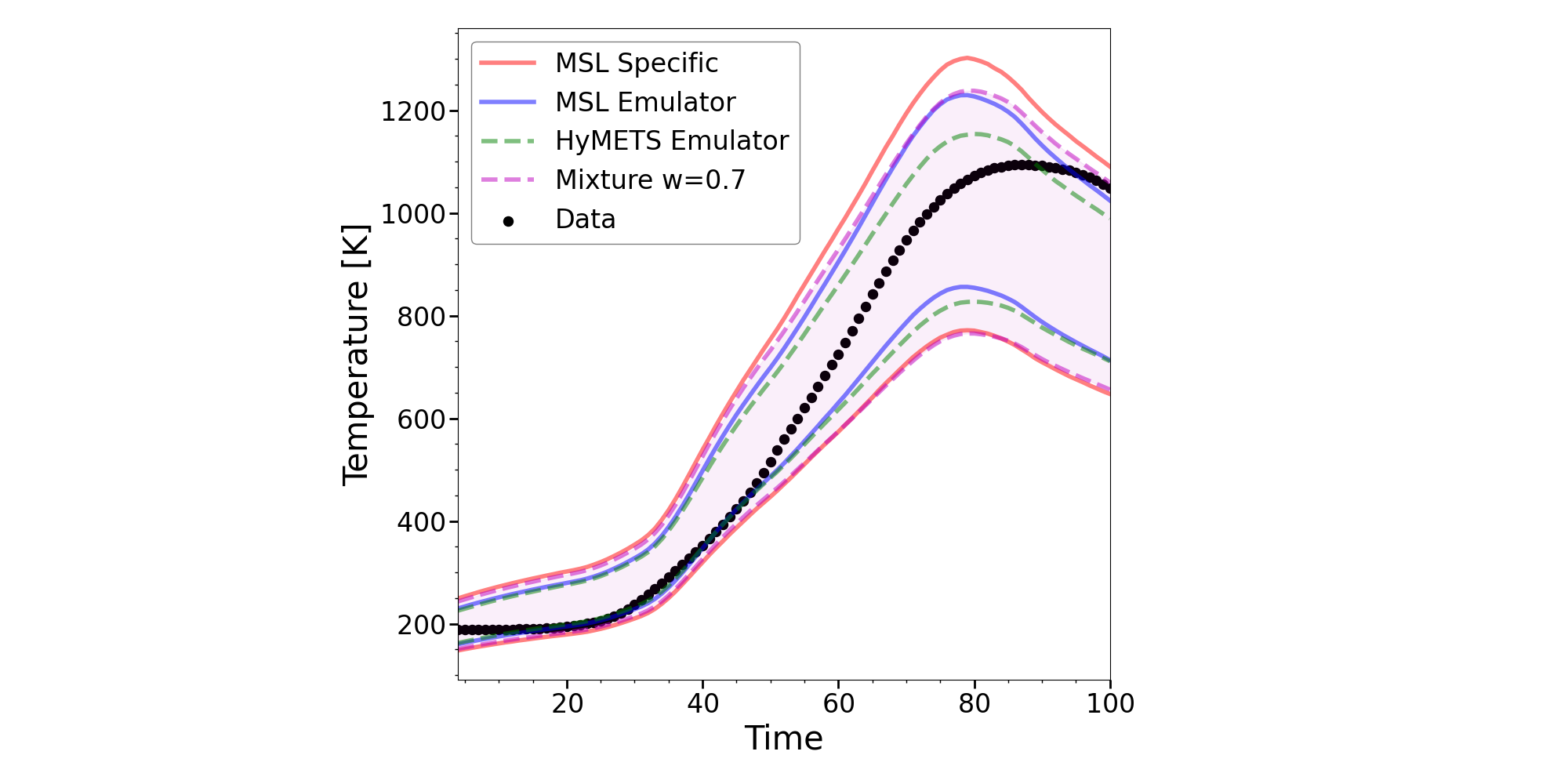}
    \caption{}
    \label{f:optimal_mixture_a0.95_w0.7_tc1}
  \end{subfigure}
  \hfill
  \begin{subfigure}{0.49\linewidth}
    \centering
    \includegraphics[width=\linewidth, trim=10cm 0cm 12cm 0cm, clip]{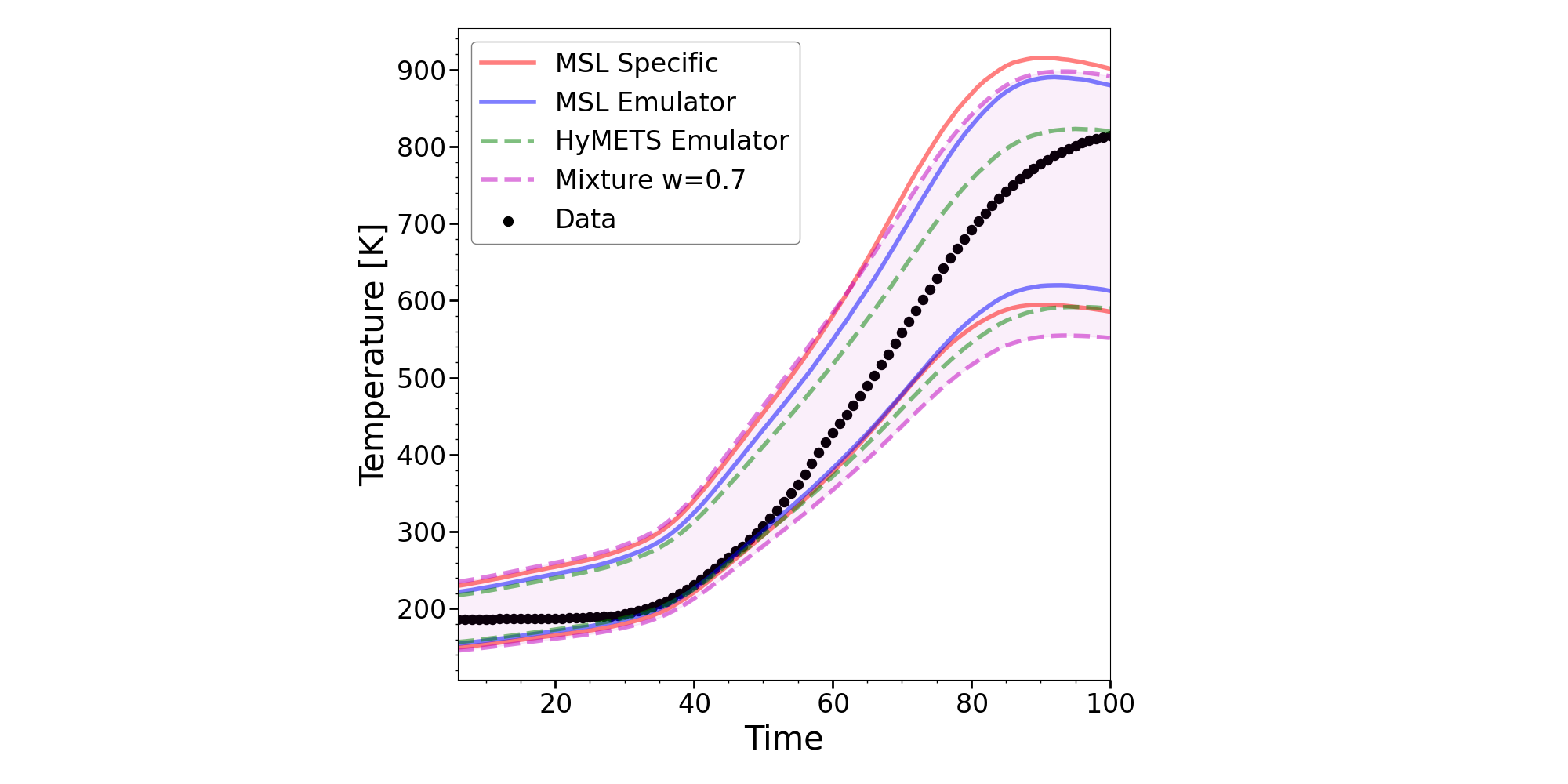}
    \caption{}
    \label{f:optimal_mixture_a0.95_w0.7_tc2}
  \end{subfigure}
  \medskip
  \begin{subfigure}{0.49\linewidth}
    \centering
    \includegraphics[width=\linewidth, trim=10cm 0cm 12cm 0cm, clip]{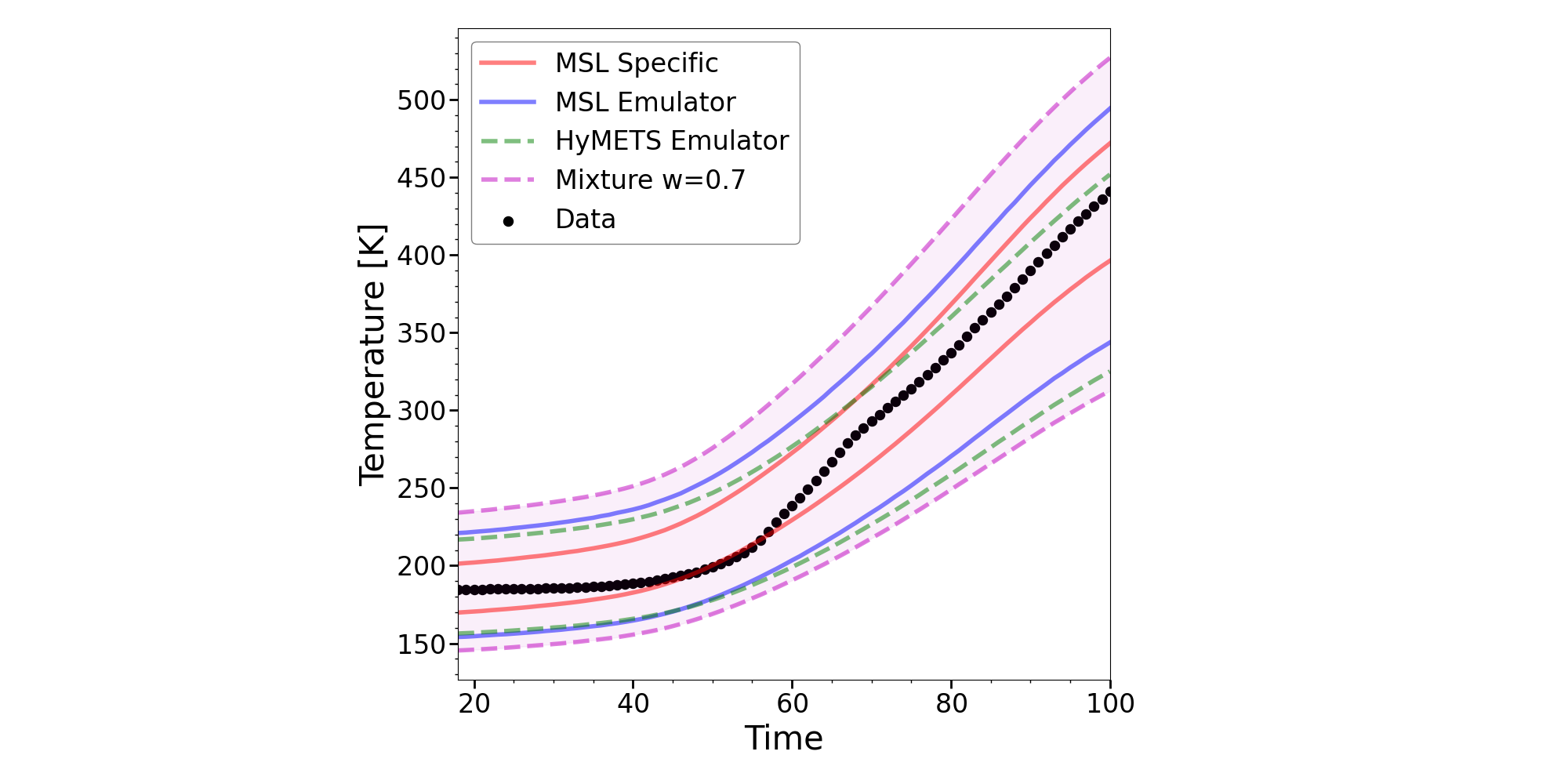}
    \caption{}
    \label{f:optimal_mixture_a0.95_w0.7_tc3}
  \end{subfigure}
  \hfill
  \begin{subfigure}{0.49\linewidth}
    \centering
    \includegraphics[width=\linewidth, trim=10cm 0cm 12cm 0cm, clip]{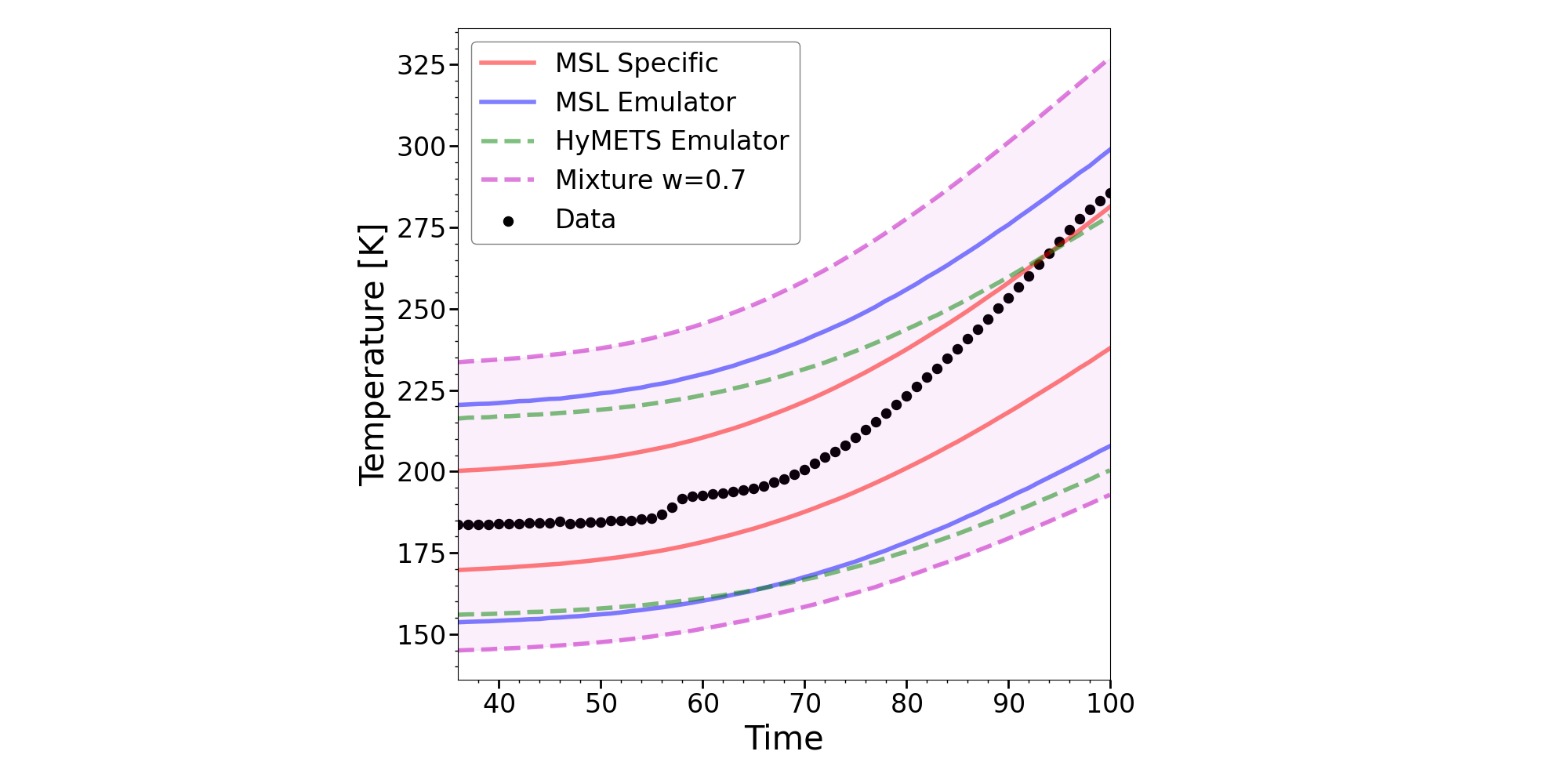}
    \caption{}
    \label{f:optimal_mixture_a0.95_w0.7_tc4}
  \end{subfigure}
  \caption{Comparison of 95\% prediction intervals corresponding to the predictive inference result with scenario specific likelihood, MSL emulator trained with MSL data, unmodified HyMETS data based emulator distribution forward propagated for the MSL scenario, and the prior predictive distribution given the mixture approach with w=0.7 for TC1-TC4 locations (a-d) within the MISP-4 assembly.}
  \label{f:optimal_mixture_a0.95_w0.7}
\end{figure}

However, it is imperative to consider the effects of the tail regions of the distributions under comparison on obtained Jeffrey's and KL divergence values and consequently on the obtained optimal value of $w$ in the present context. One likely scenario consists of the effects of the logarithmic term in the computation of the divergence in regions of the domain where the reference distribution is near-zero. At the same time, the value of the PDF of the other counterpart can decrease slowly and is non-negligible in those parts of the domain concerning the integrand but still small overall. The behavior in the distribution's tails in such a case would become dominant in the a priori determination of the mixing hyperparameter. These effects are especially relevant when kernel density estimators are utilized to approximate both distributions and effects of ``bumps'' in the tails, as some samples can reside far away from the mean and median distribution measures and can cause significant fluctuations in the value of the involved integrand. Under these circumstances, it may be worthwhile to consider computing divergence values of truncated distributions falling inside the union of 95\% PI of the two distributions. The two distributions would need to be rescaled to account for the truncation. This process was carried out following the same divergence computation procedure where an optimal value of $w=0.7$ was determined using both Jeffrey's divergence and conservative KL divergence values. Corresponding 95\% PI of the non-truncated distributions for all simulated thermocouple responses are shown in Figure \ref{f:optimal_mixture_a0.95_w0.7}, where the same intervals obtained with the emulator approach calibrated with MSL data are consistently captured. Prediction intervals originating from scenario-specific likelihood calibration are also predominantly captured and only underestimated slightly for the simulated TC1 response near the peak temperature. 

The mixture approach with the optimal $w$ determined using Jeffrey's divergence values in sum significantly outperforms the emulator approach alone. The optimal mixture yields 99\% and 95\% prediction bounds that consistently encapsulate the same level bounds associated with the emulator solution trained with MSL data and flight data across all thermocouple locations studied in this work. The mixture also captures 99\% and 95\% prediction intervals obtained from the scenario-specific inference procedure's posterior predictive solution while slightly underestimating the upper bound ranges for parts of the response domain for shallower thermocouple locations. However, as shown in Figure \ref{f:optimal_mixture_a0.99_w0.8}, using the more conservative results obtained with $D_\mathrm{KL} \left( \mathrm{P}_\mathrm{MSL} || \mathrm{P}_\mathrm{Mixture} \right)$ yields prediction intervals that capture the MSL-specific bounds completely without significant overcompensation. The mixture approach improves upon predictive capabilities based on these results in estimating uncertainty in flight predictions. The application of additional uncertainty in this methodology is constrained by both the information encoded in the prior and models implemented as part of the material response framework. The determined value of $w$ then sets the degree to which to balance information extracted from the ground facility scenario and the knowledge expressed through the prior.

\section{Conclusions}
\label{sec:conclusions}

A non-deterministic framework was presented in this work for use in the extension of the uncertainty of predictions made with ground facility data calibrated material response tools to flight conditions outside the testing envelope. The solution accounts for modeling and parametric sources of error and ground facility data discrepancy with performance measurements captured in flight. The introduced approach also trivializes the material model under study as a black-box process, increasing its generality to even non-related scientific fields where adequate estimation of uncertainty associated with prediction results is vital.

The undertaken development process included methodical sensitivity analysis and surrogate modeling studies before investigating the shortcomings of a straight-forward propagation of non-deterministic calibration results followed by the emulator and mixture-based solutions. In the forward propagation exercise of Bayesian inference results, only the parametric uncertainty could be propagated to the flight scenario where uncertainty in the non-deterministic response was inadequate. Predicted uncertainty bounds also failed to consistently capture flight measurements, resulting from the absence of contribution to total quantified uncertainty due to model inadequacy. This shortcoming was subsequently amended by a data-driven emulator approach, albeit a basic one, where the model inadequacy term was approximated by an emulator whose inputs included model output and additional tuning parameters. Removal of scenario-specific constraints for the emulator's inputs allowed for extension of model uncertainty contribution to environments other than those associated with calibration data. Non-deterministic material temperature predictions based on this addition to model output alone are better able to capture measured thermocouple data and uncertainty quantified through Bayesian inference exercises with flight data.

However, calibration of the emulator model with flight data revealed a non-negligible discrepancy in extracted information from ground facility measurements, which was later amended by applying the mixture approach detailed in this work. Based on a single applicability hyperparameter, the model averaging strategy mixes informative and non-informative components based on the relevance of ground facility data to flight conditions to construct a prior for the flight scenario. Flight measurements in this work were used solely to select an appropriate value of the weighting factor to show how proper selection of the parameter yields more reliable predictions for uncertainty quantification that take into account ground facility limitations. The value of this factor can alternatively be obtained by elicitation of field experts and applied to more disparate conditions between ground and flight scenarios in the present context like different gas mixtures, material coating presence, and others.

The present discussion utilized a single set of ground facility and flight measurements due to the limited availability of PICA performance data and aerothermal solutions for testing campaign environments inside plasma wind tunnel complexes. These limitations were accounted for by deliberate elementary formulation of the model inadequacy emulator term. Additional material performance data obtained using simulated environments inside ground complexes can be utilized to increase emulator term complexity to account for heteroscedastic phenomena and degree of char, among others. Moreover, additional flight data corresponding to distinctive scenarios can be utilized to study the optimal values for the mixing hyperparameter. The basic survey approach for the superior value of the mixture hyperparameter conducted in this work can also be further elaborated upon to yield increased performance of the epitomized framework.
\section*{Acknowledgements}

This work was supported by a NASA Space Technology Research Fellowship. The authors would like to acknowledge Dinesh Prabhu and Arnaud Borner for their efforts in obtaining solutions for the aerothermal environment of the MSL entry as well as Patricia Ventura Diaz for providing the HyMETS environment simulation results. In addition, the authors extend their gratitude to Dr. Ralph C. Smith for the invaluable input throughout the composition and derivation process of the results presented herein. Resources supporting this work were provided by the NASA High-End Computing (HEC) Program through the NASA Advanced Supercomputing (NAS) Division at Ames Research Center. Jeremie B.E. Meurisse was funded by NASA contract NNA15BB15C to Analytical Mechanics Associates (AMA), Inc.

\section*{Appendix}
\label{sec:appendix}

\renewcommand{\thesubsection}{\Alph{subsection}.}

\setcounter{table}{0}
\renewcommand{\thetable}{\Alph{subsection}.\arabic{table}}

\setcounter{figure}{0}
\renewcommand{\thefigure}{\Alph{subsection}.\arabic{figure}}

\setcounter{equation}{0}
\renewcommand{\theequation}{\Alph{subsection}.\arabic{equation}}

\subsection{Material Response}

\subsubsection*{Solution of Governing Equations}

The material response of the MSL heat shield and the PICA sample in the HyMETS facility are simulated using NASA's PATO platform. The numerical procedure of the toolbox consists of the solution of conservation equations for both solid and gas phases of the material. The decomposition of the material is further modeled using the phenomenological approach established by Goldstein \cite{goldstein:1965} that poses the advancement of pyrolysis reactions, and hence progression of the material from virgin state to char, using Arrhenius expressions concerning multiple decomposing sub-phases of the material. Details behind the derivation of the volume-averaged expressions and other embedded models can be found in Meurisee et al. \cite{meurisse:2018} and Lachaud et al. \cite{lachaud:2014, lachaud:2015, lachaud:2017}, as well as the specifics of the implementation of the analysis code. Dissemination of material models and properties of the PICA material is restricted. Instead, the efforts here are performed with the Theoretical Ablative Composite for Open Testing (TACOT) material database that provides a suitable analog for the restricted material \cite{vanetal:2014}.

As per the discussion posed in referenced literature, the recession of the material is simulated by solving surface mass and energy balance equations in combination with the B' table approach. The methodology requires the specification of the gas mixture composition generated through the pyrolysis process as well as the composition of the environment mixture. Corresponding mass fractions are listed in Table \ref{t:elemental_fractions}, where equilibrium properties of the HyMETS and MSL gas mixtures are modeled using the Johnston-Brandis' 10-species plus Argon model utilized by Diaz et al. \cite{diazetal:2021}. In turn, the mass fractions of the generated pyrolysis gas mixture are obtained assuming constant stoichiometric coefficients and 50\% of decomposing phenolic resin mass and all of H and O are converted to gas.

\begin{table}
 \begin{center}
    \caption{Elemental Mass Fractions Configurations Per Respective Gas Mixture.}
    \label{t:elemental_fractions}
    \begin{tabular}{lccccc}
      \toprule
      \toprule
      \multirow[c]{2}{*}{Gas} & \multicolumn{5}{c}{Element} \\
       & C & H & O & N & Ar \\
     \midrule
      Mars & 0.2640 & 0 & 0.7039 & 0.01742 & 0.01472 \\
      HyMETS Mars-like & 0.2419 & 0 & 0.6444 & 0.06637 & 0.04733 \\
      Pyrolysis & 0.5315 & 0.1285 & 0.3400 & 0 & 0 \\
    \end{tabular}
  \end{center}
\end{table}

\subsubsection*{Boundary Conditions}

The material response simulations for the MSL heat shield and HyMETS sample require the availability of several time-dependent quantities at the boundary. The boundary conditions for the former are those utilized by Meurisse et al. \cite{meurisse:2018} and those obtained by Borner et al. \cite{borner:2018} for earlier trajectory segments. The exact procedures with which these numerical results were obtained can be found in referenced literature and are limited to 100s following vehicle entry into the planetary atmosphere past the defined interface. Boundary conditions for the HyMETS simulations were originally obtained by Diaz et al. \cite{diazetal:2021} and detail the distribution of boundary condition quantities over the shape of the PICA material sample plug.

\subsubsection*{Grid Generation}

Simulation of the material temperature of the MSL heat shield is limited to the MISP-4 thermocouple assembly. Moreover, the simulations are performed on translating 1-D meshes obtained with the \textit{blockMesh} utility made available by \textit{OpenFOAM} consisting of 100 cells, with the cell expansion parameter value set to $r=0.1$ to emphasize the accuracy of the solution near the surface \cite{greenshields:2020}. This resolution was found to be adequate in previous work by the authors, where additional formulations for determining individual cell widths can also be found \cite{rostkowski:2022}. In contrast, the HyMETS scenario is simulated on a 2-D axisymmetric mesh detailed in Diaz et al. \cite{diazetal:2021} due to the prevalence of multidimensional phenomena in the response of the small material sample. The computational mesh is comprised of $80000$ cells with a maximum grid spacing of $10^{-5}$ m near the walls. In both cases, the recession of the material is simulated with a translating computational grid whose translation velocity is obtained as part of the B' table approach.

\setcounter{table}{0}
\renewcommand{\thetable}{\Alph{subsection}.\arabic{table}}

\setcounter{figure}{0}
\renewcommand{\thefigure}{\Alph{subsection}.\arabic{figure}}

\setcounter{equation}{0}
\renewcommand{\theequation}{\Alph{subsection}.\arabic{equation}}

\subsection{Uncertain Parameters}

The various embedded models implemented as part of the PATO toolbox introduced numerous parameters that can be included in an inversion process. A subset of selected model inputs is shown in Table \ref{t:uncertain_parameters} and includes parameters of the material decomposition model and those related to the effective transport of thermal energy in the material. In particular, the material's thermal conductivity scales with material temperature and is influenced by both solid conduction and radiative effects. The relationship of the parameter with temperature is modeled by the following polynomial expression
\begin{equation}
\label{e:k_relationship}
\mathrm{k}(\mathrm{T}) = \mathrm{k}_0 + \mathrm{k}_3 \mathrm{T}^3
\end{equation}
consisting of $\mathrm{k}_0$ and $\mathrm{k}_3$ inputs. The nominal values of these parameters are obtained by fitting the expression with standard values included in the TACOT description. The bounds of included uncertain model inputs were chosen to guarantee a significant variance in the sampled output. It is essential to acknowledge that this work focuses on the development of the extension procedure, which can then be applied in conjunction with restricted PICA models and property inputs for which uncertainty bounds are better known in contrast to the synthetic TACOT material. The MAP values of the TACOT model's parameters in this work are not intended for reliable predictions of real scenarios or to be used as part of TPS sizing studies.

\begin{table}
  \begin{center}
    \caption{Uncertain Inputs and Their Distributions.}
    \label{t:uncertain_parameters}
    \begin{tabular}{lccccc}
      \toprule
      \toprule
      Input & Units & Symbol & Distribution & Nominal Value & Calibrated \\
      \midrule
      Log Pre-exponential (2,1) & $\mathrm{s}^{-1}$ & $\log \mathrm{A}_{2,1}$ & U[6.908, 11.51] & 9.393 & \Checkmark \\
      Log Pre-exponential (2,2) & $\mathrm{s}^{-1}$ & $\log \mathrm{A}_{2,2}$ & U[16.12, 23.02] & 20.03 & \Checkmark \\
      Log Pre-exponential (2,3) & $\mathrm{s}^{-1}$ & $\log \mathrm{A}_{2,3}$ & U[16.12, 23.02] & 20.03 & \\
      \multirow[t]{2}{*}{Virgin Conductivity} & $\mathrm{W} \ \mathrm{m}^{-1} \ \mathrm{K}^{-1}$ & $\mathrm{k}_{0,t,v}$ & N[0.2294, 0.03825] & 0.2294 & \Checkmark \\
      &  & $\mathrm{k}_{3,t,v}$ & N[1.694e-11, 2.823e-12] & 1.694e-11 & \\
      \multirow[t]{2}{*}{Char Conductivity} & $\mathrm{W} \ \mathrm{m}^{-1} \ \mathrm{K}^{-1}$ & $\mathrm{k}_{0,t,c}$ & N[0.2569, 0.04283] & 0.2569 & \Checkmark \\
      &  & $\mathrm{k}_{3,t,c}$ & N[4.510e-11, 7.515e-12] & 4.510e-11 & \Checkmark \\
    \end{tabular}
  \end{center}
\end{table}
 
\clearpage

\printbibliography

\end{document}